\documentclass[paper]{JHEP3} 

\usepackage{amsfonts}
\usepackage{amsmath}
\usepackage{epsfig}
\usepackage{latexsym}
\usepackage{graphicx}
\usepackage{framed}

\usepackage{amssymb}
\numberwithin{equation}{section}

\def\be{\begin{equation}}
\def\ee{\end{equation}}
\def\bea{\begin{eqnarray}}
\def\eea{\end{eqnarray}}
\def\bequ{\begin{equation}}
\def\eequ{\end{equation}}

\renewcommand{\thefootnote}{\fnsymbol{footnote}}

\newcommand{\eq} {equation}
\newcommand{\eqa} {eqnarray}
\newcommand{\NN} {\mbox {$\nonumber$}}

\def\dag{\dagger}

\def\sitarel#1#2{\mathrel{\mathop{\kern0pt #1}\limits_{#2}}}

\title{On Perturbation theory improved by Strong coupling expansion}

\author{\large Masazumi Honda 
\vspace*{0.5cm} \\
Harish-Chandra Research Institute,\\
Chhatnag Road, Jhusi, Allahabad 211019, India\\
\vspace*{0.5cm} \\
\email{masazumihonda@hri.res.in}}

\preprint{HRI/ST/1411}

\abstract{
In theoretical physics,
we sometimes have two perturbative expansions of physical quantity 
around different two points in parameter space.
In terms of the two perturbative expansions,
we introduce a new type of smooth interpolating function consistent with the both expansions,
which includes the standard Pad\'e approximant and fractional power of polynomial method 
constructed by Sen as special cases.
We point out that
we can construct enormous number of such interpolating functions in principle
while the "best" approximation for the exact answer of the physical quantity
should be unique among the interpolating functions.
We propose a criterion to determine the "best" interpolating function,
which is applicable except some situations even if we do not know the exact answer.
It turns out that our criterion works 
for various examples including specific heat in two-dimensional Ising model,
average plaquette in four-dimensional $SU(3)$ pure Yang-Mills theory on lattice
and free energy in $c=1$ string theory at self-dual radius.
We also mention possible applications of the interpolating functions
to system with phase transition.
}


\begin{document}
\setcounter{footnote}{0}
\renewcommand{\thefootnote}{\arabic{footnote}}
\allowdisplaybreaks

\section{Introduction}
\label{sec:intro}
Perturbative expansion\footnote{
Throughout this paper, by ``perturbative expansion",
we mean power series expansion of a function around a point in parameter space.
} is ubiquitous tool in theoretical physics.
It is widely known that
perturbative expansion does not often give satisfactory understanding of physics.
Indeed we encounter non-convergent\footnote{
Strictly speaking, 
perturbative expansion would be sometimes asymptotic but Borel summable.
Then Borel resummation often gives sufficient understanding of physics.
} perturbative series in many situations. 
For example 
weak coupling perturbation theory in quantum field theory typically yields asymptotic series.
Even if perturbative expansion has a nice property,
computing higher order coefficients of the expansion is usually difficult task.

Sometimes
we have two perturbative expansions of physical quantity around different two points
in parameter space, e.g., 
in theory with S-duality, 
lattice gauge theory with weak and strong coupling expansions,
field theory with gravity dual, and 
statistical system with low and high temperature expansions, etc. 
It is nice 
if the two perturbative expansions give useful information 
on the physical quantity in intermediate region between these two points.  
As we will argue,
we can indeed construct smooth interpolating functions of the two expansions.
The interpolating functions
are consistent with the both expansions up to some orders.
One of the standard approaches is Pad\'e approximant,
which is a rational function giving the both expansions around the two points.
Recently Sen has also constructed another type of interpolating function \cite{Sen:2013oza},
which is described by fractional power of polynomial (FPP) consistent with the two expansions.
This method has been applied to S-duality improvement of string perturbation theory \cite{Sen:2013oza,Pius:2013tla}.
See also similar considerations\footnote{
In similar spirits,
the Pad\'e approximant has been applied to negative eigenvalues of the Schwarzschild black hole \cite{Asnin:2007rw} and various quantities in the $\mathcal{N}=4$ super Yang-Mills theory \cite{Banks:2013nga}.
See also application of another type of interpolating function \cite{Kleinert:2001ax} to $O(N)$ non-linear sigma model. 
} \cite{Beem:2013hha,Alday:2013bha} in $\mathcal{N}=4$ super Yang-Mills theory.
It has turned out that 
such interpolating functions usually give more precise approximation of the physical quantity
than each perturbative series in intermediate regime.
Although this situation is quite nice as first attempts,
it has been somewhat unclear 
{\it when the interpolating functions give good approximations}.

In this paper, we address this question.
First we introduce a new type of smooth interpolating function, 
which includes the Pad\'e approximant and FPP as special cases.
This interpolating function is described by ``fractional power of rational function" (FPR).
Then we argue that
we can construct enormous number of interpolating functions in principle, 
which give the same two perturbative expansions up to some orders. 
We point out that this fact leads us to ``landscape problem of interpolating functions".
Namely,
while the ``best" approximation of the physical quantity should be unique among the interpolating functions,
when we do not know the exact result,
it is unclear 
{\it which interpolating function gives the best answer}.
Finally we propose a criterion to determine the ``best" interpolation
in terms of the both expansions.
We expect that
this criterion is applicable except some situations 
even if we do not know the exact answer.
We also explicitly test our criterion  
in various examples including two-dimensional Ising model,
four-dimensional $SU(3)$ pure Yang-Mills theory on lattice
and $c=1$ string theory at self-dual radius.

The rest of this paper is organized as follows.
In section \ref{sec:interpolate},
we introduce some types of interpolating functions.
After review of the Pad\'e and FPP,
we argue that 
they are special cases of more general interpolating function (FPR). 
In the last of section \ref{sec:interpolate},
we demonstrate ``the landscape problem of interpolating functions" in a simple example.
In section \ref{sec:criterion},
we propose the criterion to determine the ``best" interpolating function.
We also mention some limitations of approximation by interpolating functions and our criterion.
In section \ref{sec:examples}, we explicitly check our criterion in various examples. 
As a result  
our criterion works 
for specific heat in the 2d Ising model,
average plaquette in the 4d $SU(3)$ pure Yang-Mills theory on lattice
and free energy in the $c=1$ string theory.
We also point out the possibility that
interpolating functions would be useful for searching critical point in discrete system.
Section \ref{sec:discussions} is devoted to conclusions and discussions.

\section{Interpolating functions}
\label{sec:interpolate}
In this section we describe some types of interpolating functions,
which are smooth, and consistent 
with the two expansions around different two points in parameter space.
After we briefly review the standard Pad\'e approximant  and
fractional power of polynomial method (FPP) previously constructed by Sen \cite{Sen:2013oza},
we introduce more general interpolating function 
including the Pad\'e and FPP as special cases.
In the last subsection,
we point out ``the landscape problem of interpolating functions" 
in terms of a simple but nontrivial example.

Suppose a function $F(g)$, 
which has the small-$g$ expansion $F_s^{(N_s )}(g)$ around $g=0$
and large-$g$ expansion $F_l^{(N_l )}(g)$ around $g=\infty$ taking the forms 
\begin{\eq}
F_s^{(N_s )}(g) = g^a \sum_{k=0}^{N_s} s_k g^k ,\quad
F_l^{(N_l )}(g) = g^b \sum_{k=0}^{N_l} l_k g^{-k} .
\label{eq:asymptotics}
\end{\eq}
Then we naively expect
\begin{\eq}
 F(g) = F_s^{(N_s )}(g) +\mathcal{O}(g^{a+N_s +1}) = F_l^{(N_l )}(g) +\mathcal{O}(g^{b-N_l -1}) .
\end{\eq}
In terms of the both expansions, 
we would like to construct smooth interpolating function\footnote{
One might have two perturbative expansions around $g=g_1$ and $g=g_2$ with $g_2 >g_1$,
and would like to study interpolating problem between $g_1$ and $g_2$.
Then if we perform a change of variable, for example, such as $x=(g-g_1)/(g_2 -g)$,
then these expansions are reduced to the small-$x$ and large-$x$ expansions.
Hence our setup does not lose generality.
},
which coincides with the small-$g$ and large-$g$ expansions up to some orders.

\subsection{Pad\'e approximant}
Let us construct the Pad\'e approximant $\mathcal{P}_{m,n}(g)$ 
with $m\leq N_s$ and $n\leq N_l$,
which realizes the small-$g$ and large-$g$ expansions up to $\mathcal{O}(g^{a+m+1})$ and $\mathcal{O}(g^{b-n-1})$, respectively.  
The Pad\'e approximant for $b-a \in \mathbb{Z}$ is given by
\begin{\eq}
\mathcal{P}_{m,n}(g)
= s_0 g^a \frac{ 1 +\sum_{k=1}^p c_k g^k}{1 +\sum_{k=1}^q d_k g^k } ,
\label{eq:Pade}
\end{\eq}
where 
\begin{\eq}
p = \frac{m+n+1 +(b-a)}{2}  ,\quad q = \frac{m+n+1 -(b-a)}{2}  .
\end{\eq}
We determine $c_k$ and $d_k$ such that
power series expansions of $\mathcal{P}_{m,n}(g)$ around $g=0$ and $g=\infty$ agree with
the small-$g$ and large-$g$ expansions up to $\mathcal{O}(g^{a+m+1})$ and $\mathcal{O}(g^{b-n-1})$, respectively. 
By construction the Pad\'e approximant satisfies $F(g) = \mathcal{P}_{m,n}(g) +\mathcal{O}(g^{a+m +1}, g^{b-n -1})$.

Note that this approach needs
\begin{\eq}
\frac{m+n-1+b-a}{2} \in \mathbb{Z} .
\label{eq:Pade_constraint}
\end{\eq}
Sometimes the denominator in \eqref{eq:Pade} becomes zero in region of interest
and the Pad\'e approximant has poles.
This situation signals limitation of approximation by the Pad\'e unless $F(g)$ has poles.
This method has been applied to the negative eigenvalue of the Schwarzschild black hole \cite{Asnin:2007rw} and various quantities 
in the 4d $\mathcal{N}=4$ super Yang-Mills theory \cite{Banks:2013nga}.

\subsection{Fractional Power of Polynomial method}
Recently Sen has constructed \cite{Sen:2013oza} a new type of interpolating function, 
which we call fractional power of polynomial method (FPP),
given by
\begin{\eq}
F_{m,n}(g)
= s_0 g^a \Biggl[ 1 +\sum_{k=1}^m c_k g^k +\sum_{k=0}^n d_k g^{m+n+1 -k} \Biggr]^{\frac{b-a}{m+n+1}} .
\label{eq:FPP}
\end{\eq}
Here the coefficients $c_k$ and $d_k$ are determined 
as in the Pad\'e approximant.
The FPP satisfies again $F(g) = F_{m,n}(g) +\mathcal{O}(g^{a+m +1}, g^{b-n -1})$ by its definition.

Note that the FPP does not have constraint such as \eqref{eq:Pade_constraint} in the Pad\'e approximant.
We sometimes encounter that
the polynomial in the parenthesis of \eqref{eq:FPP} becomes negative 
in region of interest.
Then, when the power $(b-a)/(m+n+1)$ is not integer, 
the FPP takes complex value and signals breaking of approximation.
This method has been applied to S-duality improvement of string perturbation theory \cite{Sen:2013oza,Pius:2013tla}.
See also similar considerations in the 4d $\mathcal{N}=4$ super Yang-Mills theory \cite{Beem:2013hha,Alday:2013bha}.

\subsection{Fractional Power of Rational function method}
The Pad\'e and FPP are special cases of
the following interpolating function
\begin{\eq}
F_{m,n}^{(\alpha )} (g)
= s_0 g^a \Biggl[ \frac{ 1 +\sum_{k=1}^p c_k g^k}{1 +\sum_{k=1}^q d_k g^k } \Biggr]^\alpha ,
\label{eq:FPR}
\end{\eq}
where 
\begin{\eq}
p = \frac{1}{2} \left( m+n+1 -\frac{a-b}{\alpha} \right) ,\quad
q = \frac{1}{2} \left( m+n+1 +\frac{a-b}{\alpha} \right) .
\end{\eq}
Here we determine $c_k$ and $d_k$ as in the Pad\'e and FPP.
We refer to this interpolating function as ``fractional power of rational function method" (FPR).
Note that this approach needs
\begin{\eq}
p,q \in \mathbb{Z}_{\geq 0} ,
\end{\eq}
which leads
\begin{\eq}
\alpha = \left\{ \begin{matrix}
\frac{a-b}{2\ell +1}  & {\rm for} & m+n:{\rm even} \cr
\frac{a-b}{2\ell}  & {\rm for} & m+n:{\rm odd}
\end{matrix} \right. ,\quad
{\rm with}\ \ell \in\mathbb{Z} .
\end{\eq}
If we take $2\ell+1=a-b$ for $a-b \in \mathbb{Z}$ and $m+n$ to be even, 
then this is nothing but the Pad\'e approximant
while taking $2\ell+1=m+n+1$ ($2\ell=m+n+1 $) for even (odd) $m+n$ gives the FPP.
Thus the FPR includes the Pad\'e and FPP as the special cases.
When the rational function in the parenthesis has
poles or takes negative values for non-integer $\alpha$,
then we cannot trust approximation by the FPR.
Note also that
the FPR is smooth unless the rational function has poles.
Therefore we expect that
when a physical quantity exhibits phase transition,
its approximation by the FPR is always bad as well as the Pad\'e and FPP.
We will explicitly demonstrate this in 2d Ising model 
in section \ref{sec:Ising}.

\subsection{``Landscape" of interpolating functions}
\label{sec:landscape}
In the previous subsections,
we have seen that we can construct various interpolating functions for given $(m,n)$.
Obviously, a superposition over multiple FPR's like
\begin{\eq}
\sum_i \frac{1}{n_i} F_{m,n}^{(\alpha_i )} (g) ,\quad {\rm with}\ \sum_i n_i =1 ,
\end{\eq} 
also has the same small-$g$ and large-$g$ expansions up to 
$\mathcal{O}(g^{a+m +1})$ and $\mathcal{O}( g^{b-n -1})$, respectively.
Furthermore,
if we add some terms, which cannot be fixed by the both expansions, to an interpolating function,
this also gives the same expansions up to the desired orders as the interpolating function. 
Such terms are typically proportional to $g^h$ with\footnote{
The power exponent $h$ does not often exist.
} $a+m+1 < h< b-n-1$ and $e^{-g-1/g}$. 
These facts tell us that
we can construct infinite number of interpolating functions in principle
while the best interpolating function, which is most close to the true function $F(g)$ among the interpolations,
should be unique.

The above argument leads\footnote{
This problem has been sharpened in early collaborations with Ashoke Sen and Tomohisa Takimi.
Hence we are grateful to them for this point.
} us to ``landscape problem of interpolating functions".
Namely,
given many interpolating functions,
it is unclear which interpolating function gives the best approximation of the true function $F(g)$.
Needless to say, this is trivial if we know the exact answer of $F(g)$, i.e.,
we can choose the best interpolating function just by comparing with $F(g)$. 
On the other hand, when we know its small-$g$ and large-$g$ expansions 
up to some finite orders but not the exact answer,
this problem is highly non-trivial.
{\it The goal of this paper is 
to construct a good criterion to choose the best interpolating functions
in terms of information on the both expansions.}

\begin{figure}[t]
\begin{center}
\includegraphics[width=7.4cm]{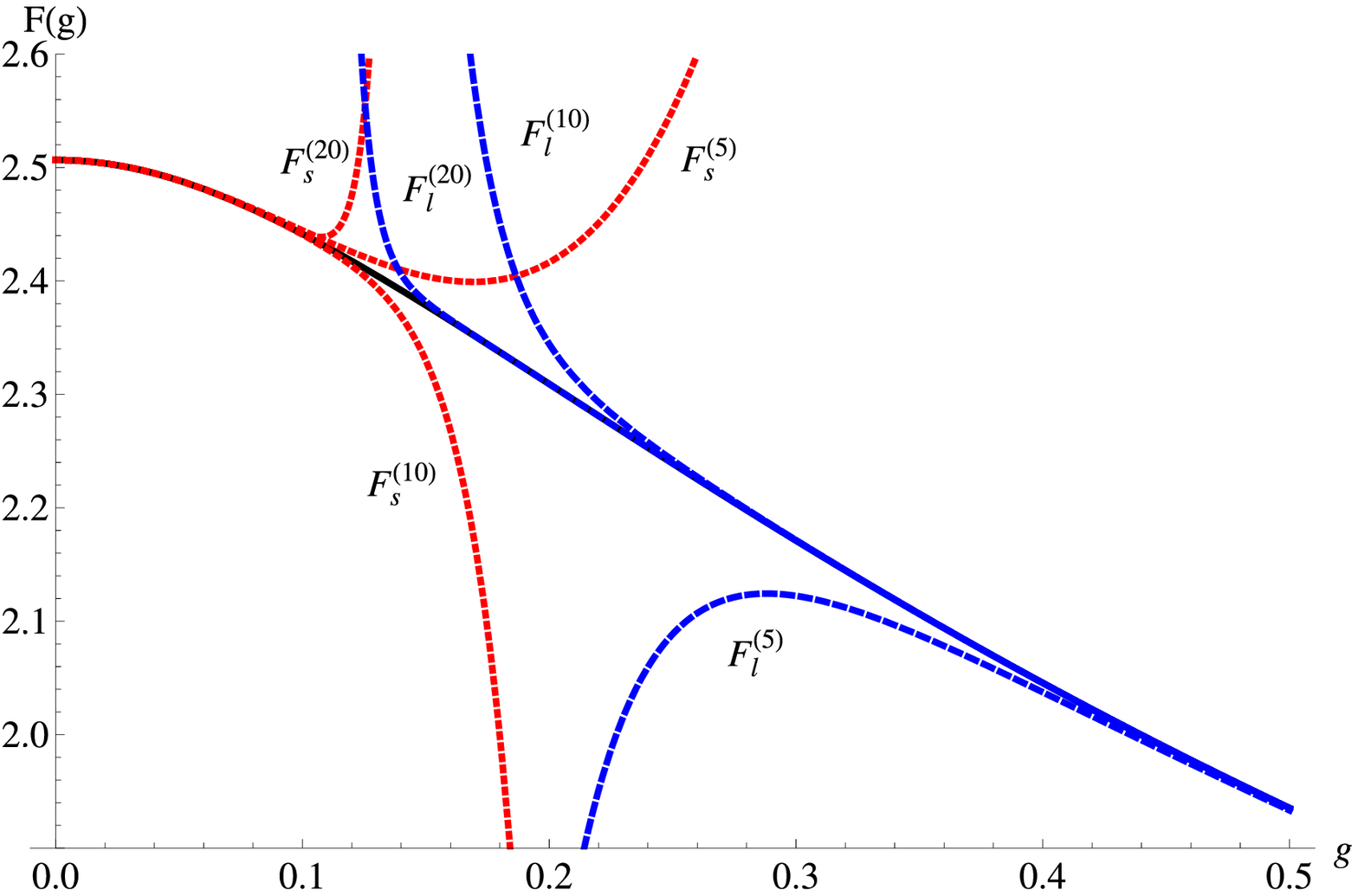}
\includegraphics[width=7.4cm]{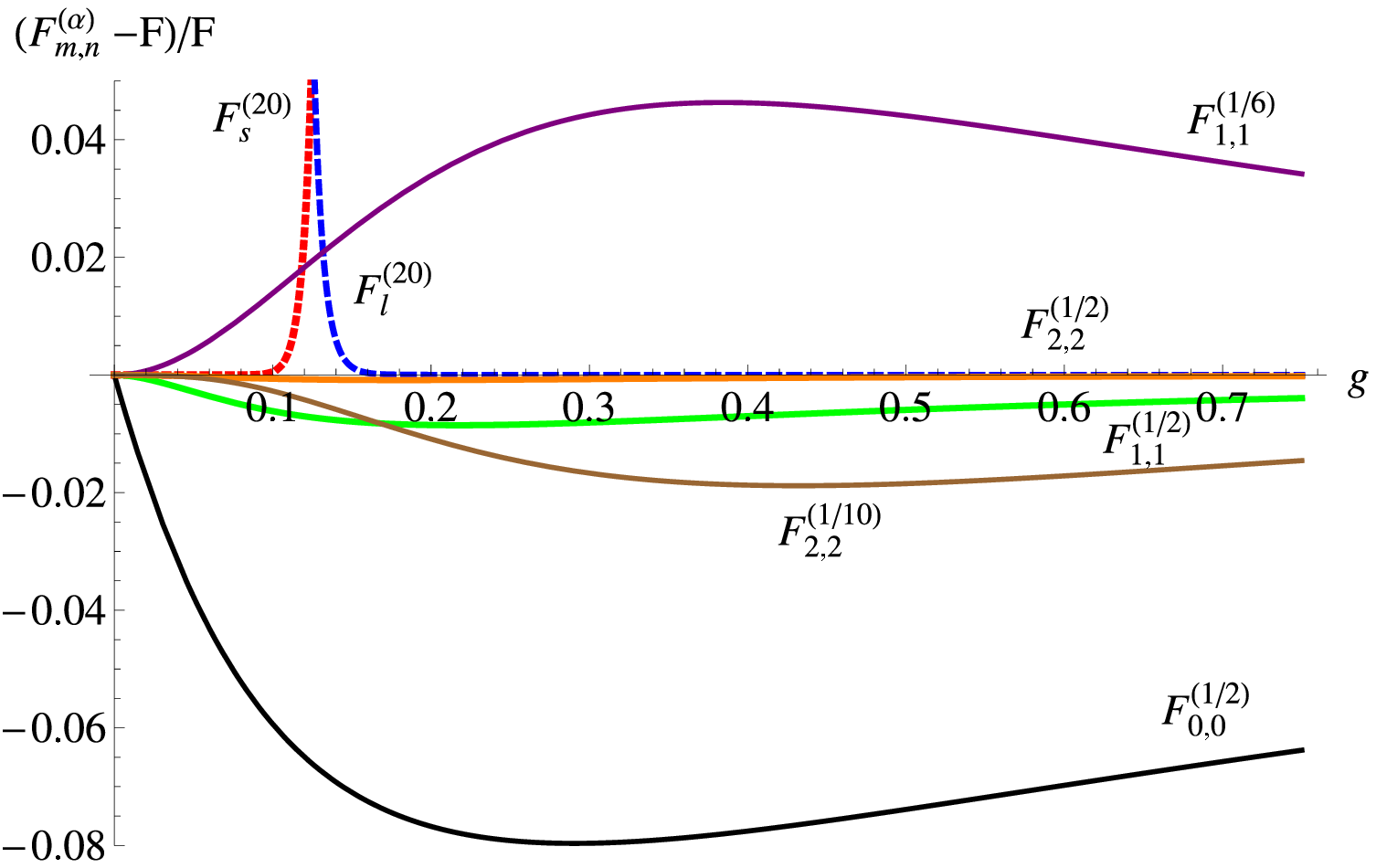}
\end{center}
\caption{
[Left] The function $F(g) = \int_{-\infty}^\infty dx\ e^{-x^2 /2 -g^2 x^4} $ (black solid),
its weak coupling expansions $F_s^{(N_s )}(g)$ (red dotted)  
and strong coupling expansions $F_l^{(N_l )}(g) $ (blue dashed) are plotted to $g$.
[Right] The differences between the interpolating functions $F_{m,n}^{(\alpha )}(g)$ and $F(g)$
normalized by $F(g)$ are plotted to $g$ (solid),
with $(F_s^{(20)}(g)-F)/F$ (red dotted) and $(F_l^{(20)}(g)-F)/F$ (blue dashed).
}
\label{fig:ex4}
\end{figure}
Let us demonstrate ``the landscape problem of interpolating functions" in the example
\begin{\eq}
F(g) = \int_{-\infty}^\infty dx\ e^{-\frac{x^2}{2} -g^2 x^4} 
=\frac{e^{\frac{1}{32 g^2}} }{2 \sqrt{2}g } K_{\frac{1}{4}}\left(\frac{1}{32 g^2}\right) ,
\label{eq:ex4}
\end{\eq}
which is the partition function of the zero-dimensional $\varphi^4$ theory.
Here $K_n (z)$ is the modified Bessel function of the second kind.
The function $F(g)$ has the following weak\footnote{
The form of the small-$g$ expansion depends on the argument of $g$.
Here we take real positive $g$.
Hence our construction of the interpolating functions does not guarantee that
the interpolating functions approximate also correct analytic property as complex functions of $g$
beyond the positive real axis.
} and strong coupling expansions:
\begin{\eqa}
&&F_s^{(N_s )}(g) =  \sum_{k=0}^{N_s} s_k g^k ,\quad
s_{2k+1}=0,\quad s_{2k} =\frac{\sqrt{2} \Gamma (2k+1/2)}{k!} (-4)^k , \NN\\
&&F_l^{(N_l )}(g) = \frac{1}{\sqrt{g}} \sum_{k=0}^{N_l} l_k g^{-k},   \quad
l_k = \frac{\Gamma \left( \frac{k}{2} +\frac{1}{4}\right)}{2k!} \left( -\frac{1}{2} \right)^k .
\label{eq:ex4coeff}
\end{\eqa}
In fig.~\ref{fig:ex4} [Left], we plot $F(g)$, $F_s^{(N_s )}(g) $ and $F_l^{(N_l )}(g)$ for some $N_s$ and $N_l$.
We see that as increasing $N_s (N_l )$, 
the weak (strong) coupling expansion blows up at smaller $g$ (larger $1/g$).
This can be understood from that 
the weak coupling expansion is asymptotic while the strong coupling expansion converges everywhere
as seen in \eqref{eq:ex4coeff}.

We can construct various interpolating functions of $F(g)$, 
whose explicit forms are given in appendix \ref{app:ex4}.
In terms of  the FPR interpolating function $F_{m,n}^{(\alpha )}(g)$,
we plot
\[
\frac{F_{m,n}^{(\alpha )}(g) -F(g)}{F(g)} ,
\]
for some values of $(m,n,\alpha )$ in fig.~\ref{fig:ex4} [Right].
At first sight, this plot shows that
the interpolating functions with larger $(m,n)$ roughly tend to give better approximations of $F(g)$
as naively expected.
However, we observe an exception of this tendency and somewhat unclear point. 
The exception is that $F_{1,1}^{(1/2)}(g)$ is closer than $F_{2,2}^{(1/10)}(g)$ to $F(g)$.
The unclear point is that 
if we consider the same value of $(m,n)$ but different values of $\alpha$,
then precisions of some interpolating functions are fairly different from each other, e.g.,
the maximal value of $|F_{2,2}^{(1/10)}(g)-F|/F$ is 0.0188286
while the one of $|F_{2,2}^{(1/2)}(g)-F|/F$ is 0.000853566.
These facts lead us to the following important question.
{\it When we do not know an explicit form of a function $F(g)$ 
but know its small-$g$ and large-$g$ expansions,
how do we choose the best approximation of $F(g)$ 
among its interpolating functions?}

In next section, 
we propose a criterion to determine the best interpolating functions,
which covers a wide class of problem. 
In section \ref{sec:examples},
we will come back to the above example and
check that our criterion correctly determines the best interpolating function
as well as other more nontrivial examples.

\section{Criterion for the ``best" interpolating function}
\label{sec:criterion}
In the previous section,
we have seen that
we can construct enormous interpolating functions of a function $F(g)$
in principle.
As demonstrated, 
then we have encountered ``the landscape problem of interpolating functions", i.e.,
it is unclear how one determines the best approximation of $F(g)$ among interpolating functions 
without knowing an explicit form of $F(g)$.
In this section, 
we construct a criterion to choose the best interpolating function
by using information on the small-$g$ and large-$g$ expansions\footnote{
One might expect that
some global properties of interpolating functions themselves are also useful for this purpose. 
Actually, in the early collaborations with Sen and Takimi,
we had adopted
minimizing curvatures of interpolating functions in intermediate coupling region as the part of criterion
but this had not been efficient in various examples.
We thank Takimi for pointing out the inefficiency of the curvatures explicitly and
emphasizing importance of information on the two expansions.
}.
We also mention some limitations of our criterion.

Suppose a function $F(g)$ and
that we know its power series expansions around $g=0$ up to $\mathcal{O}(g^{a+N_s +1})$
and around $g=\infty$ up to $\mathcal{O}(g^{b-N_l -1})$.
Namely, we have explicit forms of $F_s^{(N_s )}(g)$ and $F_l^{(N_l )}(g)$.
Then we consider a test function $G(g)$,
which would approximate $F(g)$ such as interpolating functions.
Given the test function $G(g)$,
let us introduce the two quantities 
\begin{\eq}
I_s [G] = \int_0^{g_s^\ast} dg \left| G(g) -F_s^{(N_s^\ast )}(g) \right|  ,\quad
I_l [G] = \int_{g_l^\ast}^\Lambda dg \left| G(g) -F_l^{(N_l^\ast )}(g) \right|  ,
\label{eq:criterion}
\end{\eq}
where the parameter $\Lambda$ in $I_l [G]$ is the cutoff of the integration
to make $I_l [G]$ well-defined and hence taken to be large. 
Throughout this paper, we take $\Lambda =1000$.   
The parameters $g_s^\ast ,g_l^\ast , N_s^\ast$ and $N_l^\ast$ will be defined shortly.
Roughly speaking, 
these are taken such that
$F_s^{(N_s^\ast )}(g)$ and $F_l^{(N_l^\ast )}(g)$ are sufficiently close to $F(g)$ 
for $0\leq g\leq g_s^\ast$ and $g\geq g_l^\ast $, respectively.
Then we propose that
 $I_s [G]$ and $I_l [G]$ measure
precision of approximation of $F(g)$ by the test function $G(g)$.
In other words,
we expect that
the ``best" interpolating function $G_{\rm best}(g)$ minimizes 
a value of  $I_s [G]$ plus $I_l [G]$
among candidates of interpolating functions:
\begin{framed}
\begin{\eq}
I_s [G_{\rm best} ] +I_l [G_{\rm best} ] 
=\min{\left\{ I_s [G ] +I_l [G] \right\}} .
\label{eq:criterion2}
\end{\eq}
\end{framed}

Now we define the parameters $g_s^\ast ,g_l^\ast , N_s^\ast$ and $N_l^\ast$.
Actually we differently take these values depending on properties of 
the small-$g$ and large-$g$ expansions. 
We classify the expansions into the following four types.
\begin{enumerate}
\item The both small-$g$ and large-$g$ expansions are convergent.

\item The small-$g$ expansion is asymptotic while the large-$g$ expansion is convergent.

\item The small-$g$ expansion is convergent while the large-$g$ expansion is asymptotic.

\item The both small-$g$ and large-$g$ expansions are asymptotic.

\end{enumerate}
Although we cannot determine whether the expansions are convergent or not for finite $N_s$ and $N_l$ in principle,
we can often extrapolate the finite data of the coefficients to infinite $N_s$ and $N_l$ 
as we will demonstrate in some examples.
In the rest of this section,
we explain how to choose the parameters $g_s^\ast ,g_l^\ast , N_s^\ast$ and $N_l^\ast$ for the above four types.

\subsubsection*{Type 1: convergent small-$g$ and large-$g$ expansions}
Suppose that 
the small-$g$ and large-$g$ expansions are convergent inside the circle $|g|=g_s^c$ and
outside the circle $|g|=g_l^c$, respectively.
First, we assume that $N_s,$ and $N_l$ are sufficiently large for a simplicity of explanation.
Then we will relax this condition later.

For $N_s \gg 1$, 
the small-$g$ expansion $F_s^{(N_s )}(g)$ should be 
very close to $F(g)$ for $|g|<g_s^c$ up to non-perturbative effect in a sense of $g$ 
such as $\sim e^{-1/g}$.
Correspondingly, for $N_l \gg 1$,  
the large-$g$ expansion $F_l^{(N_l )}(g)$ is also very close to $F(g)$ for $|g|>g_l^c$ up to non-perturbative effect in a sense of $1/g$ 
such as $\sim e^{-g}$.
Hence if we take $0<g_s^\ast <g_s^c$ and $N_s^\ast =N_s$ in \eqref{eq:criterion},
then sufficiently small value of $I_s [G]$ implies that
$G(g)$ is almost $F(g)$ in the region $0\leq g\leq g_s^\ast$ up to 
the non-perturbative effect of the $g$-expansion.
Similarly, taking $g_l^\ast >g_l^c$ and $N_l^\ast =N_l $,
the condition $I_l [G]\ll1 $ means that
$G(g)$ approximates $F(g)$ very well for $g\geq g_l^\ast$ up to 
the non-perturbative effect of the $1/g$-expansion.
Thus if we also impose values of $g_s^\ast$ and $g_l^\ast$ such that the non-perturbative effects are negligible for $0\leq g\leq g_s^\ast$ and $g\geq g_l^\ast$,
then the test function $G(g)$ satisfying $I_s [G]\ll 1$ and $I_l [G]\ll1$
is sufficiently close to the function $F(g)$ in these domains.
Note that although imposing only $I_s [G]\ll 1$ has the non-perturbative ambiguity 
in the sense of the small-$g$ expansion,
this ambiguity would be partially fixed by imposing $I_l [G]\ll 1$,
and vice versa as demonstrated in the last of this subsection.

\begin{figure}[t]
\begin{center}
\includegraphics[width=7.0cm]{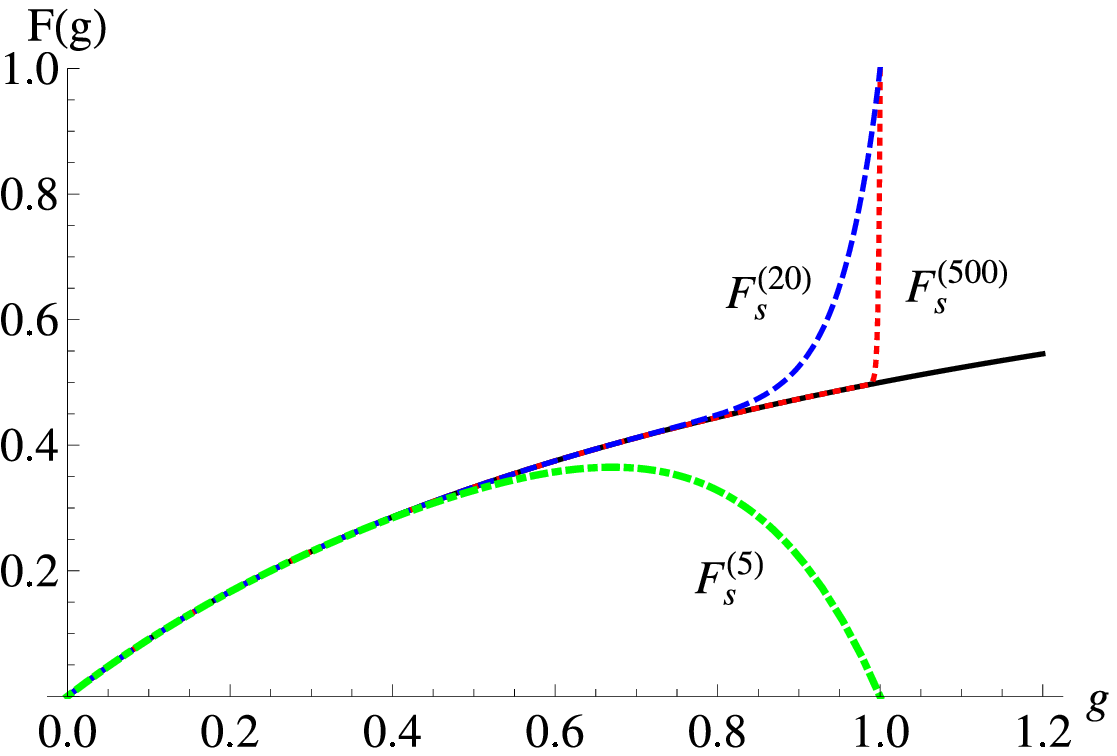}
\includegraphics[width=7.0cm]{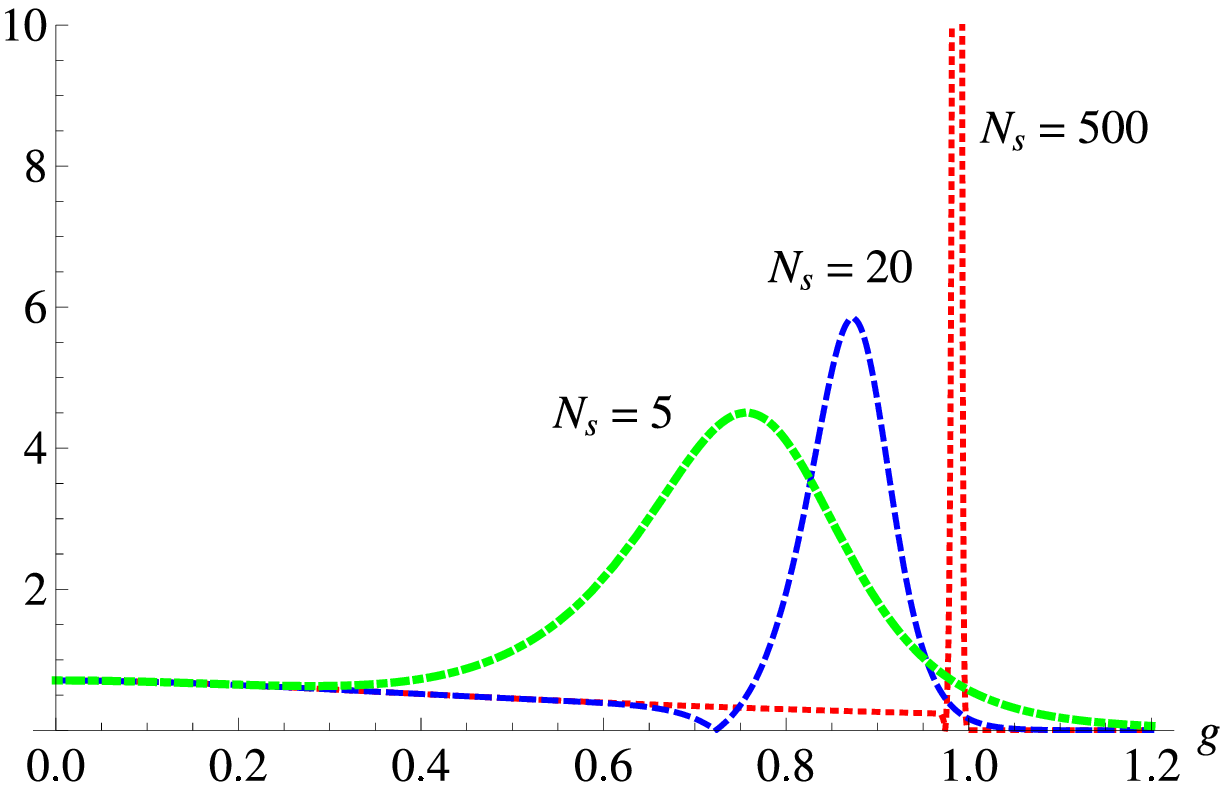}
\end{center}
\caption{
[Left] The function $F(g)=g/(1+g) $ and 
its small $g$ expansions $F_s^{(N_s)}$ 
are plotted to $g$.
[Right] The magnitudes of the curvatures of $F_s^{(N_s)}$ 
are plotted against $g$.
}
\label{fig:demo_blow}
\end{figure}
Now let us relax the conditions $N_s ,N_l \gg 1$.
When $N_s$ is sufficiently large but finite,
the small-$g$ expansion $F_s^{(N_s )}(g)$ blows up at almost $g=g_s^c$.
As decreasing $N_s$,
this blow up point, say $g_s^b$, becomes smaller (see fig.~\ref{fig:demo_blow} [Left]).
In the region $g>g_s^b$,
the small-$g$ expansion is obviously no longer close to the function $F(g)$
even if we are inside of the convergent radius.
Similarly this is true also for the large-$g$ expansion.
The locations of the blow-up points $g_s^b$ and $g_l^b$ 
can be found by studying their curvatures of $F_s^{(N_s )}(g)$ and $F_l^{(N_l )}(g)$, respectively,
because these points have very large curvatures around $g=g_s^b$ and $g=g_l^b$ 
as demonstrated in fig.~\ref{fig:demo_blow} [Right].

As a conclusion,
we choose the parameters 
$N_s^\ast$ and $N_l^\ast$
as
\begin{\eq}
N_s^\ast =N_s ,\quad N_l^\ast =N_l .
\label{eq:type1}
\end{\eq}
We take $g_s^\ast$ and $g_l^\ast $ such that
\begin{itemize}
\item $g_s^\ast$ and $g_l^\ast $ satisfy
\begin{\eq}
0< g_s^\ast <g_s^b ,\quad g_s^\ast <g_s^c ,\quad  g_l^\ast > g_l^b ,\quad g_l^\ast > g_l^c .
\end{\eq}

\item The non-perturbative effects are ignorable\footnote{
Note that only this condition is beyond information on the two expansions.
For example, in quantum field theory,
we can estimate classical weight of instanton effect by WKB analysis.
This condition means that 
such weights reading from some analysis are very small in the regions $0\leq g\leq g_s^\ast$ and $g\geq g_l^\ast$.
} for $0\leq g\leq g_s^\ast$ and $g\geq g_l^\ast$.

\item The regions $0\leq g\leq g_s^\ast$ and $g\geq g_l^\ast$ are as wide as possible\footnote{
Strictly speaking,  the curvature peaks around the blow-up points have finite widths.
Hence we should take $g_s^\ast$ and $g_l^\ast$ to avoid the finite widths of peaks.
} with satisfying the above two conditions.
\end{itemize}

\subsubsection*{Type 2: asymptotic small-$g$ expansion and convergent large-$g$ expansion}
Suppose that
the small-$g$ expansion is asymptotic but the large-$g$ expansion is convergent.
Then, while we take the parameters $g_l^\ast$ and $N_l^\ast$ as in type 1,
we differently take $g_s^\ast$ and $N_s^\ast$ for this case.

Before going to explain this,
we shall recall ``optimization" of asymptotic series.
Let us consider a function $F(g)$ and its power series expansion $F_s^{(N_s )}(g)$ around $g=0$.
If this expansion is asymptotic and  non-convergent, $F_s^{(\infty )}(g)$ diverges.
We would like to know which value of $N_s$ 
optimally approximates $F(g)$ with fixed $g$.
Such optimization is usually achieved by
minimization of the last term of the asymptotic series (see e.g. \cite{Marino:2012zq}).
Namely,
the optimized value $N_s^o$ is determined by the condition
\begin{\eq}
\left. \frac{\partial}{\partial k}\log{s_k} \right|_{k =N_s^o} +\log{g} = 0.
\label{eq:opt}
\end{\eq}
We can also estimate ``error" of this optimization by the next term of the series:
\begin{\eq}
\delta_s (g) = \left| s_{N_s^o +1} g^{a+N_s^o +1} \right| .
\label{eq:error}
\end{\eq}
Thus we choose $N_s^\ast$ and $N_l^\ast$ in \eqref{eq:criterion} as
\begin{\eq}
N_s^\ast =N_s^o ,\quad N_l^\ast =N_l .
\label{eq:type2}
\end{\eq}
We take $g_s^\ast$ and $g_l^\ast $ such that
\begin{itemize}
\item $g_s^\ast$ and $g_l^\ast $ satisfy
\begin{\eq}
\delta_s (g_s^\ast  ) \leq \epsilon_s \ll 1 ,\quad g_l^\ast > g_l^b ,\quad g_l^\ast > g_l^c .
\label{eq:epsw}
\end{\eq}

\item The non-perturbative effects are ignorable for $0\leq g\leq g_s^\ast$ and $g\geq g_l^\ast$.

\item The regions $0\leq g\leq g_s^\ast$ and $g\geq g_l^\ast$ are as wide as possible.
\end{itemize}
Since $\delta_s (g)$ is monotonically increasing,
the above condition $\delta_s (g_s^\ast  ) \leq \epsilon_s$ means
$\delta_s (g  ) \leq \epsilon_s \ll 1$ for $0\leq g\leq g_s^\ast$.
Here the value of $\epsilon_s$ are taken depending on desired precision.

\subsubsection*{Type 3: convergent small-$g$ expansion and asymptotic large-$g$ expansion}
This case is quite parallel to type 2.
The optimized order $N_l^o$ of the large-$g$ expansion is determined by
\begin{\eq}
\left. \frac{\partial}{\partial k}\log{l_k} \right|_{k =N_l^o} -\log{g} = 0 ,
\end{\eq}
with the error 
\begin{\eq}
\delta_l (g) = \left| l_{N_l^o +1} g^{b-N_l^o -1} \right| .
\end{\eq}
Thus we take $N_s^\ast$ and $N_l^\ast$  as
\begin{\eq}
N_s^\ast =N_s ,\quad N_l^\ast =N_l^o .
\label{eq:type3}
\end{\eq}
The parameters $g_s^\ast$ and $g_l^\ast $ are taken such that
\begin{itemize}
\item $g_s^\ast$ and $g_l^\ast $ satisfy
\begin{\eq}
0< g_s^\ast < g_s^b  ,\quad g_s^\ast < g_s^c ,\quad \delta_l (g_l^\ast ) \leq \epsilon_l \ll 1 .
\end{\eq}

\item The non-perturbative effects are ignorable for $0\leq g\leq g_s^\ast$ and $g\geq g_l^\ast$.

\item The regions $0\leq g\leq g_s^\ast$ and $g\geq g_l^\ast$ are as wide as possible.
\end{itemize}
Since $\delta_l (g)$ is the monotonically decreasing function of $g$,
the condition $\delta_l (g_l^\ast  ) \leq \epsilon_l$ means
$\delta_l (g  ) \leq \epsilon_l \ll 1$ in the region $g\geq g_l^\ast$.
Here we take the value of $\epsilon_l$ depending on desired precision
as $\epsilon_s$.

\subsubsection*{Type 4: asymptotic small-$g$ and large-$g$ expansions}
When the both series are asymptotic,
we take
\begin{\eq}
N_s^\ast =N_s^o ,\quad N_l^\ast =N_l^o .
\label{eq:type4}
\end{\eq}
The parameters $g_s^\ast$ and $g_l^\ast $ are taken to be
\begin{itemize}
\item $g_s^\ast$ and $g_l^\ast $ satisfy
\begin{\eq}
\delta_s (g_s^\ast  ) \leq \epsilon_s   ,\quad \delta_l (g_l^\ast ) \leq \epsilon_l  .
\end{\eq}

\item The non-perturbative effects are ignorable for $0\leq g\leq g_s^\ast$ and $g\geq g_l^\ast$.

\item The regions $0\leq g\leq g_s^\ast$ and $g\geq g_l^\ast$ are as wide as possible.
\end{itemize}

\subsubsection*{Remarks}
\begin{figure}[t]
\begin{center}
\includegraphics[width=7.0cm]{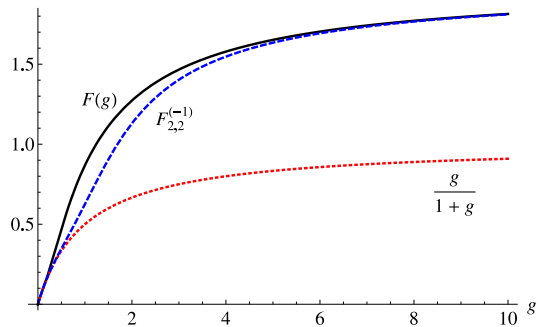}
\end{center}
\caption{
The function $F(g) =g/(1+g) +e^{-1/g}$ and
its perturbative part $g/(1+g)$ in the sense of $g$ are plotted to $g$.
Hence the deviation between them shows purely the non-perturbative effect $e^{-1/g}$.
We see that the Pad\'e approximant 
$F_{2,2}^{(-1)}(g) =g ( 4 g^2 /5 -4g/5 +1) /(2 g^3 /5 +g/5+1)$ of $F(g)$
partially knows about the non-perturbative effect. 
}
\label{fig:demo1}
\end{figure}
As a simple example, let us consider the function
\begin{\eq}
F(g) = \frac{g}{1+g} +c e^{-\frac{1}{g}} ,
\end{\eq}
with a constant $c$.
Clearly the function $F(g)$ has the convergent small-$g$ and large-$g$ expansions.
Since the second term is non-perturbative in a sense of $g$,
the small-$g$ expansion is independent of $c$.
Therefore only imposing $I_s [G]\ll 1$ with $g_s^\ast < g_s^c (=1)$ and $e^{-1/g_s^\ast}\ll 1$
is insufficient to check whether $G(g)$ nicely approximates $F(g)$ 
including the non-perturbative effect or not.
Namely, we should also require $I_l [G]\ll 1$ to fix the non-perutbative ambiguity of the approximation as depicted in fig.~\ref{fig:demo1}.

Next, suppose the example
\begin{\eq}
F(g) = \frac{g}{1+g} +c_1 e^{-\frac{1}{g}} +c_2 e^{-g}+c_3 e^{-g-\frac{1}{g}} ,
\end{\eq}
whose expansions around $g=0$ and $g=\infty$ are also convergent.
While the small-$g$ expansion has information on the 1st and 3rd terms,
the large-$g$ expansion has information on the 1st and 2nd terms.
Note that the both expansions never know about the last term.
This exhibits limitations of our interpolating functions and criterion, i.e.,
our criterion never fix the ambiguity of mixed non-perturbative effects such as $e^{-g-1/g}$,
which are non-perturbative in senses of both $g$ and $1/g$.
Although the above examples correspond to type 1,
we expect that
this feature is true also for the other types.

\section{Some examples}
\label{sec:examples}
In this section,
we explicitly check that our criterion proposed in section~\ref{sec:criterion} works 
for the partition function of the zero-dimensional $\varphi^4$ theory,
specific heat in the 2d Ising model,
average plaquette in the 4d $SU(3)$ pure Yang-Mills theory on lattice and
free energy in the $c=1$ string theory at self-dual radius.

\subsection{Partition function of zero-dimensional $\varphi^4$ theory} 
\begin{figure}[t]
\begin{center}
\includegraphics[width=7.0cm]{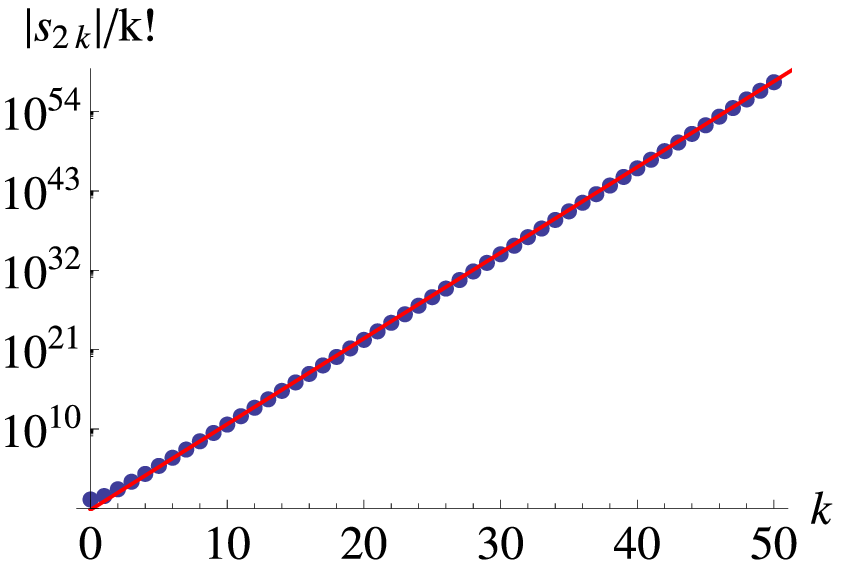}
\includegraphics[width=7.0cm]{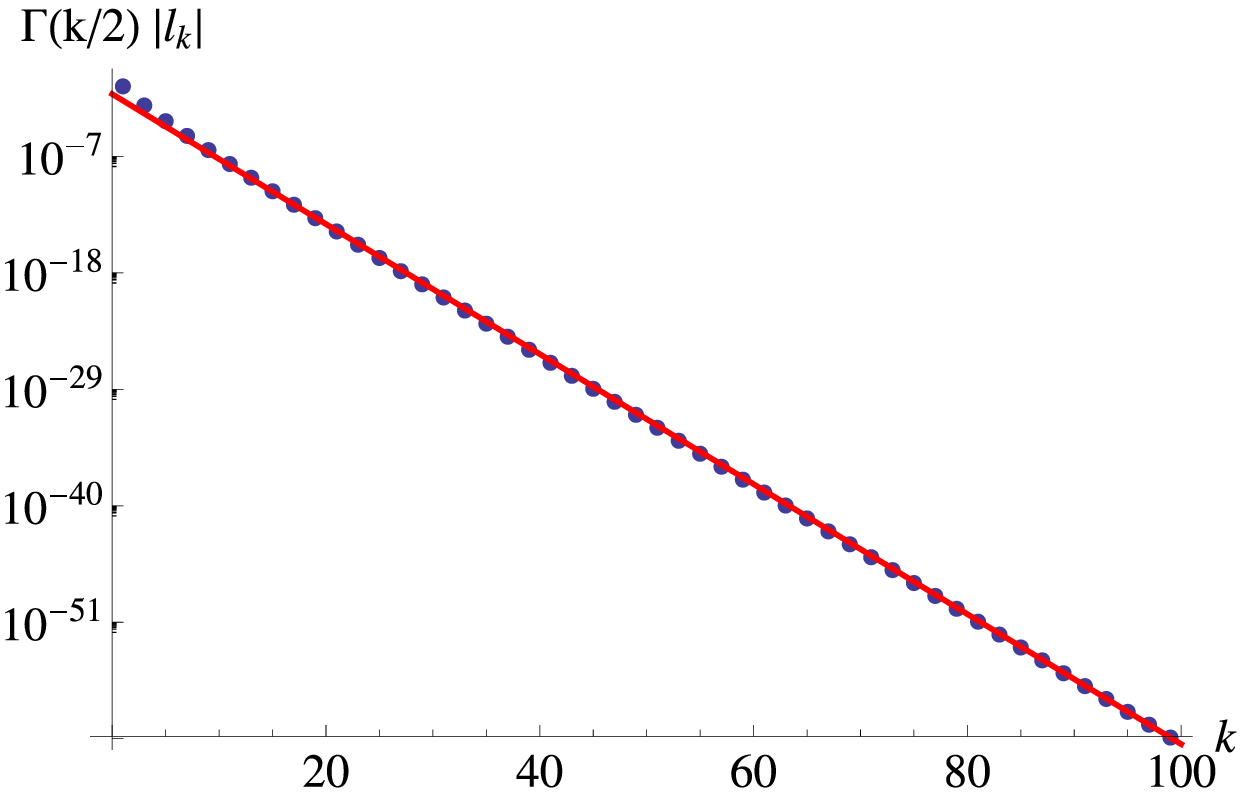}
\end{center}
\caption{
[Left] The weak coupling expansion coefficients $|s_{2k}|$ divided by $k!$
is plotted to $k$ in semi-log scale.
The straight line shows fitting by the linear function $\log{(|s_{2k}|/k!)} = \log{c_s}+k\log{A_s}$.
[Right] The strong coupling expansion coefficients $|l_k |$ 
multiplied by $\Gamma(k/2)$ is plotted against $k$ in semi-log scale.
The straight line denotes the linear fitting $\log{(\Gamma (k/2) |l_k|)} = \log{c_l}+k\log{A_l}$.
}
\label{fig:ex4c}
\end{figure}
\begin{figure}[t]
\begin{center}
\includegraphics[width=7.0cm]{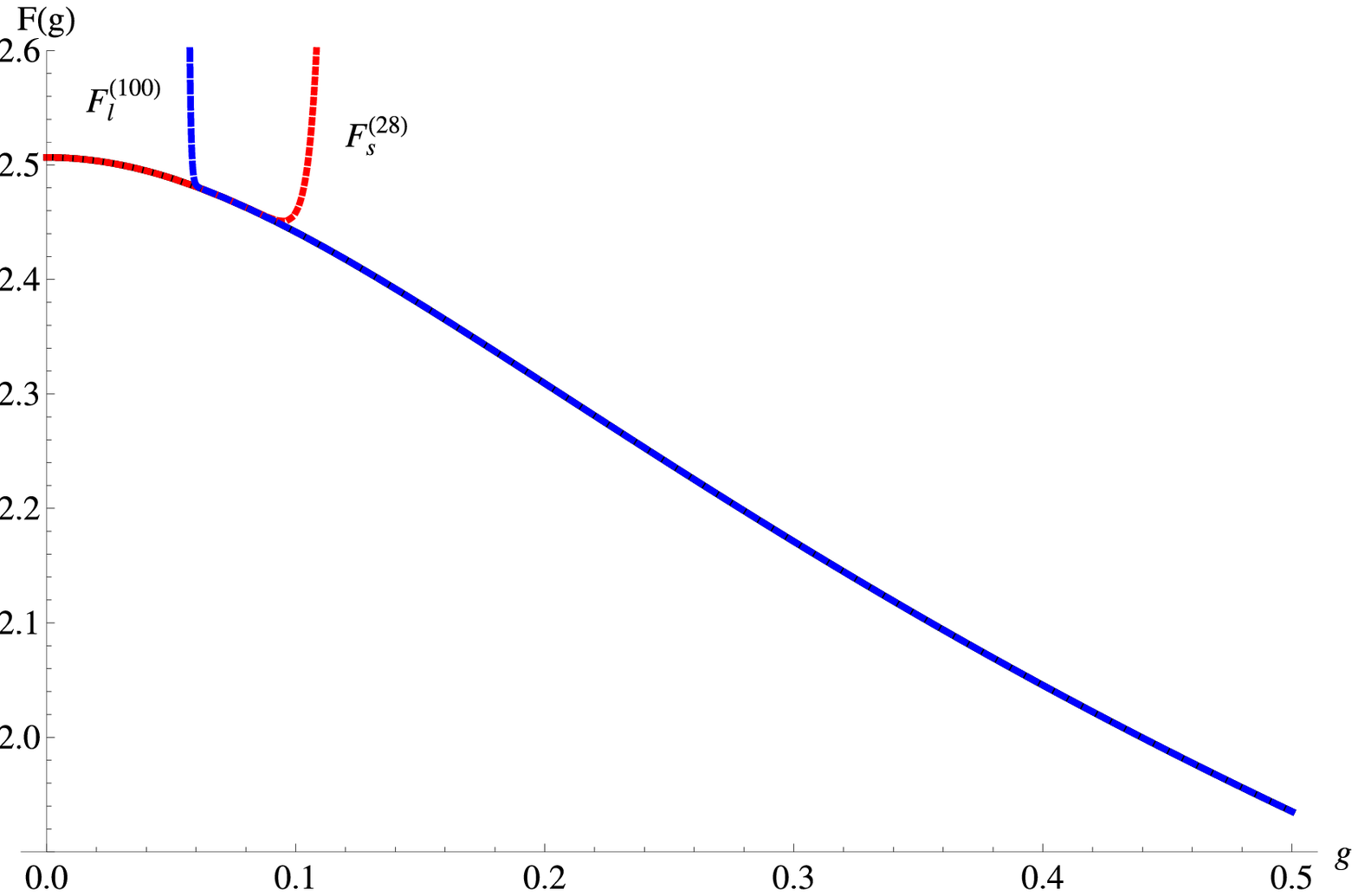}
\includegraphics[width=7.0cm]{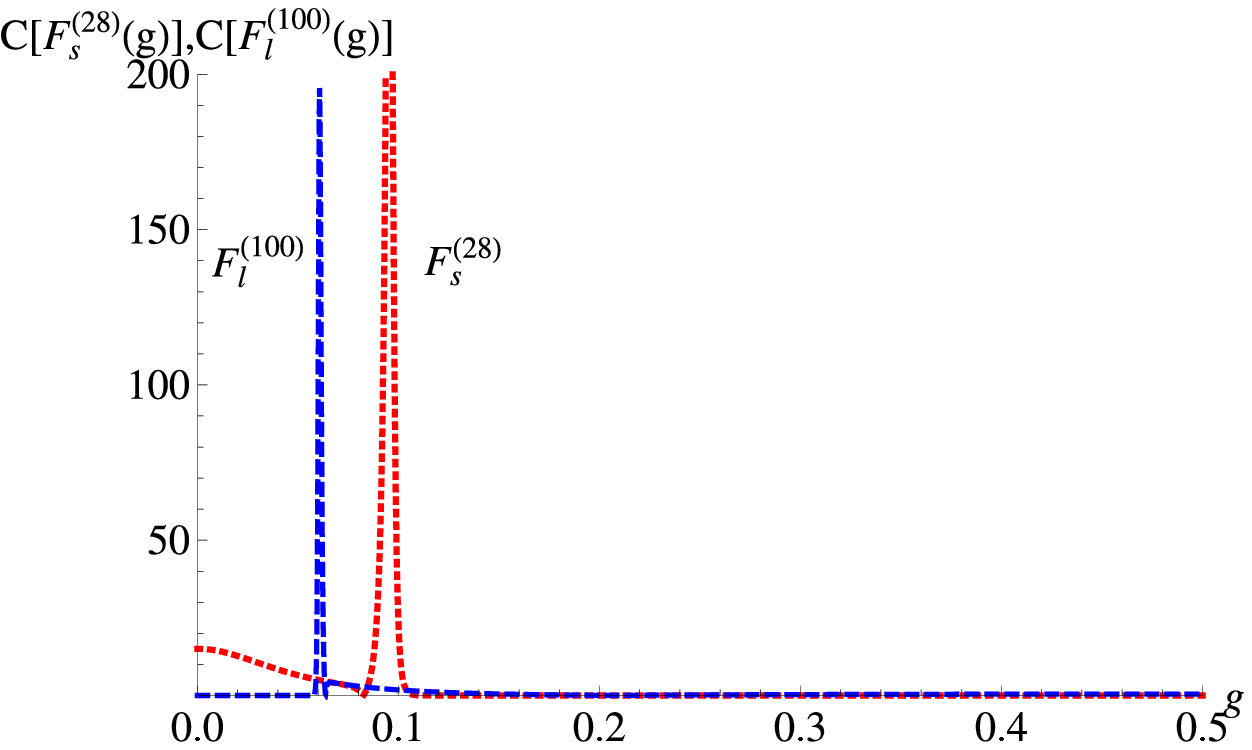}
\end{center}
\caption{
[Left] $F(g)$ (black solid), $F_s^{(28)}(g) $ (red dotted) and  
$F_l^{(100)}(g)$ (blue dashed) are plotted to $g$.
[Right] The absolute values of the curvatures of $F_s^{(28)}(g) $ (red dotted) and  
$F_l^{(100)}(g)$ (blue dashed) are plotted to $g$.
}
\label{fig:ex4blow}
\end{figure}
Let us come back to the example \eqref{eq:ex4} appearing in section \ref{sec:landscape}:
\[
F(g) = \int_{-\infty}^\infty dx\ e^{-\frac{x^2}{2} -g^2 x^4} .
\]
We can easily see from \eqref{eq:ex4coeff} that
the weak coupling expansion is asymptotic while the strong one is convergent.
Hence this problem corresponds to type 2.

Although we know all the coefficients of the both expansions for this case,
in more practical situations,
we usually have their small-$g$ and large-$g$ expansions up to only finite orders.
As an exercise of such practical situations,
let us restrict that we know only the first 101 coefficients for a practical purpose,
i.e., we have only information on $F_s^{(100)}$ and $F_l^{(100)}$.
Then we will estimate large order behaviors of the coefficients by extrapolation
in terms of the finite data.

\subsubsection*{Determination of $g_s^\ast$}
First let us study large order behavior of the weak coupling expansion.
In fig.~\ref{fig:ex4c} [Left],
we plot how the coefficients
$s_{2k}$ behave as increasing $k$.
From this plot, we find that the coefficient in large $k$ regime grows as
\begin{\eq}
\frac{|s_{2k}|}{k!} \sim c_s A_s^k ,\quad
{\rm with} \ c_s =0.0620840\ {\rm and}\ A_s =15.4189 .
\end{\eq}
Since the coefficient grows by the factorial,
the weak coupling expansion is asymptotic.
By performing optimization \eqref{eq:opt},
we find\footnote{We have used the Stirling approximation
$\log{k!}=k(\log{k}-1)+\mathcal{O}(\log{k})$.}
\begin{\eq}
N_s^o = \frac{2}{A_s g^2} ,\quad 
\delta_s (g) = c_s e^{-\frac{1}{A_s g^2}} .
\end{\eq}
Taking $\epsilon_s =10^{-7}$ in \eqref{eq:epsw},
we find
\begin{\eq}
N_s^\ast = N_s^o =28, \quad  g_s^\ast = 0.0680628 .
\end{\eq}

\subsubsection*{Determination of $g_l^\ast$}
\begin{table}[t]
\begin{center}
  \begin{tabular}{|c|c|c|c|c|}
  \hline & $\Lambda^{-1}\int dg \bigl|\frac{(F_{m,n}^{(\alpha )}-F }{F} \bigr|$ &$I_s [F_{m,n}^{(\alpha )}]$ &$I_l [F_{m,n}^{(\alpha )}]$&$I_s +I_l $ \\
  \hline\hline $F_{0,0}^{(1/2)} $   &0.000659728 &0.00446072 & 0.381344 &0.385805\\
 \hline $F_{1,1}^{(1/2)} $   & $9.27801\times 10^{-6}$ &0.000297906 &0.0142222 &0.0145201\\
 \hline $F_{1,1}^{(1/6)} $   & 0.0000760393 &0.000432581 &0.106287  &0.106720 \\
 \hline $F_{2,2}^{(1/2)} $   & $4.61177\times 10^{-7}$ &0.0000230059 &0.000849124 &0.000872130 \\
 \hline $F_{2,2}^{(1/6)} $   & $5.24010\times 10^{-6}$ &0.0000450000 &0.00905419 &0.00909919 \\
 \hline $F_{2,2}^{(1/10)} $   & 0.0000235129 &0.0000430012 &0.0373156 &0.0373586 \\
 \hline $F_{3,3}^{(1/2)} $   &$2.96944\times 10^{-8}$ &$1.94617\times 10^{-6}$&0.0000576043 &0.0000595505 \\
 \hline $F_{3,3}^{(1/6)} $   &$3.84001\times 10^{-7}$ &$5.09656\times 10^{-6}$&0.000738006 &0.000743103  \\
 \hline $F_{3,3}^{(1/14)} $   &$8.84054\times 10^{-6}$ &$2.59016\times 10^{-6}$&0.0148826 &0.0148852 \\
 \hline $\bf F_{4,4}^{(1/2)} $   &$\bf 2.17241\times 10^{-9}$ &$\bf 1.78480\times 10^{-7}$& $\bf 4.25411\times 10^{-6}$ & $\bf 4.43259\times 10^{-6}$ \\
 \hline $F_{4,4}^{(1/6)} $   &$2.85852\times 10^{-8}$ &$5.50786\times 10^{-7}$&0.0000577750 &0.0000583258\\
 \hline $F_{4,4}^{(1/10)} $   &$5.77057\times 10^{-7}$ &$1.52640\times 10^{-6}$&0.00111431 &0.00111584 \\
 \hline  $ F_{4,4}^{(1/18)} $   &$ 3.17581\times 10^{-6}$ &$ 8.72352\times 10^{-7}$
&  0.00549043 & 0.00549131 \\
\hline
  \end{tabular}
\caption{
Result on the zero-dimensional $\varphi^4$ theory.
}
\label{tab:ex4}
\end{center}
\end{table}

Next we estimate large order behavior of the strong coupling expansion.
In fig.~\ref{fig:ex4c} [Right],
we plot $\Gamma (k/2)|l_k|$ against $k$ in semi-log scale.
Then fitting the data leads us to
\begin{\eq}
\Gamma (k/2)|l_k| \sim c_l A_l^k ,\quad
{\rm with} \ c_l =0.0769534\ {\rm and}\ A_l =0.243235 .
\end{\eq}
This shows that the strong coupling expansion is convergent with the infinite convergent radius.
Since the strong coupling expansion $F_l^{(100)}$ behaves 
as $\sim g^{-201/2}$ in small $g$ regime,  
this blows up at some point as seen in fig.~\ref{fig:ex4blow} [Left].
In order to look for the blow-up point, 
we study the magnitude of the curvature:
\begin{\eq}
C[f(g)] = \Bigl| \frac{f''(g)}{\left( 1+( f'(g) )^2 \right)^{3/2}} \Bigr| .
\end{\eq}
In fig.~\ref{fig:ex4blow} [Right],
we plot  the curvature of $F_l^{(100)}(g)$ to $g$.
From this plot, we find
the sharp peak of $C[F_l^{(100)}]$ around $g=0.0607430$
and thus we take
\begin{\eq}
 g_l^\ast =0.1 .
\end{\eq}

\subsubsection*{Result}

Now we are ready to test our criterion for the ``best" interpolating function.
We measure precision of the interpolating function $F_{m,n}^{(\alpha )} (g)$ by
\begin{\eq}
\frac{1}{\Lambda} \int_0^\Lambda dg \Bigl| \frac{F_{m,n}^{(\alpha )}(g) -F(g)}{F(g)} \Bigr| , 
\end{\eq}
which should be minimized by the best interpolating function.
Comparing this with values of $I_s [F_{m,n}^{(\alpha )}]$ and $I_l [F_{m,n}^{(\alpha )}]$,
we can test our criterion \eqref{eq:criterion2} proposed in the previous section.
We summarize our result in tab.~\ref{tab:ex4}.
We easily see that 
the interpolating function $F_{4,4}^{(1/2)}(g)$ gives
the most precise approximation of $F(g)$ among the interpolating functions in the table.
Note that $F_{4,4}^{(1/2)}(g)$ has also the minimal values of $I_s$ and $I_l$ 
among the candidates\footnote{
Actually we can construct better interpolating function
by considering a superposition of the interpolating functions.
For example,  $G(g)=\frac{9}{10}F_{4,4}^{(1/2)}(g) +\frac{1}{10}F_{4,4}^{(1/6)}(g)$ gives
$\Lambda^{-1} \int dg |(G-F)/F|=1.33651\times 10^{-9}$, $I_s [G]=1.05554\times 10^{-7}$ and $I_l [G]=2.42407\times 10^{-6}$.
This also supports our criterion.
}.
This shows that our criterion correctly chooses the best interpolating function 
in this example.

\subsection{Specific heat in two-dimensional Ising model}
\label{sec:Ising}
\begin{figure}[t]
\begin{center}
\includegraphics[width=7.0cm]{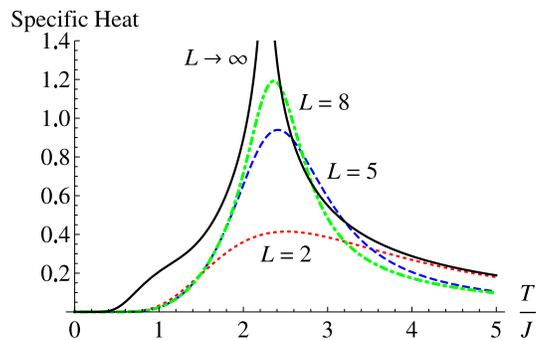}
\end{center}
\caption{
The specific heat in the 2d Ising model is plotted to $T/J$ for various lattice sizes. 
}
\label{fig:specific}
\end{figure}

Let us consider the 2d Ising model 
on the $L\times L$ square lattice with periodic boundary condition.
We can study interpolation problem between high and low temperature expansions in this model.
It is well known that
this system in the $L\rightarrow\infty$ limit exhibits the second order phase transition and
the specific heat has a singularity as depicted in fig.~\ref{fig:specific}.
On the contrary,
our interpolating function \eqref{eq:FPR} gives a smooth function
unless this has poles.
Hence we expect that
our interpolating function fails to approximate the specific heat 
in the infinite volume limit
while it should work for finite $L$.
Thus it is interesting to study 
how properties of interpolating function change as increasing $L$.

The Hamiltonian of the 2d Ising model is given by
\begin{\eq}
H = -J \sum_{(\mathbf{x},\mathbf{y})\in {\rm neighbor}} \sigma_{\mathbf{x}} \sigma_{\mathbf{y}} ,
\end{\eq}
where $\sigma_{\mathbf{x}}$ is the spin variable at the position $\mathbf{x}$ taking the values $\pm 1$
and $J$ is the coupling constant.
In terms of the temperature $T$, we define the partition function of this model by 
\begin{\eq}
Z_L (K) = \sum_{\{ \rm state \}} e^{-\frac{1}{T} H} 
= \sum_{\{ \rm state \}} e^{K \sum_{(\mathbf{x},\mathbf{y})} \sigma_{\mathbf{x}} \sigma_{\mathbf{y}}}  ,\quad
{\rm with}\ K=\frac{J}{T}.
\end{\eq}
The exact solution of the partition function is given by (see e.g. \cite{Kastening})
\begin{\eqa}
&&Z_L (K) = \frac{1}{2} \left( S_{11} (K)  +2S_{10} (K) - S_{00} (K)  \right) ,\NN\\
&& S_{\sigma_1 \sigma_2} (K) = 2^{L^2} \prod_{p,q=0}^{L-1} \Bigl[ 
\cosh^2{(2K)} -\sinh{(2K)} \Bigl( \cos{\frac{(2p+\sigma_1 )\pi}{L}} +\cos{\frac{(2q+\sigma_2 )\pi}{L}} \Bigr)
\Bigr]^{\frac{1}{2}} . 
\end{\eqa}
Let us introduce the quantity 
\begin{\eq}
C_L (K) =  \frac{1}{L^2} \frac{\partial^2}{\partial K^2} \log{Z_L (K)} ,
\label{eq:CL}
\end{\eq}
which gives the specific heat by $K^2 C_L (K)$.
Since the partition function is a power series of $e^K$,
we see that 
high temperature expansion around $K=0$ of $C_L (K)$ is a power series of $K$
while low temperature expansion around $K=\infty$ gives a power series of $e^{-K}$.
In order to make the both expansions the power series form \eqref{eq:asymptotics},
we introduce the parameter $g$ by
\begin{\eq}
e^{2K} = 1+g .
\end{\eq}
Then the high and low temperature expansions of $C_L (g)$ become
power series of $g$ and $1/g$, respectively.
Namely, the small-$g$ and large-$g$ expansions in the previous sections correspond to 
the high and low temperature expansions, respectively.
Note that 
$C_L (g)$ is a rational function of $g$ for arbitrary finite $L$.
Therefore 
one of the Pad\'e approximants should give the exact\footnote{
In app.~\ref{app:anotherIsing},
we perform a similar analysis under another parametrization $e^{8K}=1+g^2$,
where all FPR interpolating functions \eqref{eq:FPR}
do not give the exact answer.
} answer for finite $L$ in principle.

Below we consider interpolation problem for $L=2,5,8$ and $\infty$.
Denoting the high and low temperature expansions as
\begin{\eq}
C_s^{(N_s )} =g^a \sum_{k=0}^{N_s} s_k g^k ,\quad
C_l^{(N_l )} =g^b \sum_{k=0}^{N_l} l_k g^{-k} ,
\end{\eq}
we also assume that
we have information only on the high temperature expansion $C_s^{(50)}(g)$ and 
low temperature expansion $C_l^{(50)}(g)$.
For all the cases below,
we explicitly write down 
explicit formulas for interpolating functions $C_{m,n}^{(\alpha )}(g)$ 
taking the FPR form \eqref{eq:FPR} in app.~\ref{app:ising}.

\subsubsection{$2\times 2$ lattice}
\begin{figure}[htbp]
\begin{center}
\includegraphics[width=7.3cm]{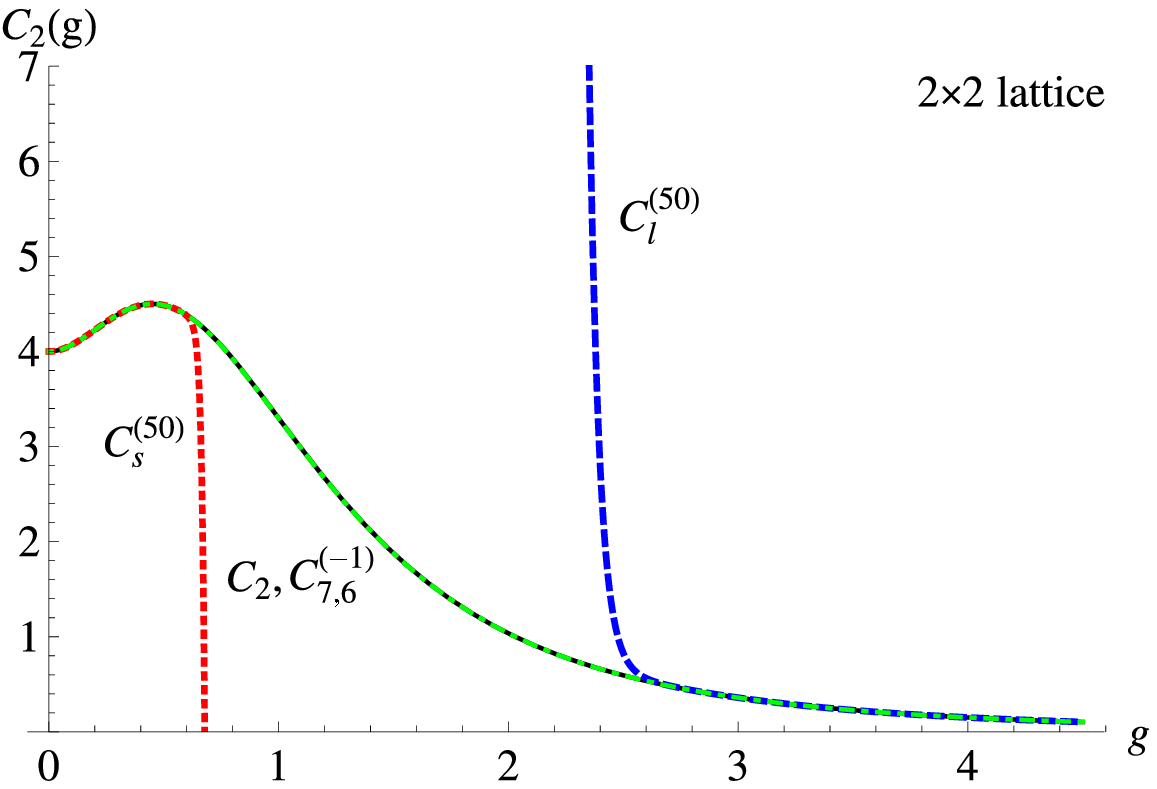}
\includegraphics[width=7.3cm]{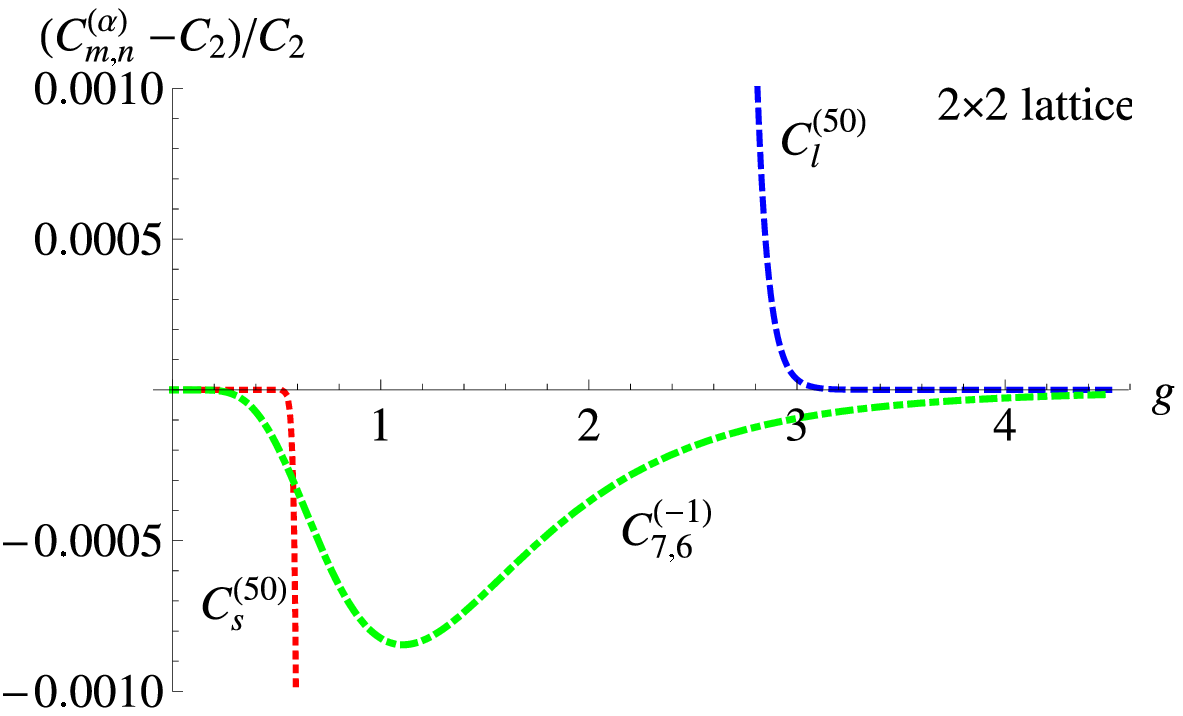}\\
\includegraphics[width=7.3cm]{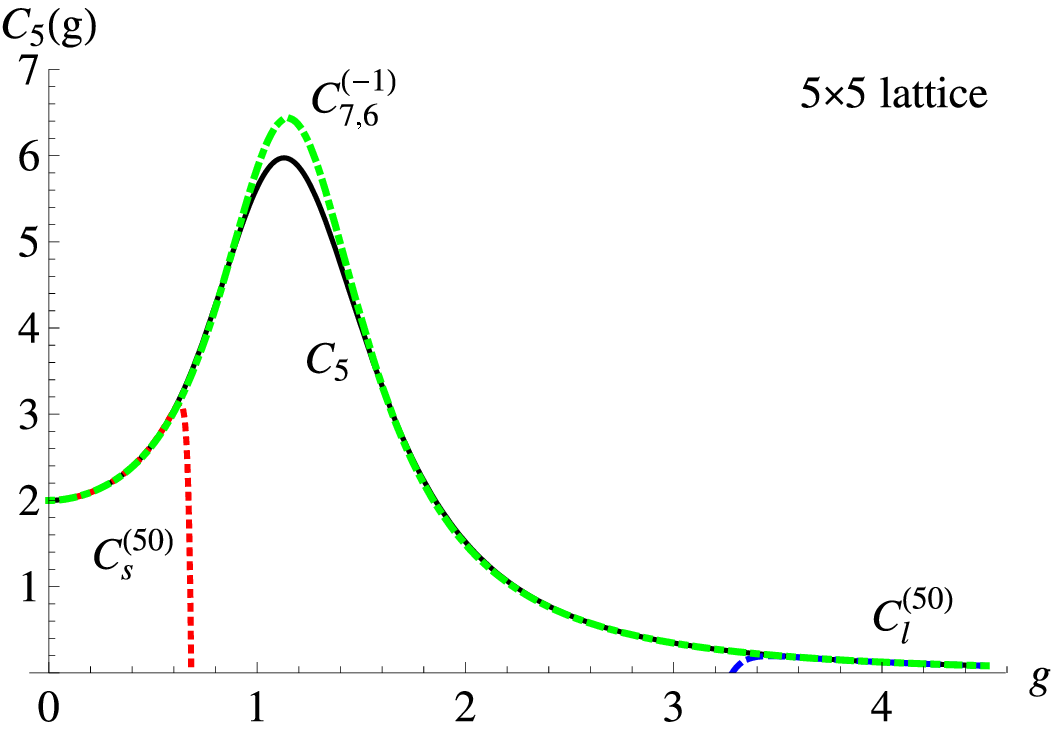}
\includegraphics[width=7.3cm]{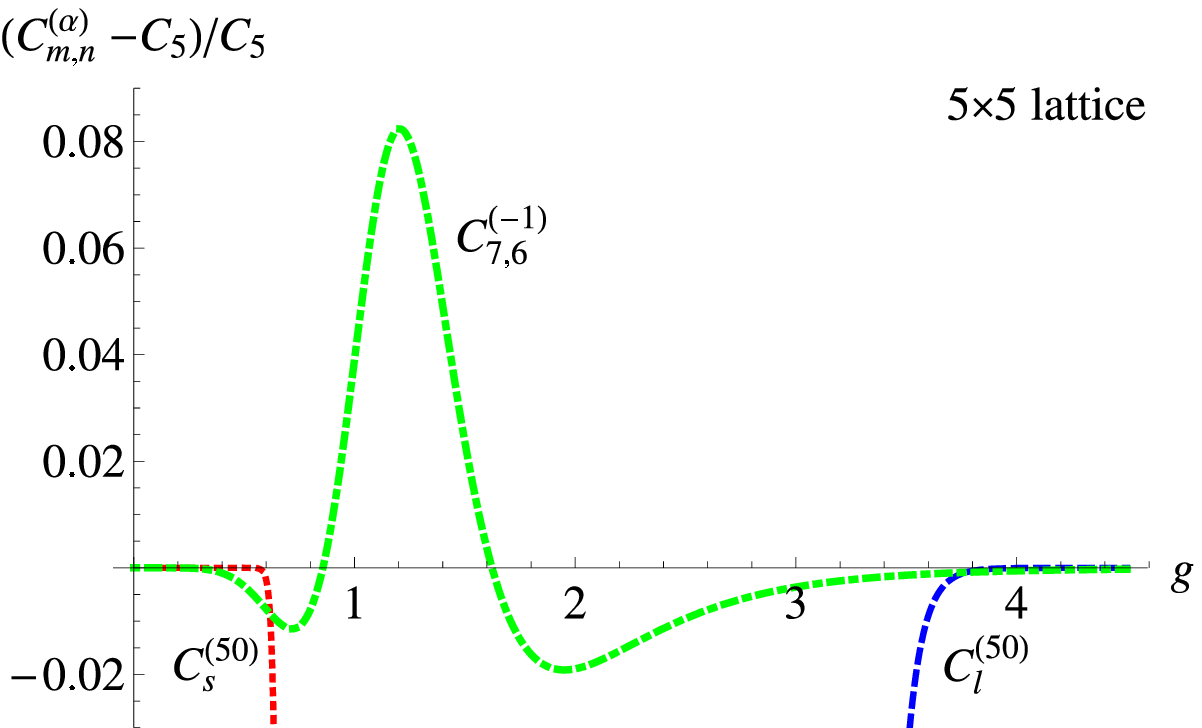}\\
\includegraphics[width=7.3cm]{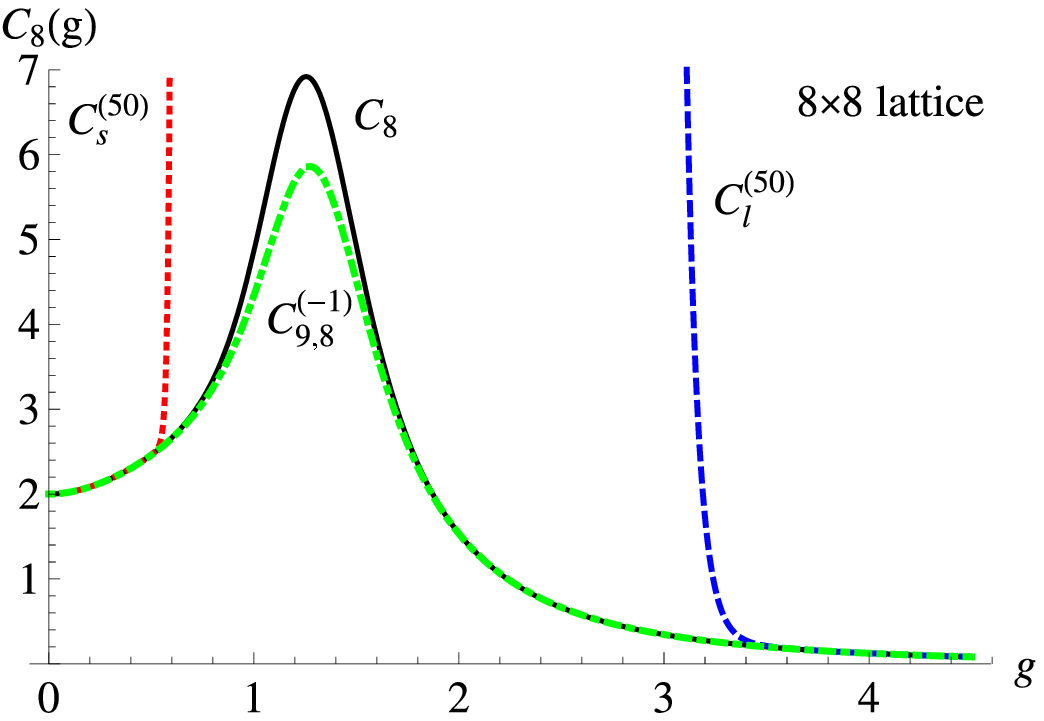}
\includegraphics[width=7.3cm]{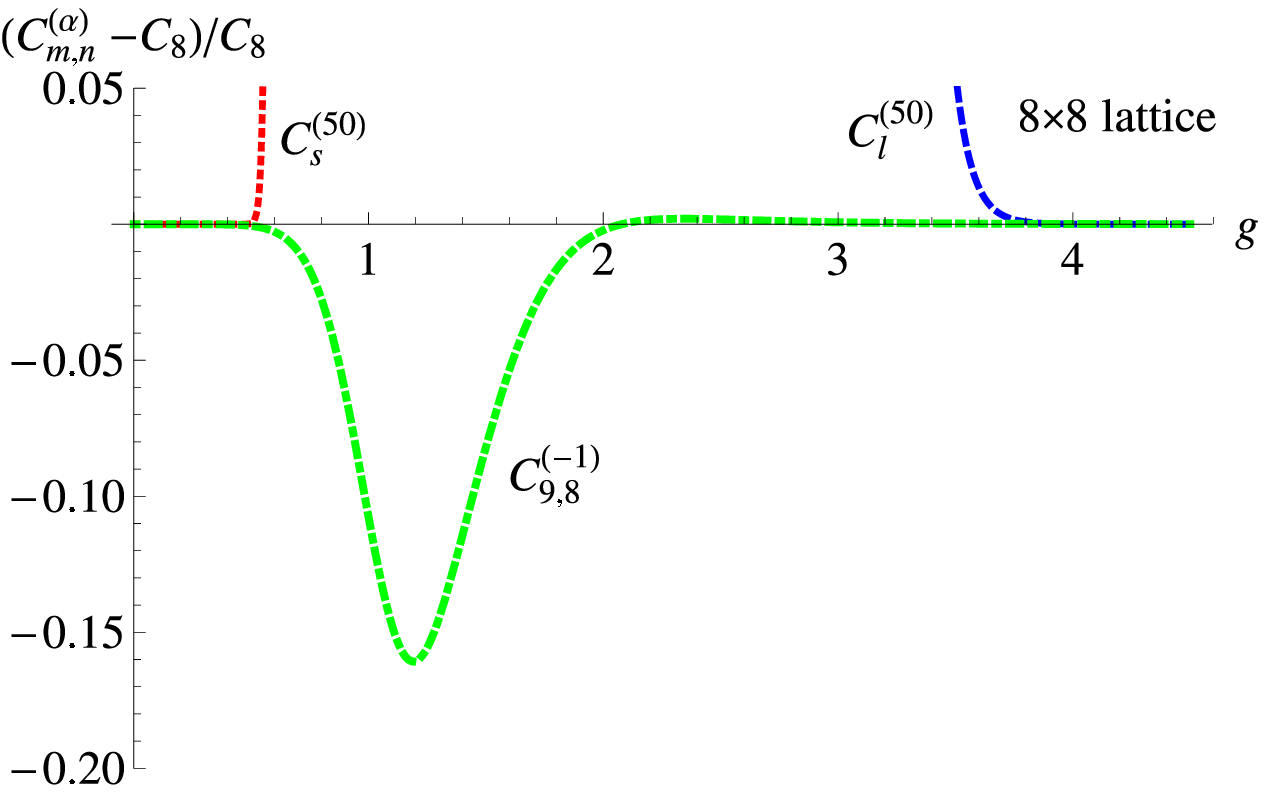}\\
\includegraphics[width=7.3cm]{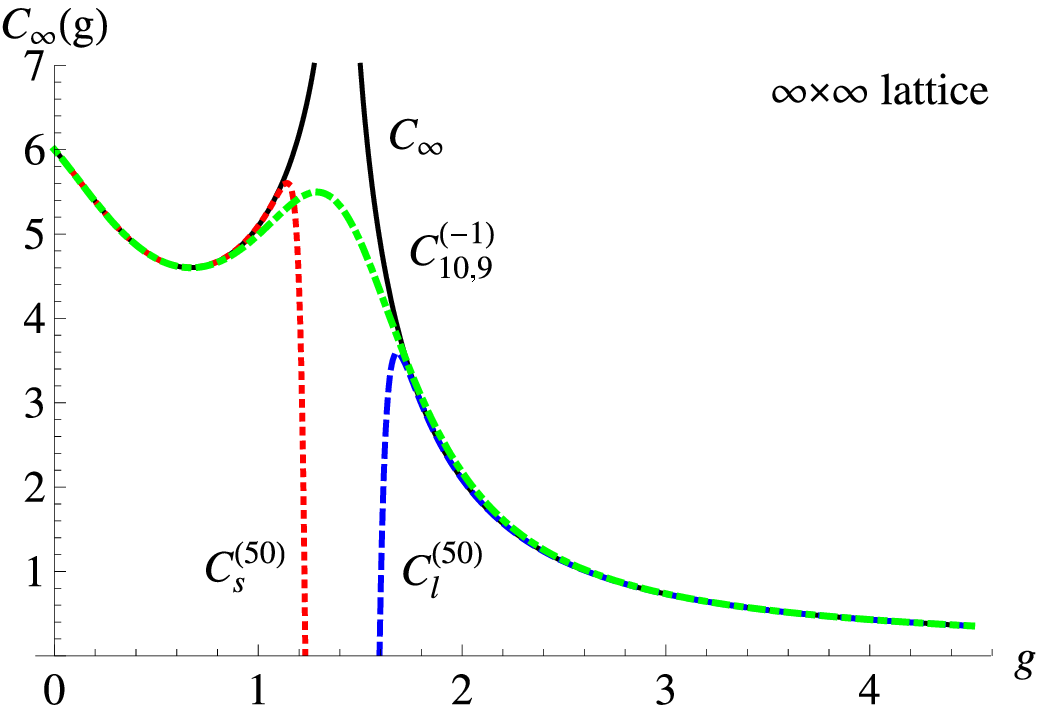}
\includegraphics[width=7.3cm]{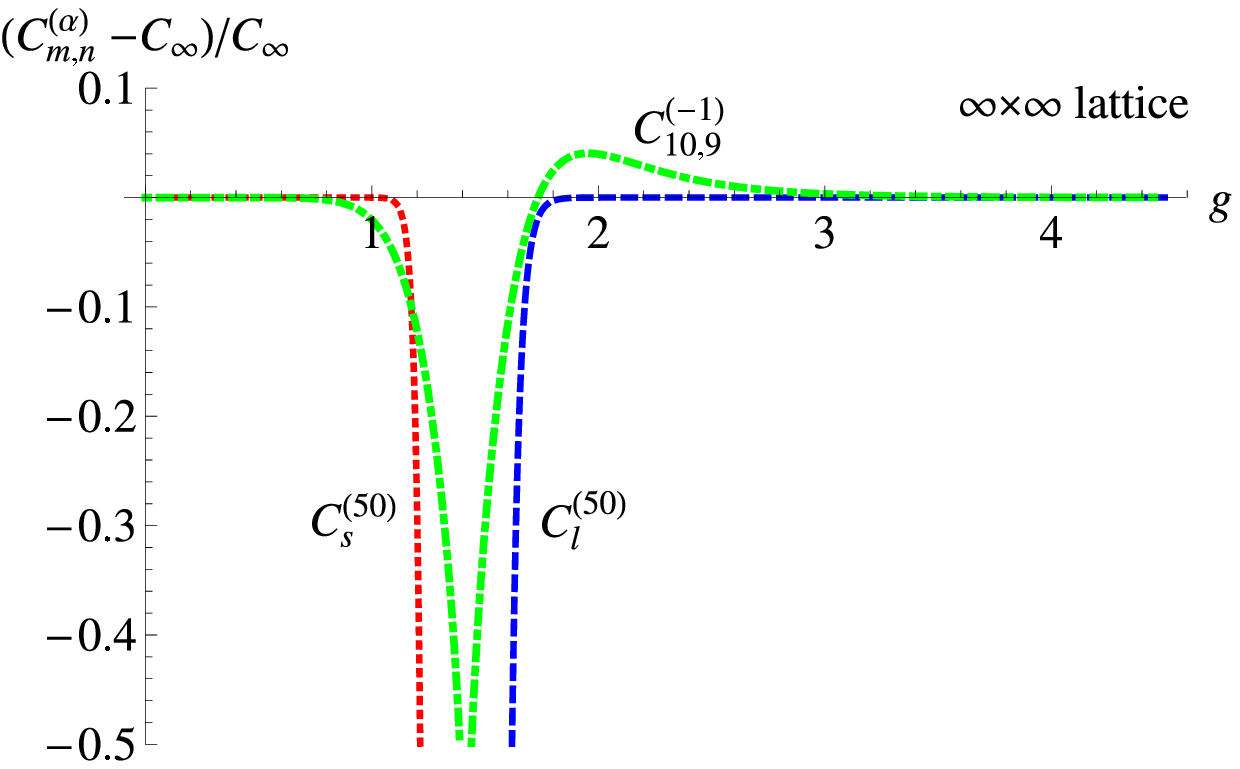}
\end{center}
\caption{
[Left] The function $C_L (g)$ (black solid), 
the best interpolating function (green dot-dashed),
its high (red dotted) and low (blue dashed) temperature expansions 
are plotted to $g$ for each $L$. 
[Right] $|(C_{m,n}^{(\alpha )}-C_L )/C_L |$ (green dot-dashed), 
$|(C_s^{(50 )}-C_L )/C_L |$ (red dotted) and $|(C_l^{(50 )}-C_L )/C_L |$ (blue dashed)
are plotted against $g$ for each $L$.
}
\label{fig:ising2d}
\end{figure}

For $L=2$, the function $C_2 (g)$ has the following expansions
\begin{\eqa}
C_2 (g) 
&=& 4 +8g^2 -8g^3 -30g^4 +\mathcal{O}(g^5 ) \NN\\
&=& g^{-4}\left( 96 -384g^{-1} +960g^{-2} -1920g^{-3} +2272g^{-4}   +\mathcal{O}(g^{-5} ) \right) .
\end{\eqa}
In terms of these expansions,
we can construct various interpolating functions for $C_2 (g)$.
As in the previous example, we read large order behaviors of the both expansions by extrapolation.
By fitting the data of $s_k$ and $l_k$ for $k=10\sim 50$,
we find
\begin{\eqa}
&& |s_k | \sim c_s A_s^k ,\quad {\rm with}\ c_s =8.38769 \ {\rm and}\  A_s =1.4614 ,\NN\\
&& |l_k | \sim c_l A_l^k ,\quad    {\rm with}\     c_l =154.976 \ {\rm and}\  A_l =2.4487 .
\end{\eqa}
Hence the high and low temperature expansions seem to be convergent
for $|g|< A_s^{-1} =0.684275$ and $|g|>A_l $, respectively.
We also find the blow-up points of $F_s^{(50)}(g)$ and $F_l^{(50)}(g)$
at $g=g_s^b =0.453215$ and $g=g_l^b =2.59736$, respectively.
Thus we take 
\begin{\eq}
N_s^\ast =50 ,\quad g_s^\ast =0.3,\quad N_l^\ast =50,\quad g_l^\ast =2.8.
\end{\eq}
We summarize our result in fig.~\ref{fig:ising2d} and tab.~\ref{tab:L2}.
We find that 
our criterion correctly determines the best interpolating function $C_{7,6}^{(-1)} (g)$
among the candidates as in the previous example.

\begin{table}[h]
\begin{center}
  \begin{tabular}{|c|c|c|c|c|}
  \hline & $\Lambda^{-1}\int dg \bigl| \frac{C_{m,n}^{(\alpha )}-C_2}{C_2} \bigr|$ &$I_s [C_{m,n}^{(\alpha )}]$ &$I_l [C_{m,n}^{(\alpha )}]$ & $I_s +I_l$\\
  \hline\hline $C_{1,1}^{(-4)} $   &0.00224809 &0.0912750 &0.102638 &0.193913 \\
\hline
$C_{1,1}^{(-4/3)} $   &0.000817041 &0.0617656 &0.0193989 &0.0811645 \\
\hline
 $C_{1,2}^{(-2)} $   &0.00100228 &0.0751219 &0.0466568 & 0.121779\\
\hline
 $C_{1,2}^{(-1)} $   &0.000286070 &0.058021  &0.00257514 &0.0605961 \\
\hline
 $C_{2,2}^{(-4)} $   &0.000173889  &0.00768450   &0.00852503 &0.0162095 \\
\hline
 $C_{2,3}^{(-2)} $   &0.000158806 &0.00224879  &0.00550386 &0.00775265\\
\hline
 $C_{3,2}^{(-2)} $   &0.000322997  &0.00658097  &0.0116865 &0.0182674\\
\hline
 $C_{3,4}^{(-1)} $   &0.0000147709  &0.000148814  &0.000258785 &0.000407598\\
\hline
 $C_{4,3}^{(-2)} $   &0.000168121  &0.00345741 &0.00565649 &0.0091139\\
\hline
 $C_{4,3}^{(-1)} $   &0.0000651441 &0.00156065  &0.00170668 &0.00326733\\
\hline
 $C_{5,4}^{(-1)} $   &0.0000207392  &0.000525855  &0.000327812 &0.000853666\\
\hline
 $C_{6,5}^{(-1)} $   &0.0000119340  &0.000174690 &0.000192164 &0.000366854\\
\hline
 $\bf C_{7,6}^{(-1)} $   &$\bf 1.22853\times 10^{-6}$  &$\bf 3.39663\times 10^{-6}$ 
&$\bf 0.0000523107$ &\bf 0.0000557073\\
\hline
 $C_{6,7}^{(-1)} $   &0.0000128648  &0.000129274  &0.0000598797 &0.000189154\\
\hline
  \end{tabular}
\caption{Result for $L=2$}
\label{tab:L2}
\end{center}
\end{table}
\clearpage
\clearpage

\subsubsection{$5\times 5$ lattice}
\begin{table}[t]
\begin{center}
  \begin{tabular}{|c|c|c|c|c|}
  \hline & $\Lambda^{-1}\int dg \bigl| \frac{C_{m,n}^{(\alpha )}-C_5}{C_5}\bigr|$ &$I_s [C_{m,n}^{(\alpha )}]$ &$I_l [C_{m,n}^{(\alpha )}]$ & $I_s +I_l$\\
  \hline\hline $C_{1,1}^{(-4)} $   &0.00447397  &0.0908906 &0.0573872 &0.148278 \\
\hline
$C_{1,1}^{(-4/3)} $   &0.00228464  &0.0639680 &0.0237879 & 0.0877559\\
\hline
 $C_{1,2}^{(-2)} $   &0.00265831 &0.0726585 &0.0320523 &0.104711 \\
\hline
 $C_{1,2}^{(-1)} $   &0.00164725  &0.0540871 &0.0113251 &0.0654122 \\
\hline
 $C_{2,2}^{(-4)} $   &0.00142494  &0.0137917  &0.0102147 &0.0240064 \\
\hline
 $C_{2,3}^{(-2)} $   &0.00112491 &0.00956534  &0.00538250 &0.0149478 \\
\hline
 $C_{2,3}^{(-1)} $   &0.00137710 &0.0372464  &0.00707500 &0.0443214 \\
\hline
 $C_{3,2}^{(-2)} $   &0.000961824  &0.00689287 &0.00279760 &0.00969046 \\
\hline
$C_{3,3}^{(-4)} $   &0.00113015   &0.00776624   &0.00559209 &0.0133583 \\
\hline
 $C_{3,3}^{(-4/7)} $   &0.00143699   &0.000319958   &0.0237685 &0.0240885\\
\hline
 $C_{3,4}^{(-2)} $   &0.00100163  &0.00699082  &0.00374339 &0.0107342\\
\hline
 $C_{3,4}^{(-1)} $   &0.000721790   &0.00555444   &0.00130205 &0.00685649 \\
\hline
 $C_{4,3}^{(-1)} $   &0.000469692  &0.00359619  &0.00281633 &0.00641253 \\
\hline
 $C_{4,4}^{(-4/9)} $   &0.00294461   &0.00712618  &0.0459126 &0.0530387\\
\hline
 $C_{4,5}^{(-1)} $   &0.000596309   &0.00398025   &0.000648710 &0.00462896\\
\hline
 $C_{6,5}^{(-1)} $   &0.000440902   &0.00134450   &0.000398831 &0.00174333\\
\hline
 $\bf C_{7,6}^{(-1)} $   &$\bf 0.0000555327$  &$\bf 0.000123232$  &$\bf 0.0000416657$ &\bf 0.000164898\\
\hline
 $C_{6,7}^{(-1)} $   &0.0000898635  &0.000325997  &0.0000649638 &0.000390961\\
\hline
  \end{tabular}
\caption{Result for $L=5$}
\label{tab:L5}
\end{center}
\end{table}

Next we consider $5\times 5$ lattice.
For this case, $C_5 (g)$ has the following expansions
\begin{\eqa}
C_5 (g)
&=& 2+\frac{5 g^2}{2}-\frac{3 g^3}{2}+\frac{17 g^4}{8} +\mathcal{O}\left(g^5\right) ,\NN\\
&=&g^{-4}\left( 64-256g^{-1} +928g^{-2} -3008g^{-3} +9440g^{-4} 
+\mathcal{O}(g^{-5} )  \right) .
\end{\eqa}
By fitting the data of $s_k$ and $l_k$ for $k=10\sim 50$,
we again find
\begin{\eqa}
 |s_k | \sim c_s A_s^k ,\quad &&{\rm with}\ c_s =0.980413 \ {\rm and}\  A_s =1.60585 ,\NN\\
 |l_k | \sim c_l A_l^k ,\quad    &&{\rm with}\ c_l =98.4806 \ {\rm and}\  A_l =3.21044 .
\end{\eqa}
Hence the high and low temperature expansions seem to be convergent
for $|g|< A_s^{-1} =0.622724$ and $|g|>A_l $, respectively.
We also find the blow-up points $g_s^b =0.629103$ and $g_l^b =3.37946$.
Thus we take the parameters in \eqref{eq:criterion} as
\begin{\eq}
N_s^\ast =50 ,\quad g_s^\ast =0.4,\quad N_l^\ast =50,\quad g_l^\ast =3.8.
\end{\eq}
Our result is summarized in fig.~\ref{fig:ising2d} and tab.~\ref{tab:L5}.
As in the $L=2$ case, 
we find that 
our criterion correctly chooses the relatively best interpolating function $C_{7,6}^{(-1)} (g)$.

There is an important difference from the $L=2$ case.
Although our criterion works also for the $L=5$ case,
the approximation by the best interpolation function becomes worse compared with the $L=2$ case.
Indeed we have about $8\%$ discrepancy around the peak of $C_5 (g)$.
This implies that
we should include information on higher order terms of the both expansions
to obtain sufficiently good interpolating function.
However, note that
the location of the peak of the interpolating function $C_{7,6}^{(-1)} (g)$ is almost the one of $C_5 (g)$.
We actually find that 
$C_{7,6}^{(-1)} (g)$ and $C_5 (g)$ have the peaks at $g=1.14765$ and $1.12930$, respectively.
This point would be physically important as follows.
It is known that
the peak location approaches to the second order phase transition point
as increasing $L$. 
Hence, if interpolating function for finite $L$ precisely gives the peak location of $C_L (g)$,
then the interpolating function would be useful for finding the phase transition point.
We will see in the last of this subsection that 
extrapolating the peak locations of the best interpolating functions for finite $L$ to infinite $L$
precisely gives the known phase transition point.

\subsubsection{$8\times 8$ lattice}
\begin{table}[t]
\begin{center}
  \begin{tabular}{|c|c|c|c|c|}
  \hline & $\Lambda^{-1}\int dg \bigl|\frac{C_{m,n}^{(\alpha )}-C_8}{C_8} \bigr|$ &$I_s [C_{m,n}^{(\alpha )}]$ &$I_l [C_{m,n}^{(\alpha )}]$ & $I_s +I_l$ \\
  \hline\hline $C_{1,1}^{(-4)} $   &0.00444636  &0.0831909  &0.0636770 &0.146868 \\
\hline
 $C_{1,1}^{(-4/3)} $   &0.00225426  &0.0562683   &0.0269999 &0.0832682\\
\hline
 $C_{1,2}^{(-2)} $   &0.00262870   &0.0649589   &0.0361830 &0.101142 \\
\hline
 $C_{1,2}^{(-1)} $   &0.00161692   &0.0463875   &0.0130628  &0.0594503\\
\hline
 $C_{2,2}^{(-4)} $   &0.00138694    &0.00609203   &0.0117589 &0.0178510\\
\hline
 $C_{2,3}^{(-2)} $   &0.00108717   &0.00193670   &0.00627777 &0.00821447 \\
\hline
 $C_{2,3}^{(-1)} $   &0.00134557   &0.0295467   &0.00824463 &0.0377913 \\
\hline
 $C_{3,2}^{(-1)} $   &0.00184791    &0.00144487   &0.0337078 &0.0351526\\
\hline
 $C_{3,3}^{(-4)} $   &0.00128215   &0.00500135    &0.00958245 &0.0145838\\
\hline
 $C_{3,3}^{(-4/7)} $   &0.00256497   &0.000826246   &0.0447810 &0.0456072\\
\hline
 $C_{3,4}^{(-1)} $   &0.000784632   &0.00198832   &0.00201671 &0.00400503\\
\hline
 $C_{4,3}^{(-1)} $   &0.000587861  &0.000536292  &0.00191111 &0.00244740\\
\hline
 $C_{4,4}^{(-4/9)} $   &0.00249203   &0.00386555   &0.0380123 &0.0418779\\
\hline
 $C_{4,5}^{(-1)} $   &0.000673718   &0.000729868   &0.00107283 &0.00180270\\
\hline
 $C_{5,6}^{(-1)} $   &0.000559509   &0.000312770   &0.000391225 &0.000703995\\
\hline
 $C_{7,6}^{(-1)} $   &0.000367060   &0.0000657636   &0.0000453988 &0.000111162\\
\hline
 $C_{6,7}^{(-1)} $   &0.000433024  &0.0000980238  &0.000125464 &0.000223488\\
\hline
 $C_{7,8}^{(-1)} $   &0.000388259   &0.0000680322   &0.0000822403 &0.000150272\\
\hline
 $C_{8,7}^{(-1)} $   &0.000438208  &0.000175849  &0.000128240 &0.000304089\\
\hline
$C_{8,9}^{(-1)} $   &0.000300809  &0.0000323440   &0.0000499792 &0.0000823232\\
\hline
 $\bf C_{9,8}^{(-1)} $   &$\bf 0.0000987414$ &$\bf 8.01947\times 10^{-6}$  
&$\bf 0.0000319896$ &\bf 0.0000400091\\
\hline
 $C_{9,10}^{(-1)} $   &0.000209750   &0.0000112848   &0.0000418221 &0.0000531069\\
\hline
  \end{tabular}
\caption{Result for $L=8$}
\label{tab:L8}
\end{center}
\end{table}

For $8\times 8$ lattice, 
the high and low temperature expansions of $C_8 (g)$ are given by
\begin{\eqa}
C_8 (g)
&=&2+\frac{5 g^2}{2}-\frac{5 g^3}{2}+\frac{29 g^4}{8} +\mathcal{O}(g^5 ) ,\NN\\
&=&g^{-4}\left( 64-256g^{-1} +928g^{-2} -3008g^{-3} +9440g^{-4} 
+\mathcal{O}(g^{-5} )  \right) .
\end{\eqa}
Fitting the data of $s_k$ and $l_k$ from $k=10$ to $k=50$ leads us to
\begin{\eqa}
 |s_k | \sim c_s A_s^k ,\quad &&{\rm with}\ c_s =0.620989 \ {\rm and}\  A_s =1.76359 ,\NN\\
 |l_k | \sim c_l A_l^k ,\quad    &&{\rm with}\ c_l =20.5985 \ {\rm and}\  A_l =3.44257 .
\end{\eqa}
Hence the high and low temperature expansions seem to converge
for $|g|< A_s^{-1} =0.567027$ and $|g|>A_l $, respectively.
The blow-up points are given by $g_s^b =0.522265$ and $g_l^b =3.41772$.
Thus we take the parameters in \eqref{eq:criterion} as
\begin{\eq}
N_s^\ast =50 ,\quad g_s^\ast =0.4,\quad N_l^\ast =50,\quad g_l^\ast =3.7.
\end{\eq}
According to fig.~\ref{fig:ising2d} and tab.~\ref{tab:L8},
we see again that 
our criterion correctly determines the best interpolating function $C_{9,8}^{(-1)} (g)$
among the candidates.

We also find 
the peak locations of $C_{9,8}^{(-1)} (g)$ and $C_8 (g)$ 
at $g=1.27341$ and $g=1.25509$, respectively.
Although the values at the peaks are different by about $16\%$,
these locations are very close to each other.

\subsubsection{Infinite lattice}
\begin{table}[t]
\begin{center}
  \begin{tabular}{|c|c|c|c|c|}
  \hline & $\Lambda^{-1}\int dg \bigl|\frac{C_{m,n}^{(\alpha )}-C_\infty}{C_\infty} \bigr|$ &$I_s [C_{m,n}^{(\alpha )}]$ &$I_l [C_{m,n}^{(\alpha )}]$ & $I_s +I_l$ \\
  \hline\hline $C_{1,1}^{(-2/3)} $   &0.00275349 &0.800907 &0.481499 &1.28241 \\
\hline
 $C_{2,1}^{(-1)} $   &0.00313148 &0.707807 &0.514001  &1.22181 \\
\hline
 $C_{2,2}^{(-2/5)} $   &0.00157317 &0.986636 &0.462886 &1.44952 \\
\hline
 $C_{2,3}^{(-1)} $   &0.00165066  &0.960887  &0.652703 &1.61359 \\
\hline
 $C_{3,4}^{(-1)} $   &0.00105518  &0.610940   &0.357507 &0.968448\\
\hline
 $C_{4,4}^{(-2/9)} $   &0.00141961  &0.927616  &0.538109 &1.46572 \\
\hline
 $C_{4,5}^{(-1)} $   &0.000963535   &0.548868   &0.308996 &0.857864 \\
\hline
 $C_{5,5}^{(-2/11)} $   &0.00145344   &0.901255   &0.563952 &1.46521 \\
\hline
 $C_{5,6}^{(-1)} $   &0.000715206   &0.306906   &0.194745 &0.501651\\
\hline
 $C_{6,6}^{(-2/13)} $   &0.00149621  &0.921327   &0.582883 &1.50421 \\
\hline
 $C_{6,7}^{(-1)} $   &0.000400117  &0.0916321   &0.0596090 &0.151241 \\
\hline
 $C_{7,6}^{(-1)} $   &0.000304098   &0.0566016  &0.0188506 &0.0754521\\
\hline
 $C_{7,7}^{(-2/15)} $   &0.00149432  &0.899986  &0.584099 &1.48409\\
\hline
 $C_{7,8}^{(-1)} $   &0.000405540  &0.0955453 &0.0614392  &0.156985\\
\hline
 $C_{9,8}^{(-1)} $   &0.000361400  &0.0533636  &0.0521376 &0.105501\\
\hline
 $C_{9,10}^{(-1)} $   &0.000239733  &0.0222130  &0.0123199  &0.0345328\\
\hline
   $\bf C_{10,9}^{(-1)} $   &$\bf 0.000170155$   &$\bf 0.00916161$  &$\bf 0.0250438$ &$\bf 0.0342054$\\
\hline
  \end{tabular}
\caption{Result for $L\rightarrow\infty$}
\label{tab:Linf}
\end{center}
\end{table}

In the $L\rightarrow \infty$ limit, $C_\infty (g)= \lim_{L\rightarrow\infty} C_L (g)$ is given by
\begin{\eqa}
C_\infty (g)
&=& \frac{16 (g+1)}{\pi  g (g+2)} 
\Biggl[ K\left(\frac{4 g (g+1) (g+2)}{\left(g^2+2 g+2\right)^2}\right) 
-E\left( \frac{4 g (g+1) (g+2)}{\left(g^2+2 g+2\right)^2}\right) \NN\\
&& -\left( \frac{2 (g+1)}{ (g+1)^2+1} \right)^2
 \left\{ \frac{g^4+4g^3-8 g-4}{\left(g^2+2 g+2\right)^2} K\left( \frac{4 g (g+1) (g+2)}{\left(g^2+2 g+2\right)^2}\right) +\frac{\pi}{2} \right\} 
\Biggr] ,
\end{\eqa}
where $K(z)$ and $E(z)$ are 
the complete elliptic integrals of the first and second kinds\footnote{
These are defined by
$K(z) = \int_0^{\pi /2} dt ( 1-z\sin^2{t} )^{-1/2}$,
$E(z) = \int_0^{\pi /2} dt ( 1-z\sin^2{t} )^{1/2}$.
}, respectively.
The function $C_\infty (g)$ has the expansions
\begin{\eqa}
C_\infty (g)
&=&6-\frac{11 g}{4}-\frac{15 g^2}{4}+\frac{3265 g^3}{256}-\frac{651 g^4}{64}
 +\mathcal{O}(g^5 ) ,\NN\\
&=&g^{-2}\left( 16 -72g^{-1} +164g^{-2} +15g^{-3}  -\frac{3087}{4}g^{-4} 
+\mathcal{O}(g^{-5} )  \right) .
\end{\eqa}
By fitting the data of $s_k$ and $l_k$ for $k=10\sim 50$,
we find
\begin{\eqa}
 |s_k | \sim c_s A_s^k ,\quad &&{\rm with}\ c_s =183.036 \ {\rm and}\  A_s =0.762512 ,\NN\\
 |l_k | \sim c_l A_l^k ,\quad    &&{\rm with}\ c_l =545.697 \ {\rm and}\  A_l =1.51424 .
\end{\eqa}
Hence the high and low temperature expansions converge
for $|g|< A_s^{-1} =1.31145$ and $|g|>A_l $, respectively.
We also find the blow-up points $g_s^b =1.14147$ and $g_l^b =1.68176$.
Thus we take the parameters in \eqref{eq:criterion} as
\begin{\eq}
N_s^\ast =50 ,\quad g_s^\ast =1,\quad N_l^\ast =50,\quad g_l^\ast =2.
\end{\eq}
From tab.~\ref{tab:Linf}, we see that
the best interpolating function $C_{10,9}^{(-1)} (g)$ minimizes $I_s +I_l$.
Hence we conclude that our criterion works as in the finite $L$ cases.

However,
we observe from fig.~\ref{fig:ising2d} that
the best interpolating functions do not reproduce the phase transition and
have the infinite discrepancies at the critical point.
This is true for all the interpolating functions listed in tab.~\ref{tab:Linf}
since our interpolating functions are smooth by construction
unless they have poles.
Therefore, even if our criterion to choose the relatively good interpolating function works,
the approximation itself by the interpolating function is failed.
This indicates that
our interpolating scheme cannot be {\it directly} used for approximation
when we have a phase transition. 

\begin{figure}[t]
\begin{center}
\includegraphics[width=7.4cm]{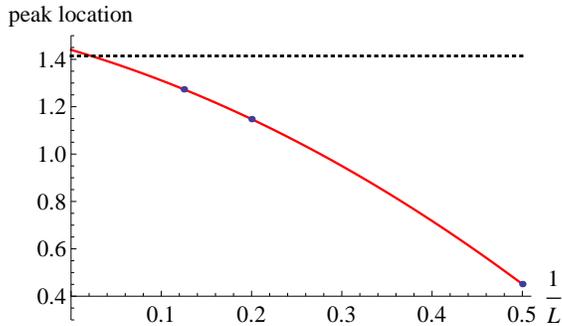}
\end{center}
\caption{
The peak location $g_p (L)$ of the best interpolating functions for $L=2,5$ and $8$ 
is plotted to $1/L$ (symbol). 
The solid line shows fitting by the ansatz $g_p (L) =p_0 +p_1/L +p_2 /L^2$ 
with $p_0 =1.44004$, $p_1 =-1.1182$ and $p_2 =-1.71877$.
The dotted line denotes ``$\sqrt{2}$", which is the phase transition point 
in the $L\rightarrow\infty$ limit.
}
\label{fig:peak}
\end{figure}

On the other hand, 
the Pad\'e approximant with some $(m,n)$ should be exact for finite $L$
since the exact answer is the rational function of $g$.
This implies that
even if we have a phase transition in discrete system in infinite volume limit,
our interpolating scheme would be {\it indirectly} useful to extract information on phase transition.
Indeed we have seen that
the best interpolating functions precisely give the peak locations of the exact answers for finite $L$
although we need higher $(m,n)$ to correctly reproduce the values at the peaks except $L=2$.
Indeed if we perform fitting of the peak locations of the best interpolating functions as a function of $L$,
then we find that the peak location in the $L\rightarrow\infty$ is given by $g=1.44004$
as depicted\footnote{
Note that the success of the fitting itself is trivial since this is the three parameter fitting in terms of the three points
but the value of the intercept is nontrivial.
} in fig.~\ref{fig:peak}.
This value is fairly close to the phase transition point $g=\sqrt{2}$. 
We expect that
this feature is true also for more general discrete systems.

\subsection{Average plaquette in pure $SU(3)$ Yang-Mills theory on lattice}
In this subsection\footnote{
Historically, this work originated in studying FPP in this example in the early collaborations with Sen and Takimi.
Then we encountered the landscape problem of interpolating functions.
We thank them for this point.
}, we study average plaquette in 4d pure $SU(3)$ Yang-Mills theory on lattice.
Although we can compute this by preforming Monte Carlo simulation as first performed by Creutz \cite{Creutz:1980wj},
exact analytical result is unknown.
\subsubsection*{Set up}
Let us consider the pure $SU(3)$ Yang-Mills theory on 4d hypercubic lattice
with the standard Wilson action \cite{Wilson:1974sk}
\begin{\eq}
S
=\beta \sum_{\mu <\nu} \sum_{\mathbf{x}} \Bigl[ 1
 -\frac{1}{3} {\rm ReTr} U_{\mathbf{x} ,\mu} U_{\mathbf{x} +\hat{\mu} ,\nu}   U_{\mathbf{x} +\hat{\nu} ,\mu}^\dag U_{\mathbf{x}  ,\nu}^\dag \Bigr] ,
\end{\eq}
where $U_{\mathbf{x} ,\mu}$ is the link variable at the position $\mathbf{x}$ with direction $\mu$ and
$\hat{\mu}$ is the unit vector along $\mu$-direction.
The parameter $\beta$ is related to the bare gauge coupling $g_0^2$ by $\beta =6/g_0^2$.
In terms of the link variables, we introduce average plaquette by
\begin{\eqa}
P (\beta )
= \Bigl\langle 1
 -\frac{1}{3}{\rm Tr} U_{\mathbf{x} ,\mu} U_{\mathbf{x} +\hat{\mu} ,\nu}   U_{\mathbf{x} +\hat{\nu} ,\mu}^\dag U_{\mathbf{x}  ,\nu}^\dag \Bigr\rangle  .  
\end{\eqa}
Although we can obtain numerical value of this by performing Monte Carlo simulation,
there is no known analytical result of this quantity for arbitrary $\beta$.
However, we have analytical result of strong coupling expansion and
numerical result of weak coupling expansion with high precision on infinite lattice.

\begin{figure}[t]
\begin{center}
\includegraphics[width=7.4cm]{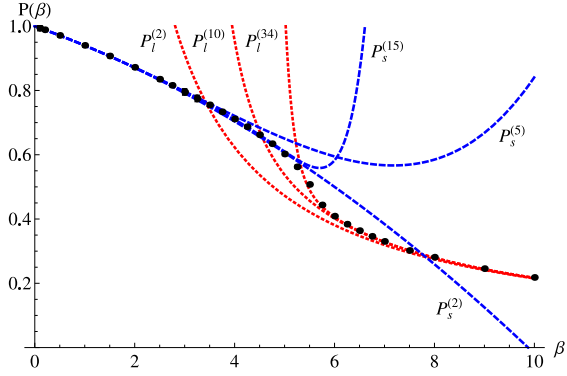}
\end{center}
\caption{
The average plaquette $P(\beta )$ in the 4d pure $SU(3)$ Yang-Mills theory is plotted to $\beta$.
}
\label{fig:SU3_0}
\end{figure}

The result of the strong coupling expansion around $\beta =0$ is given by (see e.g.~\cite{Balian:1974xw})
\begin{\eq}
P_s^{(15)} (\beta ) = \sum_{k=0}^{15} s_k \beta^k ,
\end{\eq}
where the coefficients are explicitly given by
\begin{\eqa}
&& s_0 = 1,\quad s_1 = -\frac{1}{18} ,\quad s_2 =  -\frac{1}{216} ,\quad s_3 =0,\quad 
s_4 =\frac{5}{93312} ,\quad  s_5 =\frac{49}{15116544},\NN\\
&&s_6 = -\frac{1309 }{906992640} ,\quad s_7 =-\frac{2131}{5441955840} ,\quad s_8 = -\frac{1091 }{43535646720} ,\NN\\
&&s_9 =  \frac{179081}{21158324305920} ,\quad s_{10} =\frac{1277749}{592433080565760} ,\quad 
s_{11}=\frac{93151153 }{11516899086198374400} ,\NN\\
&&s_{12} =   -\frac{3052831769 }{34550697258595123200} ,\quad 
s_{13}=-\frac{6757393949 }{414608367103141478400} ,\NN\\
&& s_{14}=\frac{1932793007}{3198407403367091404800} ,\quad s_{15}=\frac{16029793987553 }{21761963972509689918259200} .
\end{\eqa}
The result of the weak coupling expansion around $\beta =\infty$ 
is \cite{Bali:2014fea} (see also \cite{DiRenzo:1995qc,DiRenzo:2000ua,Horsley:2012ra,Di Renzo:2004ge})
\begin{\eq}
P_l^{(34)} (\beta ) = \frac{1}{\beta} \sum_{k=0}^{34} l_k \beta^{-k} ,
\end{\eq}
where\footnote{
Actually these values have errors and we are using just their center values.
See \cite{Bali:2014fea} for details.
}
\begin{\eqa}
&&l_0 =2,\quad l_1 = 1.22084,\quad l_2 =2.96043,\quad 
l_3 =9.40538,\quad l_4 =34.3245,\NN\\
&&l_5 =136.471,\quad l_6 =575.197,\quad l_7 =2528.31,\quad l_8 =11474.2,\quad
l_9 =53450.1 ,\NN\\
&& l_{10} =254138,\quad l_{11}=1.22928\times 10^6 ,\quad l_{12}=6.02757\times 10^6,\quad
l_{13}=2.99220\times 10^7,\NN\\ 
&& l_{14}=1.50156\times 10^8 ,\quad l_{15}=7.60422\times 10^8 ,\quad
l_{16}=3.88155\times 10^9 ,\NN\\
&& l_{17} =1.99751\times 10^{10} ,\quad  l_{18}=1.03474\times 10^{11},\quad 
l_{19}=5.38793\times 10^{11},\NN\\
&& l_{20}=2.82781\times 10^{12},\quad l_{21}=1.4928\times 10^{13},\quad
l_{22}=7.94067\times 10^{13} ,\NN\\
&& l_{23}=4.25644\times 10^{14},\quad l_{24}=2.30516\times 10^{15},\quad
l_{25} =1.26685\times 10^{16} ,\NN\\
&& l_{26}=7.12124\times 10^{16},\quad l_{27}=4.13752\times 10^{17},\quad
l_{28}=2.5183\times 10^{18} ,\NN\\
&& l_{29}=1.63433\times 10^{19},\quad l_{30}=1.15935\times 10^{20},\quad
l_{31}=8.72907\times 10^{20},\NN\\
&& l_{32}=6.86167\times 10^{21},\quad l_{33}=5.82436\times 10^{22},\quad
l_{34}=5.09837\times 10^{23} .
\end{\eqa}
The plot in fig.~\ref{fig:SU3_0} shows 
the result of our Monte Carlo simulation\footnote{
We have used hybrid Monte Carlo algorithm (see e.g.~\cite{Rothe:1992nt}).
We are also assuming that the $10^4$ lattice is sufficiently large 
to determine the best interpolating function in terms of the numerical data.
} on $10^4$ lattice, $P_s^{(N_s )} (\beta )$ and $P_l^{(N_l )} (\beta )$ (for some values of $N_s$ and $N_l$)
against $\beta$.
Now we determine
the parameters $\beta_s^\ast$, $N_s^\ast$, $\beta_l^\ast$ and $N_l^\ast$.

\subsubsection*{Determination of $\beta_s^\ast$}
\begin{figure}[t]
\begin{center}
\includegraphics[width=7.0cm]{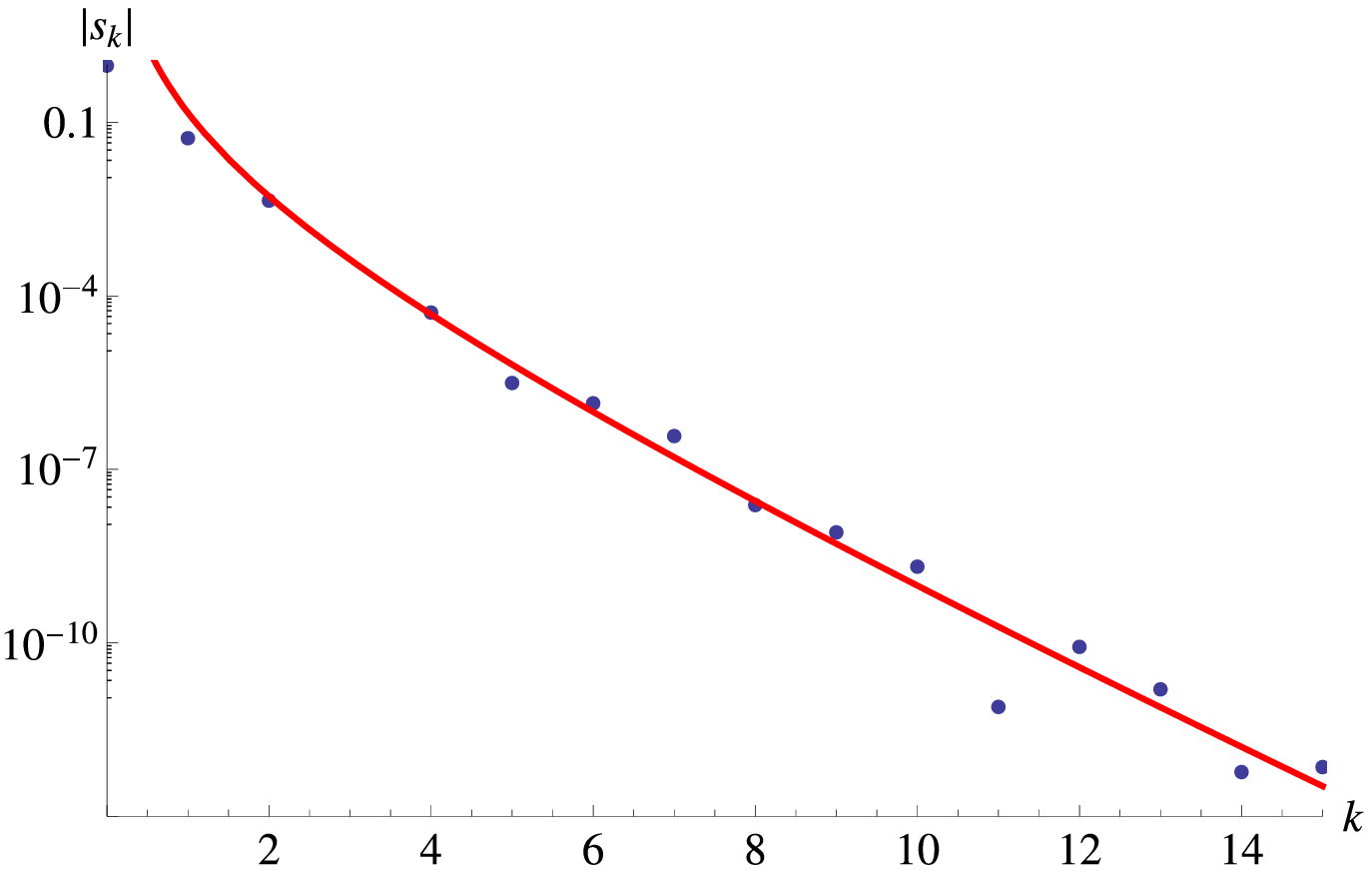}
\includegraphics[width=7.0cm]{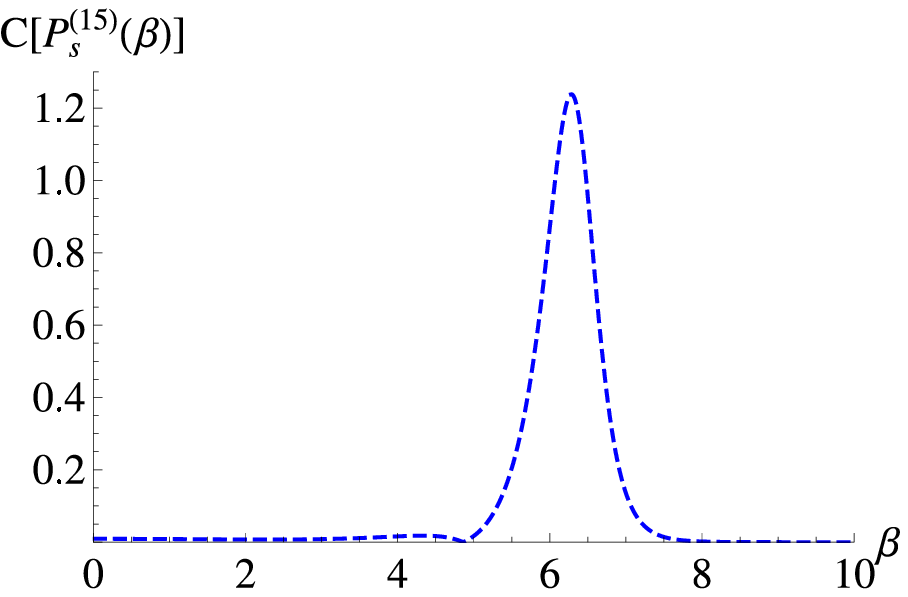}
\end{center}
\caption{
[Left] The absolute value of the $k$-th coefficient of the strong coupling expansion is plotted against $k$ in semi-log scale.
The solid line shows the fitting function $\log{|s_k |}=\log{c_s} +a_s \log{k} +k\log{A_s}$.
[Right] The absolute value of the curvature of $P_s^{(15)}(\beta )$ is plotted to $\beta$.
}
\label{fig:SU3_sc}
\end{figure}
Let us study large order behavior of the strong coupling expansion.
In fig.~\ref{fig:SU3_sc} [Left],
we plot $|s_k|$ against $k$ in semi-log scale.
By fitting this by the ansatz 
\begin{\eq}
|s_k | = c_s k^{a_s} A_s^k ,
\end{\eq}
we obtain
\begin{\eq}
c_s = 0.576352 ,\quad a_s = -2.82846,\quad A_s = 0.254563.
\end{\eq}
This implies that the strong coupling expansion is convergent inside the circle\footnote{
Convergence of strong coupling expansion in lattice gauge theory 
has been proven in \cite{Osterwalder:1977pc}.
Here we are not stating that we precisely determine the convergent radius of the strong coupling expansion.
We are understanding that this is fairly rough estimate and
this problem itself is the big issue in the subject of quantum field theory.
We just adopt this as the reference value for determining $\beta_s^\ast$.
} $|\beta |\sim A_s^{-1} =3.9283$.
We also plot the curvature of $P_s^{(15)}$ to $\beta$ in fig.~\ref{fig:SU3_sc} [Right] in order to find blow-up point.
Then 
we find that $P_s^{(15)}$ blows up at $\beta =\beta_s^b =6.28417$.
Thus we take the values of $N_s^\ast$ and $\beta_s^\ast$ as
\begin{\eq}
N_s^\ast = 15,\quad  \beta_s^\ast =3.9 .
\end{\eq}

\subsubsection*{Determination of $\beta_l^\ast$}
\begin{figure}[t]
\begin{center}
\includegraphics[width=7.0cm]{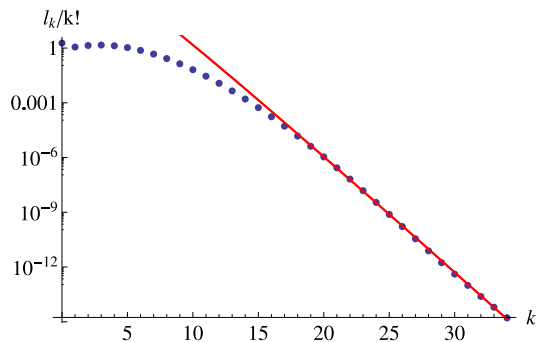}
\end{center}
\caption{
The weak coupling expansion coefficient $l_k$ divided by $k!$
is plotted to $k$ in semi-log scale.
The line denotes fitting by the function $\log{(l_k /k!)} =\log{c_l} +a_l \log{k} +k\log{A_l}$.
}
\label{fig:SU3_wc}
\end{figure}
Next we study large order behavior of the weak coupling expansion.
Fig.~\ref{fig:SU3_wc} plots $l_{k}/k!$ against $k$ in semi-log scale.
From this plot we find that the coefficient in large $k$ regime grows as
\begin{\eq}
\frac{l_k}{k!} \sim c_l k^{a_l} A_l^k ,\quad
{\rm with} \ c_l =0.0000946747,\ a_l =11.1583\ {\rm and}\ A_l =0.14984.
\end{\eq}
This exhibits that the coefficient grows by the factorial and 
hence the weak coupling expansion is asymptotic.
By performing optimization,
we find
\begin{\eq}
N_l^o= \frac{\beta}{A_l} ,\quad 
\delta_l (\beta ) = c_l \left( \frac{\beta}{A_l} +1\right)^{a_l} e^{-\frac{\beta}{A_l}} .
\end{\eq}
If we take $\epsilon_l  = 10^{-4}$,
then we should take $N_l^o =40$ and $\beta_l^\ast =5.99358 $.
However, since we know the only first 35 coefficients of the weak coupling expansion, 
we cannot perform this optimization and instead we demand
\begin{\eq}
\left. \left( c_l k^{a_w} A_l^k (\beta_l^\ast )^{-1-k} \right) \right|_{k=35} = \epsilon_l  .
\end{\eq}
Thus we take
\begin{\eq}
N_l^\ast =34,\quad \beta_l^\ast = 6.13706 .
\end{\eq}

\subsubsection*{Result}

\begin{table}[t]
\begin{center}
  \begin{tabular}{|c|c|c|c|c|}
  \hline & $\frac{1}{29}\sum_i \bigl| \frac{P_{m,n}^{(\alpha )}  -P}{P} \bigr|$ &$I_s [P_{m,n}^{(\alpha )}]$ &$I_l [P_{m,n}^{(\alpha )}]$ &$I_s +I_l$ \\
  \hline\hline 
$P_{1,1}^{(-1)}$ & 0.228616  &0.634296 &0.222215 &0.856510 \\
\hline
$P_{1,1}^{(-1/3)}$ & 0.115055 &0.206451 &0.070088 &0.276539 \\
\hline
$P_{2,2}^{(-1)}$ &0.158456  &0.380170 &0.0924484 &0.472619\\
\hline
$P_{3,3}^{(-1)}$ &0.119927  &0.247194  &0.0472852  &0.294479\\
\hline
$P_{4,4}^{(-1)}$ &0.0956988  &0.168693 &0.0272632 &0.195956 \\
\hline
$P_{5,5}^{(-1)}$ &0.0790835  &0.118552 &0.0169992 &0.135551\\
\hline
$P_{6,6}^{(-1)}$ &0.0670207  &0.0848353  &0.0112119  &0.0960472\\
\hline
$P_{7,7}^{(-1)}$ &0.0579211  &0.0614099  &0.00772215 &0.0691320 \\
\hline
$P_{8,8}^{(-1)}$ &0.0508609   &0.0447886   &0.00550651 &0.0502934 \\
\hline
$P_{9,9}^{(-1)}$ &0.0452512  &0.0328091 &0.00403859 &0.0368477 \\
\hline
$P_{10,10}^{(-1)}$ &0.0406960  &0.0240752  &0.00303056 &0.0271057 \\
\hline
$P_{11,11}^{(-1)}$ &0.0369267  &0.0176544 &0.00231792 &0.0199723 \\
\hline
$P_{12,12}^{(-1)}$ &0.0337611 &0.0129187  &0.00180261 &0.0147214\\
\hline
$P_{13,13}^{(-1)}$ &0.0310727  &0.00942950 &0.00142323 &0.0108527\\
\hline
$P_{14,14}^{(-1)}$ &0.0287673  &0.00686572  &0.00113935 &0.00800507 \\
\hline
$\bf P_{15,15}^{(-1)}$ &$\bf 0.0267697$  &$\bf 0.00498586$ &$\bf 0.000923484$ &$\bf 0.00590935$ \\
\hline
  \end{tabular}
\caption{
Result on the 4d $SU(3)$ pure Yang-Mills theory on lattice.
}
\label{tab:SU3}
\end{center}
\end{table}

\begin{figure}[t]
\begin{center}
\includegraphics[width=7.4cm]{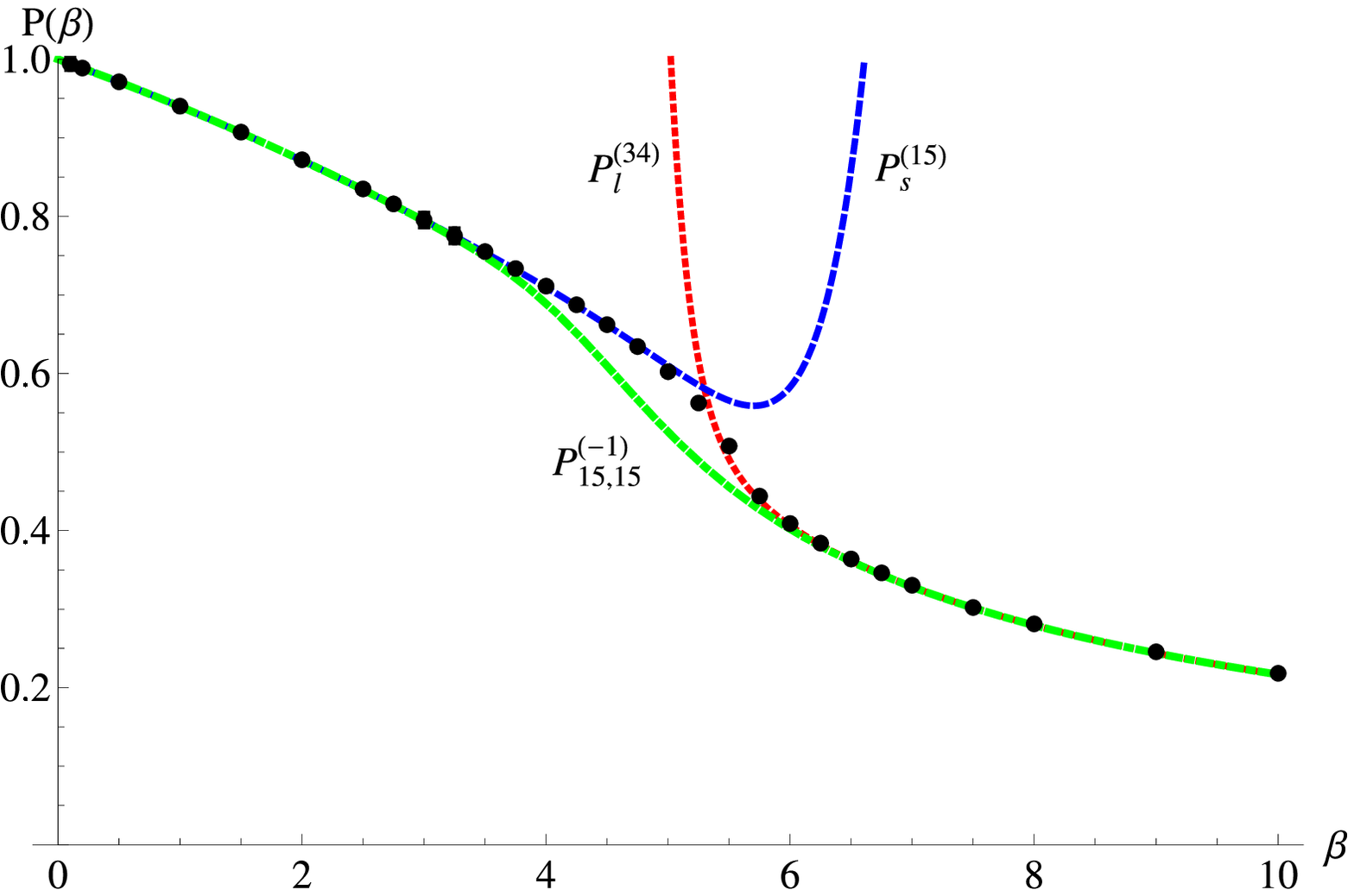}
\includegraphics[width=7.4cm]{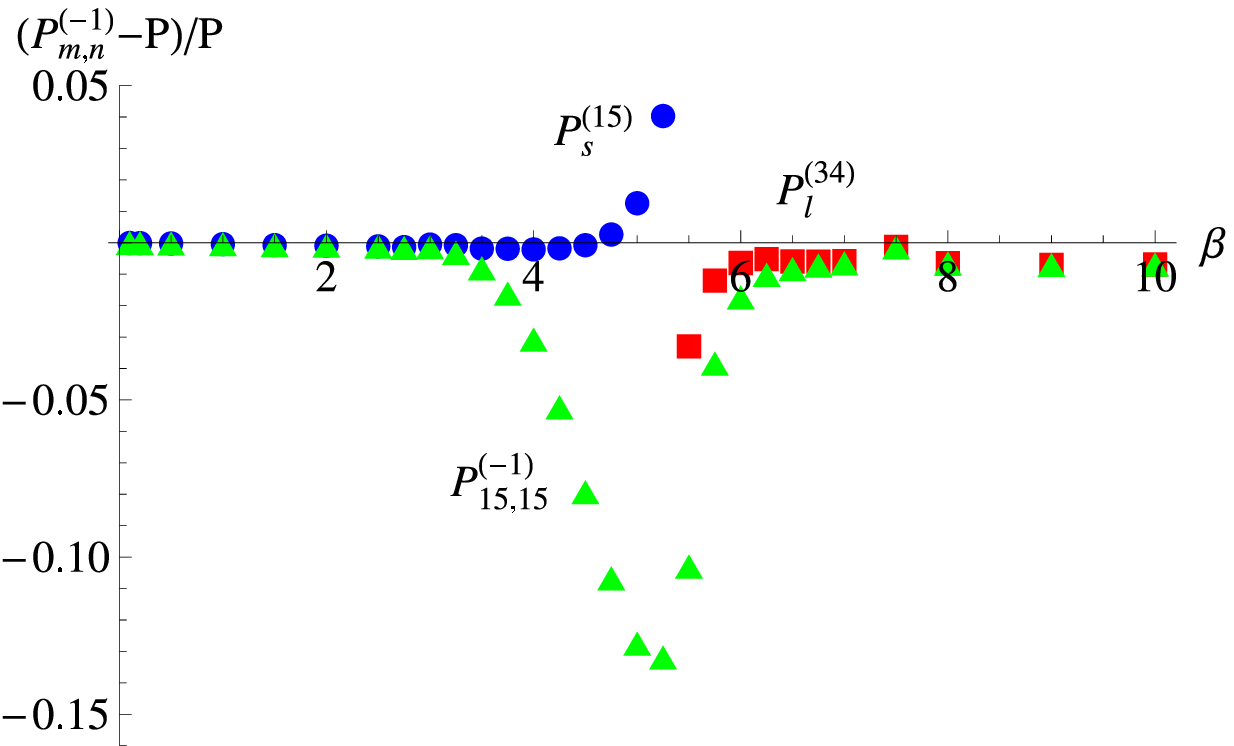}
\end{center}
\caption{
[Left]  The best interpolating function $P_{15,15}^{(-1)}(\beta )$ (green dot-dashed),
the strong coupling expansion $P_s^{(15)}(\beta )$ (blue dashed) and
the weak coupling expansion $P_l^{(34)}(\beta )$ (red dotted) are plotted to $\beta$
with the Monte Carlo simulation result (symbol).
[Right] The ratios $(P_{15,15}^{(-1)}(\beta ) -P(\beta ))/P(\beta )$ (green triangle),
$(P_s^{(15)}(\beta ) -P(\beta ))/P(\beta )$ (blue circle) and 
$(P_l^{(34)}(\beta ) -P(\beta ))/P(\beta )$ (red square) 
are plotted against $\beta$.
Here $P(\beta )$ means the Monte Carlo result.
}
\label{fig:SU3_final}
\end{figure}

Let us test our criterion in this example.
In terms of the strong and weak coupling expansions,
we can construct interpolating functions $P_{m,n}^{(\alpha )}(\beta )$ 
of the average plaquette $P(\beta )$ taking the FPR form \eqref{eq:FPR}.
Since we do not know exact analytical result for the plaquette,
we measure precision of approximation in terms of the Monte Carlo result by using\footnote{
We take $(\beta_1 ,\cdots ,\beta_{29})=$ $(0.1,$ $0.2,$ $0.5,$ $1.0,$ $1.5,$ $2.0,$ $2.5, $ $2.75,$ $3.0,$ $3.25,$ $3.5,$ $3.75,$ $4.0,$ $4.25,$ 
$4.5,$ $4.75,$ $5.0,$ $5.25,$ $5.5,$ $5.75,$ $6.0,$ $6.25,$ $6.5,$ $6.75,$ $7.0,$ $7.5,$ $8.0,$ $9.0,$ $10)$.
}
\begin{\eq}
\frac{1}{29}\sum_{i=1}^{29} \left| \frac{P_{m,n}^{(\alpha )}(\beta_i )  -P(\beta_i )}{P(\beta_i )} \right| .
\end{\eq}
This quantity should be minimized by the best interpolating function.
Compared with values of $I_s [P_{m,n}^{(\alpha )}]$ and $I_l [P_{m,n}^{(\alpha )}]$,
we test validity of our criterion.
We summarize our result in tab.~\ref{tab:SU3}.
Explicit formula for the interpolating functions can be found in app.~\ref{app:YM}.
We easily see that 
the interpolating function $P_{15,15}^{(-1)}(\beta )$ gives
the most precise approximation of $P(\beta )$,
which minimizes $I_s$ and $I_l$ among the interpolating functions.
This shows that our criterion correctly chooses the best interpolating function for this case.
We also plot the best interpolating function $P_{15,15}^{(-1)}(\beta )$ and
its normalized difference from the Monte Carlo result in fig.~\ref{fig:SU3_final}.
We find that 
$P_{15,15}^{(-1)}(\beta )$ approximates the average plaquette 
with about $13\%$ error at worst.
Because finding interpolating functions becomes so heavy for larger $m+n$,
we have stopped this up to $m+n=30$. 
It is interesting 
if we use our full knowledge about the weak and strong coupling expansions 
to construct better interpolating functions.

\subsection{Free energy in $c=1$ string theory at self-dual radius}
Let us consider so-called $c=1$ string theory at self-dual radius (for excellent reviews, see \cite{Klebanov:1991qa,Ginsparg:1993is,Nakayama:2004vk}).
The free energy in the $c=1$ string theory has the following integral representation \cite{Gross:1990ay,Gross:1990ub,Pasquetti:2009jg}
\begin{\eq}
\tilde{F}(\mu ) =
\frac{\mu^2}{2}\log{\mu} -\frac{1}{12}\log{\mu}
+\frac{1}{4}\int_0^\infty \frac{ds}{s}
\left( \frac{1}{\sinh^2{s}} -\frac{1}{s^2} +\frac{1}{3} \right) e^{-2\mu s} ,
\end{\eq}
where $\mu$ is the cosmological constant related to the string coupling constant $g_s$ roughly by $\mu \sim g_s^{-1}$.
This behavior can be also seen in critical behavior of topological string at the conifold point \cite{Ghoshal:1995wm}.
Beside the contexts of string theory, 
the last term in the above equation has also an interpretation from Schwinger effect \cite{Schwinger:1951nm,Dunne:2004nc}
as noted in \cite{Pasquetti:2009jg}.

\begin{figure}[t]
\begin{center}
\includegraphics[width=7.4cm]{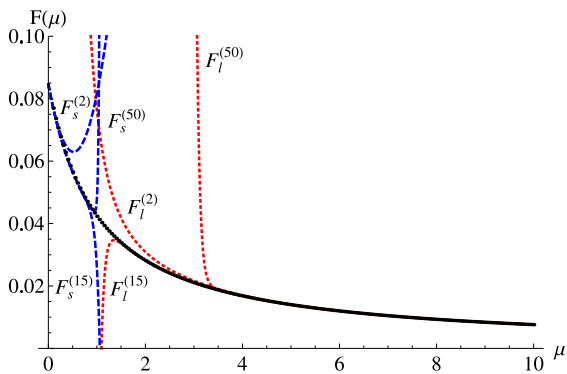}
\end{center}
\caption{
The free energy $F(\mu )$ in the $c=1$ string theory at self-dual radius 
is plotted to $\mu$.
}
\label{fig:string0}
\end{figure}

We can perform large $\mu$ and small $\mu$ expansions of $\tilde{F}(\mu )$.
The large $\mu$ expansion, which we call weak coupling expansion, is given by 
\begin{\eq}
\tilde{F}(\mu ) =
\frac{\mu^2}{2}\log{\mu} -\frac{1}{12}\log{\mu}
-\sum_{p\geq 2} \frac{B_{2p} }{2p(2p-2)} \mu^{2-2p}  ,
\end{\eq}
while the small $\mu$ expansion, which we refer to as strong coupling expansion, is given by
\begin{\eqa}
\tilde{F}(\mu ) 
= \mu^2 \log{\mu} -\frac{1}{6}\log{\mu}
+\frac{1}{12} -\log{A}  +2\mu 
     +\mu^2 (-1 +2\gamma )
+ \sum_{p\geq 3} \frac{(-\mu )^p}{p} \zeta (p-1)  .
\end{\eqa}
Here $A$ and $\gamma$ are Glaisher constant and Euler constant, respectively.
Since the two expansions of $\tilde{F}(\mu ) $ do not take the power series forms,
we redefine the free energy as
\begin{\eq}
F(\mu ) 
=\tilde{F}(\mu ) -\left( \frac{\mu^2}{2} -\frac{1}{12}\right) \log{\mu}
+\left( \frac{\mu^2}{2} -\frac{1}{12}\right) \log{\frac{\mu +1}{\mu}} -\frac{\mu}{2} +\frac{1}{4}.
\end{\eq}
Then we find that $F(\mu )$ has the following two power series expansions
\begin{\eqa}
F(\mu ) 
&=& \left(\frac{1}{3}-\log{A}\right)-\frac{\mu }{12}+\left(\frac{\gamma }{2}-\frac{5}{24}\right) \mu
   ^2+\left(\frac{17}{36}-\frac{\pi ^2}{18}\right) \mu ^3+\left(\frac{\zeta (3)}{4}-\frac{11}{48}\right) \mu^4 +\mathcal{O}\left(\mu ^5\right) \NN\\
&=& \mu^{-1} \left( \frac{1}{12}-\frac{19}{240 }\mu^{-1}
+\frac{13}{180}\mu^{-2} -\frac{4}{63}\mu^{-3} +\frac{23}{420}\mu^{-4}
+\mathcal{O}\left( \mu^{-5}  \right) \right) .
\end{\eqa}
Let us call the above two expansions $F_s^{(N_s )}(\mu )$ (up to $\mathcal{O}(\mu^{N_s})$) 
and $F_l^{(N_l )}(\mu )$ (up to $\mathcal{O}(\mu^{-1-N_l})$):
\begin{\eq}
F_s^{(N_s )}(\mu ) = \sum_{k=0}^{N_s} s_k \mu^k ,\quad
F_l^{(N_l )}(\mu ) = \mu^{-1} \sum_{k=0}^{N_l} l_k \mu^{-k} .
\end{\eq}
We assume again that
we have information only on $F_s^{(50 )}(\mu )$ and $F_l^{(50 )}(\mu )$, and
read large order behaviors of $s_k$ and $l_k$ by extrapolation.
In fig.~\ref{fig:string0},
we plot the result of numerical integration, $F_s^{(N_s )}(\mu )$ and $F_l^{(N_l )}(\mu )$
($N_s ,N_l =2,15,50$) against $\mu$.

\subsubsection*{Determination of $\mu_s^\ast$}
By fitting the data of $s_k$ for $k=15\sim 50$,
we find
\begin{\eq}
 |s_k | \sim c_s k^{a_s} ,\quad {\rm with}\ c_s =0.415148\ {\rm and}\ a_s =-0.920281 .
\end{\eq}
Thus we expect that
the strong coupling expansion is convergent for $|\mu |<1$.
We also find a blow-up point of $F_s^{(50)}(\mu )$ around $\mu = 1.01862$ and hence take
\begin{\eq}
N_s^\ast =50 ,\quad \mu_s^\ast =0.8 .
\end{\eq}

\subsubsection*{Determination of $\mu_l^\ast$}
Fitting the data of $l_k$ for $k=10\sim 50$ shows\footnote{
The even order coefficient $l_{2k}$ has the contribution only from
$(\mu^2 /2 -1/12)\log{(1 +1/\mu)}$, whose large $\mu$ expansion is convergent.
}
\begin{\eq}
 \frac{|l_{2k+1}|}{(2k)!} \sim c_l A_l^k ,\quad 
{\rm with}\ c_l =0.220021\ {\rm and}\ A_l =0.0284273 . 
\end{\eq}
Thus the weak coupling expansion is asymptotic and
its optimization leads us to
\begin{\eq}
N_l^o = \frac{\mu}{\sqrt{A_l}} +1 ,\quad
\delta_l (\mu ) = \frac{c_l}{\mu} e^{-\frac{\mu}{\sqrt{A_l}}} . 
\end{\eq}
Taking $\epsilon_l =10^{-9}$, we find
\begin{\eq}
 N_l^o =N_l^\ast  =21 ,\quad \mu_l^\ast =3.37208 .
\end{\eq}

\subsubsection*{Result}
\begin{table}[t]
\begin{center}
  \begin{tabular}{|c|c|c|c|c|}
  \hline & $\frac{1}{199}\sum_i  \bigl|\frac{F_{m,n}^{(\alpha )}-F}{F} \bigr|$ &$I_s [F_{m,n}^{(\alpha )}]$ &$I_l [F_{m,n}^{(\alpha )}]$ & $I_s +I_l$ \\
  \hline\hline $F_{1,1}^{(-1)} $   &0.000136491& $6.53166\times 10^{-6}$ & $9.73821\times 10^{-6}$ & 0.0000162699 \\
\hline
 $F_{1,1}^{(-1/3)} $   &0.0000731361 & $2.70362\times 10^{-6}$ & $5.61321\times 10^{-6}$ & $8.31684\times 10^{-6}$ \\
\hline
 $F_{2,2}^{(-1)} $   &$4.19545\times 10^{-6}$ & $1.51251\times 10^{-7}$ & $2.44485\times 10^{-7}$ & $3.95735 \times 10^{-7}$ \\
\hline
 $F_{2,2}^{(-1/3)} $   &0.0000238415 & $8.52636\times 10^{-7}$ & $1.46650\times 10^{-6}$ & $2.31914 \times 10^{-6}$ \\
\hline
 $F_{2,2}^{(-1/5)} $   &$6.3568\times 10^{-6}$ & $5.54772\times 10^{-7}$ & $2.10621\times 10^{-7}$ & $7.65393 \times 10^{-7}$ \\
\hline
 $F_{3,3}^{(-1)} $   &$2.68298\times 10^{-7}$ & $8.50196\times 10^{-9}$ & $1.36795\times 10^{-8}$ & $2.21815 \times 10^{-8}$ \\
\hline
 $F_{3,3}^{(-1/3)} $   &$3.80959\times 10^{-7}$ & $1.86495\times 10^{-8}$ & $1.30998\times 10^{-8}$ & $3.17493 \times 10^{-8}$ \\
\hline
 $F_{3,3}^{(-1/7)} $   &$1.50436\times 10^{-6}$ & $9.96742 \times 10^{-8}$ & $4.56374\times 10^{-8}$& $1.45312 \times 10^{-7}$ \\
\hline
 $F_{4,4}^{(-1)} $   &$5.60104\times 10^{-9}$ & $1.19446\times 10^{-9}$ & $1.25679\times 10^{-10}$ & $1.32014 \times 10^{-9}$ \\
\hline
 $F_{4,4}^{(-1/3)} $   &$5.48471\times 10^{-8}$ & $2.58923\times 10^{-9}$ & $1.67657\times 10^{-9}$ & $4.26580\times 10^{-9}$ \\
\hline
 $F_{4,4}^{(-1/9)} $   &$4.17108\times 10^{-8}$ & $2.65563\times 10^{-9}$ & $1.88948\times 10^{-9}$ & $4.54512 \times 10^{-9}$ \\
\hline
 $\bf F_{5,5}^{(-1)} $   &$\bf 5.40195\times 10^{-9}$ & $\bf 1.18856\times 10^{-9}$ & $\bf 1.20303\times 10^{-10}$ & $\bf 1.30886 \times 10^{-9}$ \\
\hline
 $F_{5,5}^{(-1/11)} $   &$6.11560\times 10^{-8}$ & $6.08949\times 10^{-9}$ & $7.90103\times 10^{-10} $& $6.87959 \times 10^{-9}$ \\
\hline
  \end{tabular}
\caption{Result on the $c=1$ string theory at self-dual radius}
\label{tab:string}
\end{center}
\end{table}

\begin{figure}[t]
\begin{center}
\includegraphics[width=7.4cm]{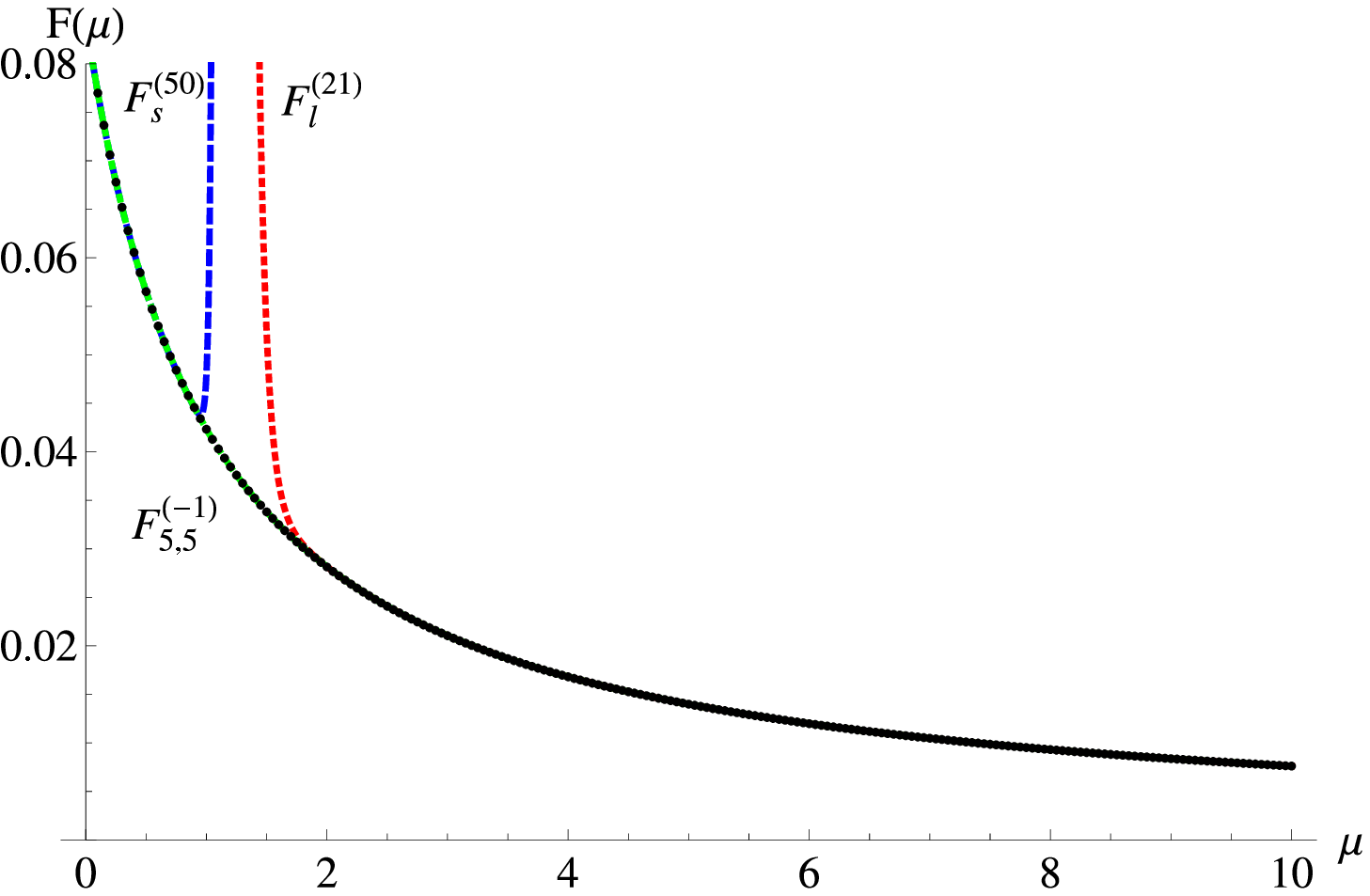}
\includegraphics[width=7.4cm]{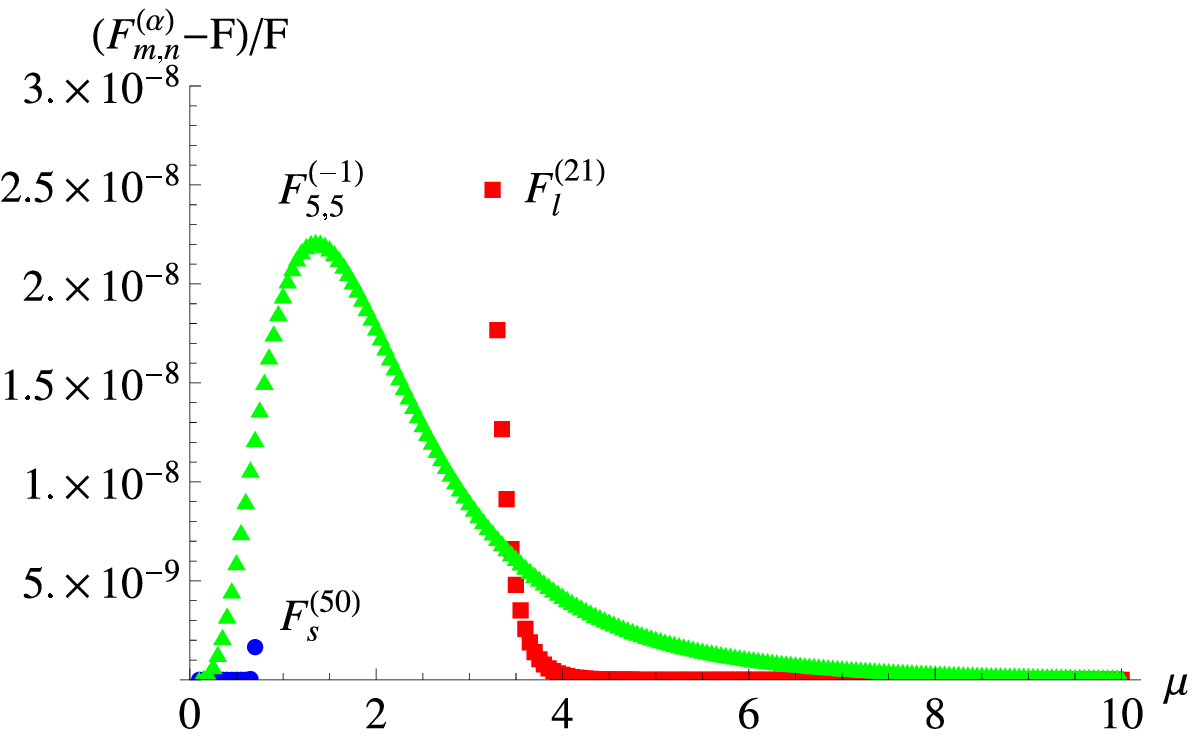}
\end{center}
\caption{
[Left]  The best interpolating function $F_{5,5}^{(-1)}(\mu )$ (green dot-dashed),
the strong coupling expansion $F_s^{(50)}(\mu )$ (blue dashed) and
the weak coupling expansion $F_l^{(21)}(\mu )$ (red dotted) are plotted to $\mu$
with the result of numerical integration (symbol).
Note that $F_{5,5}^{(-1)}(\mu )$ is almost the numerical integration result. 
[Right] The ratios $(F_{5,5}^{(-1)}(\mu ) -F(\mu ))/F(\mu )$ (green triangle),
$(F_s^{(50)}(\mu ) -F(\mu ))/F(\mu )$ (blue circle) and 
$(F_l^{(21)}(\mu ) -F(\mu ))/F(\mu )$ (red square) 
are plotted against $\mu$.
Here $F(\mu )$ means the result of the numerical integration.
}
\label{fig:string_final}
\end{figure}

Let us check our criterion in this example.
We can construct interpolating functions $F_{m,n}^{(\alpha )}(\mu )$ of $F(\mu )$ 
in terms of the two expansions (see appendix \ref{app:string} for explicit formula).
We measure precision of approximation in terms of the numerical integration result $F(\mu_i )$ by
\begin{\eq}
\frac{1}{199}\sum_{i=1}^{199} \left| \frac{F_{m,n}^{(\alpha )}(\mu_i )  -F(\mu_i )}{F(\mu_i )} \right| ,
\end{\eq}
where $\mu_i = i/200$.
Comparing this with values of $I_s [F_{m,n}^{(\alpha )}]$ and $I_l [F_{m,n}^{(\alpha )}]$,
we can test our criterion as in the other examples.
We summarize our result in tab.~\ref{tab:string}.
We easily see that 
the interpolating function $F_{5,5}^{(-1)}(\mu )$ gives
the most precise approximation of $F(\mu )$,
which minimizes $I_s$ and $I_l$ among the candidates.
This shows that our criterion correctly chooses the best interpolating function for this example.
We also plot the best interpolating function $F_{5,5}^{(-1)}(\beta )$ and
its normalized difference from the numerical integration result in fig.~\ref{fig:string_final}.
We also observe that
the best interpolating function gives the very precise approximation.

\section{Discussions}
\label{sec:discussions}
In this paper 
we have studied interpolating function,
which is smooth and consistent with two perturbative expansions of physical quantity around different two points.
First we have proposed the new type of interpolating function,
which is described by fractional power of rational function, and
includes the Pad\'e approximant and FPP constructed by Sen \cite{Sen:2013oza} 
as the special cases.
After introducing the interpolating functions,
we have pointed out that
we can construct enormous number of such interpolating functions in principle
while the ``best" approximation of the exact answer should be unique among the interpolating functions.
Then we have proposed the criterion \eqref{eq:criterion2} 
to determine the ``best" interpolating function
without knowing the exact answer of the physical quantity.
This criterion depends on convergence properties of the small-$g$ and large-$g$ expansions.
We have explicitly checked that our criterion works 
for various examples including the specific heat in the 2d Ising model,
the average plaquette in the 4d $SU(3)$ pure Yang-Mills theory on lattice
and free energy in the $c=1$ string theory at self-dual radius.
We expect that
our criterion is applicable
unless a problem of interest does not have mixed non-perturbative effect,
which is non-perturbative in the both senses of the small-$g$ and large-$g$ expansions.

Besides cases with mixed non-perturbative effects,
we have also found another limitation of approximation by our interpolating function.
In the 2d Ising model on infinite lattice,
we have observed that
interpolating functions do not correctly describe the phase transition
although our criterion correctly chooses the relatively best approximation.
This is obvious since our interpolating functions are smooth by construction unless they have poles.
This indicates that
our interpolating scheme cannot be {\it directly} used for approximation
when we have a phase transition. 
On the other hand, 
we have seen that
our best interpolating functions precisely give the peak locations of the exact answers 
for finite lattice size $L$,
which converges to the phase transition point in the $L\rightarrow\infty$ limit.
Indeed fitting the peak locations as the function of $L$ leads 
the peak location in the $L\rightarrow\infty$ limit,
which is very close to the phase transition point.
This implies that
even if we have a phase transition in discrete system in infinite volume limit,
our interpolating scheme would be {\it indirectly} useful to extract information on phase transition.
We expect that
this feature is true also for more general discrete systems.

One of obvious possible applications of our work is
to apply our interpolating scheme to physical quantities in diverse physical systems.
Although we have worked on the examples in this paper, where
we know the exact answers or numerical results for the whole regions,
we can study more nontrivial system by using our criterion.
We expect that our results are useful in various context of theoretical physics.

In this paper,
we have focused on how to choose the relatively best interpolating functions
among candidates in fixed problems.
Conversely,
it is also interesting to classify problems, 
which is approximated very well 
by the FPR \eqref{eq:FPR} with each fixed $\alpha$.

Although our criterion works in the various examples,
our criterion might be too naive, need slight modifications or have some other exceptions.
It is very illuminating 
if we can perform more rigorous treatment and 
precisely find necessary or sufficient conditions for validity of our criterion.

Also, 
our criterion requires information on large order behaviors of perturbative expansions.
From more practical point of view,
it is very nice 
if there exists an alternative equivalent criterion, 
which uses fewer information on the expansions.

We close by mentioning possible relation to recent progress on resurgence 
in quantum field theory \cite{Argyres:2012vv,Argyres:2012ka,Dunne:2012zk,Dunne:2012ae,Schiappa:2013opa,Dunne:2013ada,Cherman:2013yfa,Aniceto:2013fka,Basar:2013eka,Dunne:2014bca,Cherman:2014ofa} (see 
also recent Borel analysis \cite{Drukker:2011zy,Russo:2012kj,Hanada:2012si,Grassi:2014cla,Kallen:2014lsa} in extended supersymmetric field theories).
While ordinary weak coupling expansion 
(namely the one around trivial saddle point)
generically has non-perturbative ambiguity,
we expect that the ambiguity is partially fixed by interpolation to strong coupling expansion
as far as we focus on real positive coupling at least.
Also our interpolating scheme seems to have no obstruction 
even if perturbative expansions are non-Borel summable.
It is interesting to study interpolation problem in theories concerned with resurgence
and make clear such a relation.

\subsection*{Acknowledgment}
We are grateful to Ashoke Sen and Tomohisa Takimi 
for their early collaborations, many valuable discussions and reading the manuscript. 
We thank Sinya Aoki and Yuji Tachikawa 
for careful reading manuscript and many useful comments.
 
\appendix
\section{Analysis of two-dimensional Ising model by another parameter}
\label{app:anotherIsing}
In sec.~\ref{sec:Ising},
we have analyzed the interpolation problem in the 2d Ising model
by using the parameter $e^{2K}=1+g$.
Then we have seen that 
the exact solution of the quantity $C_L (g)$ is
the rational function of $g$ for finite $L$ and
hence one of the Pad\'e approximants gives the exact solution.
It would be interesting to repeat this analysis in terms of another parameter choice,
where all the FPR interpolating function \eqref{eq:FPR} do not give the exact solution.

In this appendix\footnote{We thank Ashoke Sen for suggesting this analysis.}, 
we repeat a part of analysis in section \ref{sec:Ising} by using the parameter
\begin{\eq}
e^{8K}=1+g^2 .
\end{\eq}
Then the exact solution of $C_L (g)$ with finite $L$ 
becomes a rational function of $\sqrt{1+g^2}$,
which does not take the form \eqref{eq:FPR} of FPR at least for\footnote{
For $L=2$, the exact solution becomes a rational function of $g$ again.
} $L=5$ and $L=8$.
Here we do not explicitly write down formulas of interpolating functions
but we have uploaded a Mathematica file to arXiv, which writes down the explicit forms.

\begin{figure}[t]
\begin{center}
\includegraphics[width=7.0cm]{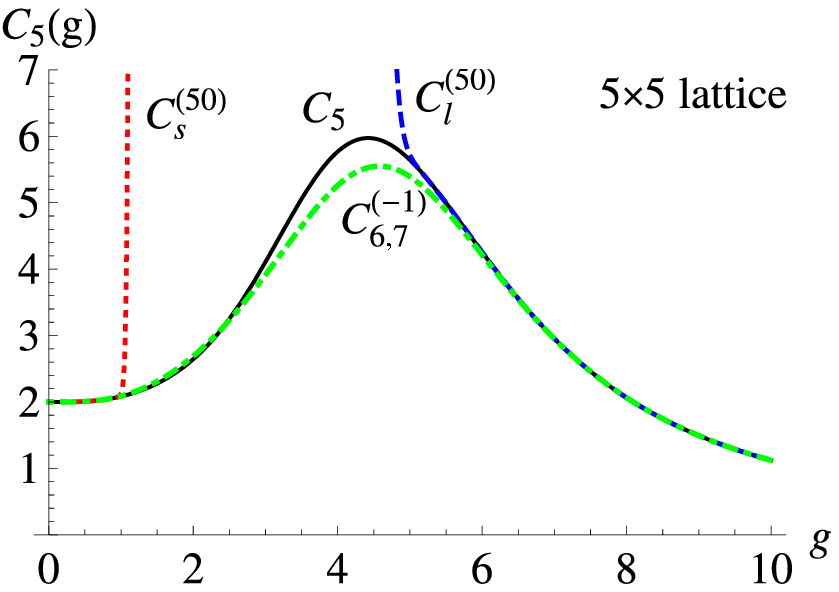}
\includegraphics[width=7.0cm]{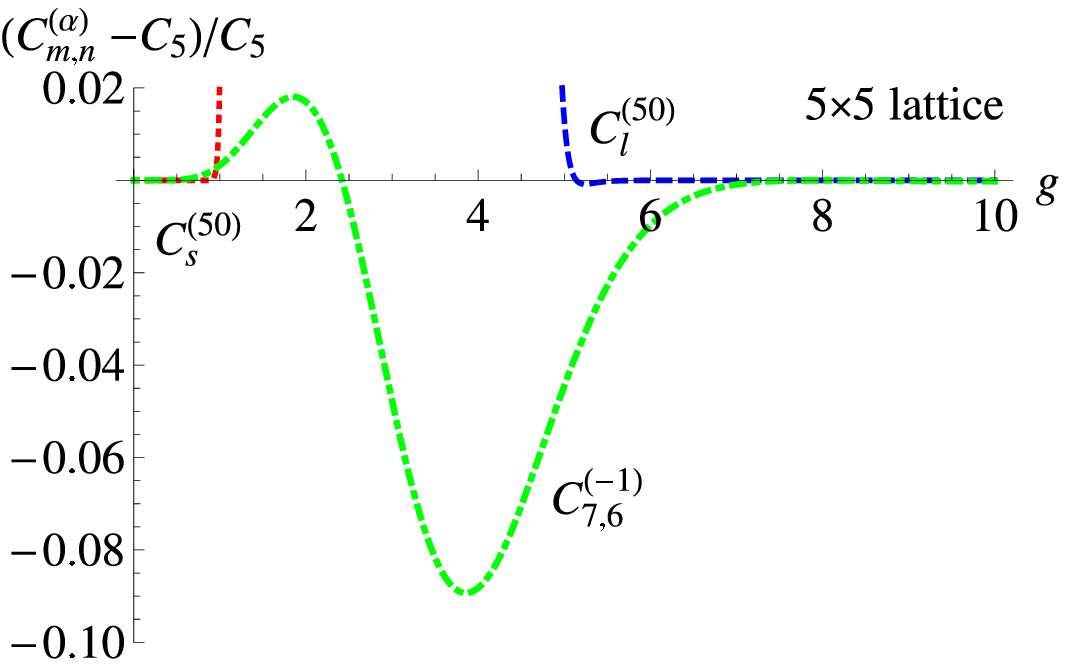}\\
\includegraphics[width=7.0cm]{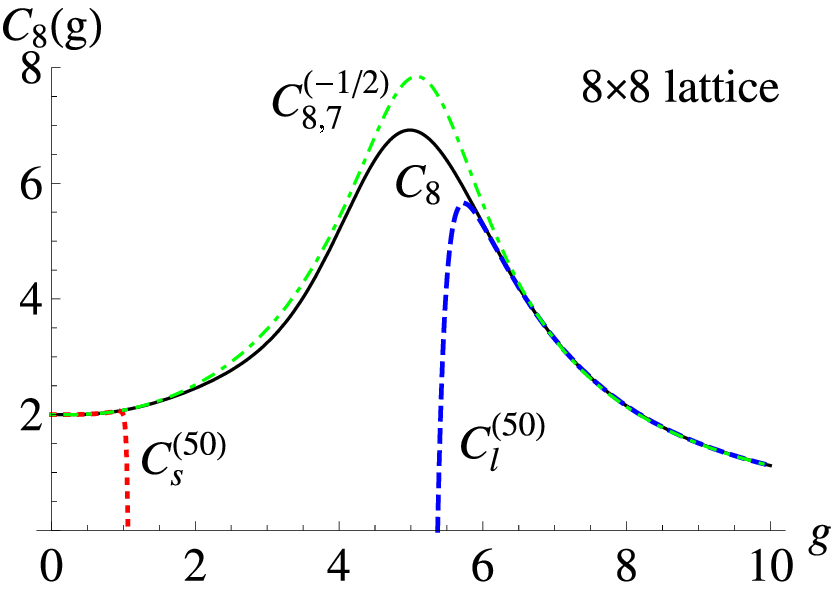}
\includegraphics[width=7.0cm]{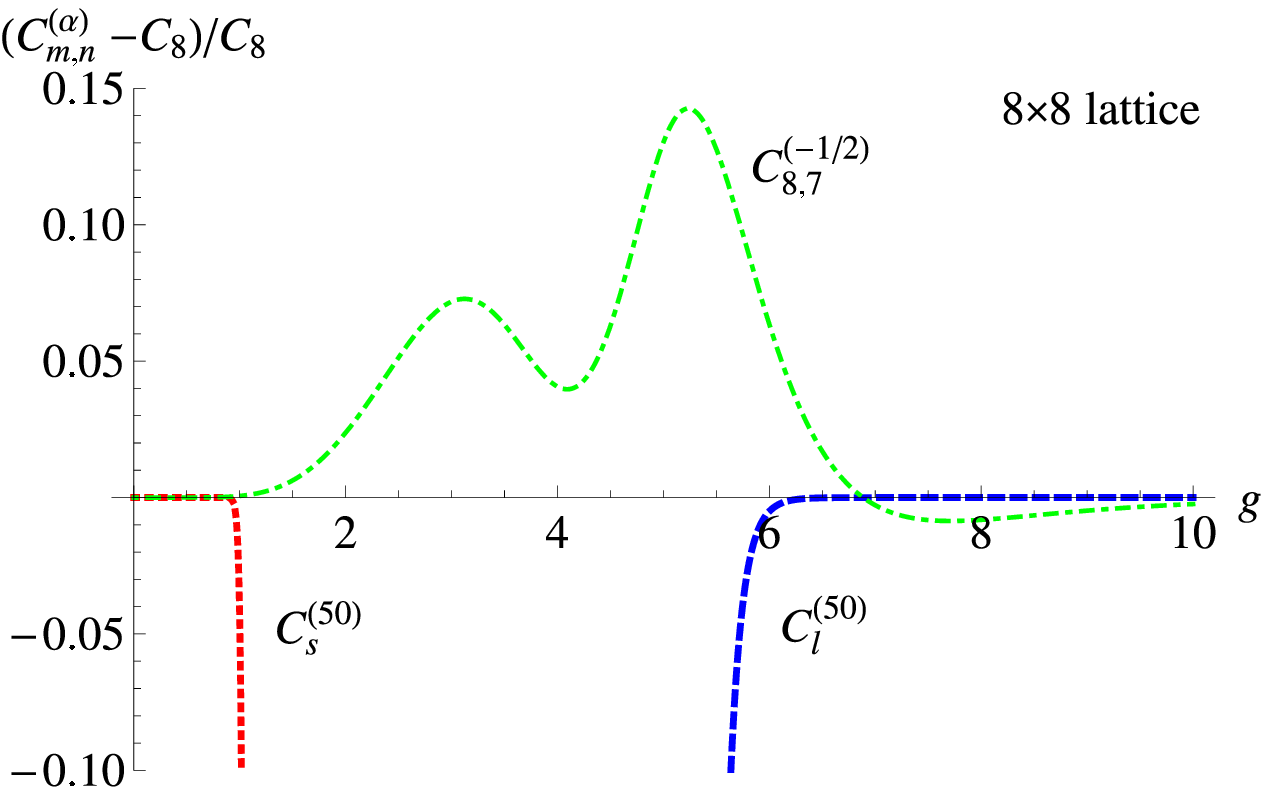}
\end{center}
\caption{
[Left] The function $C_L (g)$ (black solid), 
the best interpolating function (green dot-dashed),
its high (red dotted) and low (blue dashed) temperature expansions 
are plotted to $g$ for each $L$. 
[Right] $|(C_{m,n}^{(\alpha )}-C_L )/C_L |$ (green dot-dashed), 
$|(C_s^{(50 )}-C_L )/C_L |$ (red dotted) and $|(C_l^{(50 )}-C_L )/C_L |$ (blue dashed)
are plotted against $g$ for each $L$.
}
\label{fig:another}
\end{figure}

\subsection{$5\times 5$ lattice}
The function $C_5 (g)$ has the expansions
\begin{\eqa}
C_5 (g) 
&=& 2+\frac{5 g^4}{32} +\mathcal{O} \left(g^5\right) \NN\\
&=& g^{-2} \left( 64 +288g^{-1} +1088g^{-2} +4688g^{-3} +31936g^{-4} 
     +\mathcal{O}(g^{-5} )  \right) .
\end{\eqa}
As in section \ref{sec:Ising}, we find that
the coefficient of the expansions grow as
\begin{\eqa}
 |s_{2k} | \sim c_s A_s^k ,\quad &&{\rm with}\ c_s =0.139016 \ {\rm and}\  A_s =0.983446 ,\NN\\
 |l_k | \sim c_l A_l^k ,\quad    &&{\rm with}\ c_l =149.71 \ {\rm and}\  A_l =4.76806 ,
\end{\eqa}
and the blow-up points are 
$g_s^b= 0.972788$ and $g_l^b = 5.02376$.
Thus we take
\begin{\eq}
N_s^\ast =50 ,\quad g_s^\ast =0.8,\quad N_l^\ast =50,\quad g_l^\ast =5.5 .
\end{\eq}
Our result is summarized in fig.~\ref{fig:another} and tab.~\ref{tab:anotherL5}.
We see that
the best interpolating function $C_{6,7}^{(-1)}$ minimizes $I_s +I_l$.

\clearpage
\begin{table}[h]
\begin{center}
  \begin{tabular}{|c|c|c|c|c|}
  \hline & $\Lambda^{-1}\int dg \bigl| \frac{C_{m,n}^{(\alpha )}-C_5}{C_5} \bigr|$ &$I_s [C_{m,n}^{(\alpha )}]$ &$I_l [C_{m,n}^{(\alpha )}]$ & $I_s +I_l$\\
  \hline\hline $C_{1,1}^{(-2)} $   &0.0144367  &0.0213492  &10.3506 & 10.3719\\
\hline
$C_{1,1}^{(-2/3)} $   &0.00356604 &0.00230782 &4.09192 &4.09423 \\
\hline
$C_{1,2}^{(-1/2)} $   &0.00449818  &0.0109249  &5.46089 &5.47181 \\
\hline
$C_{2,2}^{(-2)} $   &0.00796436  &0.00837338  &8.29550 &8.30388\\
\hline
$C_{2,3}^{(-1)} $   &0.00272780  &0.00432520  &3.09134 &3.09567\\
\hline
$C_{2,3}^{(-1/2)} $   &0.00397428  &0.00752549  &4.80614 &4.81367\\
\hline
$C_{2,3}^{(-1/3)} $   &0.00275912  &0.00696200  &2.67071 &2.67767\\
\hline
$C_{3,2}^{(-1)} $   &0.00633731  &0.00740158 &7.17559  &7.18299\\
\hline
$C_{3,2}^{(-1/2)} $   &0.00375206  &0.00722903  &4.51172  &4.51895\\
\hline
$C_{3,2}^{(-1/3)} $   &0.00500575  &0.00740608  &6.16320 &6.17060\\
\hline
$C_{3,3}^{(-2)} $   &0.00594014  &0.00740588  &7.03028 &7.03768\\
\hline
$C_{3,4}^{(-1/2)} $   &0.00378784  &0.00723170 &4.55132 &4.55855\\
\hline
$C_{3,4}^{(-1/3)} $   &0.00277366  &0.00701077  &2.72476 &2.73177\\
\hline
$C_{3,4}^{(-1/4)} $   &0.00311106  &0.00735004  &3.34761 &3.35496\\
\hline
$C_{4,3}^{(-1)} $   &0.00285537 &0.00512428  &3.24096 &3.24608\\
\hline
$C_{4,3}^{(-1/2)} $   &0.00397491 &0.00754047  &4.80695 &4.81449\\
\hline
$C_{4,3}^{(-1/3)} $   &0.00276431  &0.00698137  &2.69046 &2.69744\\
\hline
$C_{4,5}^{(-1/2)} $   &0.00257637  &0.00641159  &2.49159 &2.49800\\
\hline
$C_{4,5}^{(-1/3)} $   &0.00276420   &0.00698123   &2.68889 &2.69587\\
\hline
$C_{4,5}^{(-1/4)} $   &0.00324266  &0.00741956   &3.81888 &3.82630\\
\hline
$C_{5,4}^{(-1/3)} $   &0.00277377  &0.00701331  &2.72502 &2.73203\\
\hline
$C_{5,6}^{(-1/2)} $   &0.00198535  &0.00470116 &1.35410 &1.35880\\
\hline
$\bf C_{6,7}^{(-1)} $   &$\bf 0.000206236$  &$\bf 0.000285621$  &$\bf 0.0530750$ &$\bf 0.0533606$\\
\hline
$C_{6,7}^{(-1/4)} $   &0.00370596   &0.00206143  &4.90389 &4.90595\\
\hline
$C_{7,8}^{(-1/2)} $   &0.00155076  &0.000172986  &1.10463 &1.10480\\
\hline
$C_{8,7}^{(-1)} $   &0.00116219  &0.0000226883  &1.07748 &1.07750\\
\hline
$C_{8,9}^{(-1)} $   &0.000667177  &0.0000119276  &0.353759 &0.353771\\
\hline
$C_{8,9}^{(-1/3)} $   &0.00175529  &0.0000171883  &2.16789 &2.16791\\
\hline
$C_{9,8}^{(-1)} $   &0.000932892  &0.0000183826  &0.719767 &0.719786\\
\hline
$C_{9,8}^{(-1/2)} $   &0.00157844  &0.0000190668 &3.13313 &3.13315\\
\hline
$C_{9,8}^{(-1/3)} $   &0.00210168  &0.0000232122   &2.63069 &2.63071\\
\hline
  \end{tabular}
\caption{Result for $L=5$}
\label{tab:anotherL5}
\end{center}
\end{table}
\clearpage
\begin{table}[h]
\begin{center}
  \begin{tabular}{|c|c|c|c|c|}
  \hline & $\Lambda^{-1}\int dg \bigl| \frac{C_{m,n}^{(\alpha )}-C_8}{C_8} \bigr|$ &$I_s [C_{m,n}^{(\alpha )}]$ &$I_l [C_{m,n}^{(\alpha )}]$ & $I_s +I_l$\\
  \hline\hline $C_{1,1}^{(-2)} $   &0.0143447  &0.0210712  &8.19673 &8.21780 \\
\hline
 $C_{1,1}^{(-2/3)} $   &0.00358516  &0.00215995  &2.97922 &2.98138\\
\hline
 $C_{1,2}^{(-1/2)} $   &0.00445291  &0.0106469  &3.94941 &3.96005\\
\hline
 $C_{2,1}^{(-1)} $   &0.00984413 &0.00828929  &6.80598 &6.81427\\
\hline
 $C_{2,1}^{(-1/2)} $   &0.00486112 &0.00620801  &2.61762 &2.62382\\
\hline
 $C_{2,2}^{(-2)} $   &0.00787515  &0.00809537  &6.39885 &6.40694\\
\hline
 $C_{2,2}^{(-2/5)} $   &0.00452636 &0.00760081 &4.06211 &4.06971\\
\hline
 $C_{2,3}^{(-1)} $   &0.00289778  &0.00419641 &2.55162 &2.55581\\
\hline
 $C_{2,3}^{(-1/2)} $   &0.00402696  &0.00742413  &3.52709 &3.53451\\
\hline
 $C_{2,3}^{(-1/3)} $   &0.00297962  &0.00671530  &1.98072 &1.98743\\
\hline
 $C_{3,2}^{(-1)} $   &0.00625943  &0.00712357  &5.44665 &5.45377\\
\hline
 $C_{3,2}^{(-1/2)} $   &0.00372974  &0.00695102 &3.20299 &3.20994\\
\hline
 $C_{3,3}^{(-2)} $   &0.00587186  &0.00712823  &5.31167 &5.31880\\
\hline
 $C_{3,4}^{(-1/2)} $   &0.00379867  &0.00695734  &3.27471 &3.28167\\
\hline
 $C_{3,4}^{(-1/3)} $   &0.00341859  &0.00701707  &2.73224 &2.73926\\
\hline
 $C_{3,4}^{(-1/4)} $   &0.00369312  &0.00708313  &3.17778 &3.18486\\
\hline
 $C_{4,3}^{(-1)} $   &0.00300812  &0.00492987  &2.63094 &2.63587\\
\hline
 $C_{4,3}^{(-1/2)} $   &0.00402822  &0.00746425  &3.52822 &3.53568\\
\hline
 $C_{4,3}^{(-1/3)} $   &0.00298748  &0.00673209  &1.99653 &2.00326\\
\hline
 $C_{4,5}^{(-1/2)} $   &0.00293553  &0.00614008  &2.16315 &2.16929\\
\hline
 $C_{4,5}^{(-1/3)} $   &0.00345399  &0.00715123  &2.78100 &2.78815\\
\hline
 $C_{4,5}^{(-1/4)} $   &0.00317186  &0.00694968  &2.32397 &2.33091\\
\hline
 $C_{5,4}^{(-1/3)} $   &0.00351452  &0.00837605  &2.85851 &2.86688\\
\hline
 $C_{5,6}^{(-1/4)} $   &0.00327626  &0.00722693  &2.49615 &2.50337\\
\hline
 $C_{6,5}^{(-1/2)} $   &0.000875711  &0.000194793  &0.595842 &0.596037\\
\hline
 $C_{6,5}^{(-1/3)} $   &0.00344271  &0.00621808   &2.77138 &2.77760\\
\hline
 $C_{6,7}^{(-1/3)} $   &0.00109210  &0.0000803322 &0.348460 &0.348541\\
\hline
 $C_{6,7}^{(-1/4)} $   &0.00323839  &0.00629399  &2.44618 &2.45247\\
\hline
 $C_{7,8}^{(-1/3)} $   &0.00155789  &0.0000565836  &0.762480 &0.762537\\
\hline
 $\bf C_{8,7}^{(-1/2)} $   &$\bf 0.000350404$  &$\bf 0.0000254784$  &$\bf 0.0719374$ &$\bf 0.0719629$\\
\hline
 $C_{8,7}^{(-1/3)} $   &0.00120052  &0.0000117739 &0.411958 &0.411969\\
\hline
  $C_{9,8}^{(-1/3)} $   &0.000896865 &$1.25557\times 10^{-6}$ &0.148141 &0.148142\\
\hline
  \end{tabular}
\caption{Result for $L=8$}
\label{tab:anotherL8}
\end{center}
\end{table}
\clearpage
\subsection{$8\times 8$ lattice}
The high and low temperature expansions of $C_8 (g)$ are given by
\begin{\eqa}
C_8 (g) 
&=& 2+\frac{5 g^4}{32} +\mathcal{O} \left(g^5\right) \NN\\
&=& g^{-2} \left( 64 +288g^{-1} +1088g^{-2} +4688g^{-3} +19264g^{-4} 
     +\mathcal{O}(g^{-5} )  \right) .
\end{\eqa}
As in the $L=5$ case ,
we find
\begin{\eqa}
&&c_s = 0.0614108 ,\quad A_s = 1.04269 ,\quad g_s^b =0.979314 ,\NN\\
&&c_l =99.216 ,\quad A_l =5.4084 ,\quad g_l^b = 5.73095.
\end{\eqa}
Thus we take
\begin{\eq}
N_s^\ast =50 ,\quad g_s^\ast =0.8,\quad N_l^\ast =50,\quad g_l^\ast =6.3 .
\end{\eq}
From tab.~\ref{tab:anotherL8},
we find that
the best interpolating function $C_{8,7}^{(-1/2)}$ minimizes $I_s +I_l$.

\section{Explicit formula for interpolating functions}
\label{app:explicit}
In this appendix, we explicitly write down the interpolating functions appearing in the main text.
Although we often display numerical values in 6 digits,
indeed we have worked on ``MachinePrecision" or 100 digits in Mathematica.
We have uploaded the Mathematica file to arXiv,
which writes all interpolating functions in actual precisions.
\subsection{Zero-dimensional $\varphi^4$ theory}
\label{app:ex4}
\begin{\eqa}
&& \scriptscriptstyle  
F_{0,0}^{(1/2)}(g) 
=\sqrt{2 \pi } \left(\frac{8 \pi  g}{\Gamma \left(1/4 \right)^2}+1 \right)^{-1/2} ,\quad
 F_{1,1}^{(1/2)}(g) 
=\sqrt{2 \pi  \Gamma \left( 1/4\right)} 
\left( \frac{8 \pi  g \Gamma \left( 1/4\right)
+\Gamma \left( 1/4 \right)^3+2 \pi  \Gamma \left(-1/4 \right)}{64 \pi ^2 g^2
+8 \pi  g \Gamma \left( 1/4 \right)^2+\Gamma \left( 1/4\right)^4+2 \pi  \Gamma \left(-1/4 \right) \Gamma
   \left( 1/4\right)} \right)^{1/2} ,\NN\\
&&\scriptscriptstyle F_{1,1}^{(1/6)}(g)
= 2.50663 \left( \frac{1}{6.98929 g^3+7.08691 g^2+1}  \right)^{1/6} ,\quad
 F_{2,2}^{(1/2)}(g)
= 2.50663 \sqrt{\frac{37.9117 g^2+10.1532 g+1}{72.4854 g^3+43.9117 g^2+10.1532 g+1}} ,\NN\\
&&\scriptscriptstyle F_{2,2}^{(1/10)}(g) 
=2.50663 \left( 25.5499 g^5+43.1779 g^4+32.1482 g^3+30 g^2+1 \right)^{-1/10},\NN\\
&&\scriptscriptstyle F_{3,3}^{(1/2)}(g)
=2.50663 \sqrt{\frac{324.019 g^3+110.261 g^2+16.0304 g+1}
{619.509 g^4+420.201 g^3+116.261 g^2+16.0304 g+1}} ,\NN\\
&&\scriptscriptstyle F_{3,3}^{(1/6)}(g)
=2.50663\left(\frac{28.2525 g^2+8.0997 g+1}
{197.465 g^5+256.834 g^4+145.795 g^3+46.2525 g^2+8.0997 g+1} \right)^{1/6} ,\NN\\
&&\scriptscriptstyle F_{3,3}^{(1/14)}(g)
=2.50663 \left( 93.3994 g^7+220.976 g^6+239.216 g^5+155.758 g^4+42 g^2+1 \right)^{-1/14} ,\NN\\
&&\scriptscriptstyle F_{4,4}^{(1/2)}(g)
=2.50663 \sqrt{\frac{3224.56 g^4+1303.49 g^3+238.239 g^2+22.8745 g+1}
{6165.22 g^5+4576. g^4+1440.74 g^3+244.239
   g^2+22.8745 g+1}} ,\NN\\
&&\scriptscriptstyle F_{4,4}^{(1/10)}(g)
=2.50663 \left[ \frac{ 14.4369 g^2+5.07251 g+1 }
 { 368.86 g^7+752.954 g^6+708.689 g^5  +403.106 g^4+152.175 g^3
+44.4369 g^2+5.07251 g+1 }  \right]^{1/10} ,\NN\\
&&\scriptscriptstyle F_{4,4}^{(1/18)}(g)  
=2.50663 \left( 341.428 g^9+1038.59 g^8+1475.35 g^7+1294.34 g^6+780.788 g^5 
+594 g^4 +54 g^2+1 \right)^{-1/18} .
\end{\eqa}

\subsection{Two-dimensional Ising model}
\label{app:ising}
\subsubsection{$2\times 2$ lattice}
\label{app:L2}
\begin{\eqa}
&&\scriptscriptstyle C_{1,1}^{(-4)}(g)
=\frac{48 \left(-\sqrt[4]{2} g-2 \sqrt[4]{3}+\sqrt[4]{2}\right)^4}{\left(g^2+2^{3/4} \sqrt[4]{3}
   g+2 \sqrt{6}-2^{3/4} \sqrt[4]{3}\right)^4},\quad
C_{1,1}^{(-4/3)}(g)
=4\left( \frac{g^3}{4 \sqrt[4]{2} 3^{3/4}}+\frac{1}{4} \sqrt[4]{\frac{3}{2}}
   g^2+1 \right)^{-4/3} ,\quad 
C_{1,2}^{(-1)}(g) = \frac{96}{g^4+4 g^3+6 g^2+24} ,\NN\\
&&\scriptscriptstyle C_{1,2}^{(-2)}(g)
=   \frac{576 \left(\left(2+4 \sqrt{6}\right) g+23\right)^2}{\left(2 \left(12+\sqrt{6}\right) g^3+3
   \left(16+9 \sqrt{6}\right) g^2+24 \left(1+2 \sqrt{6}\right) g+276\right)^2} ,\quad
C_{2,2}^{(-4)}(g) 
= \frac{0.0125604 \left(9.35012 g^2+4.84252 g+5.87574\right)^4}{(g+1.40614)^4 \left(g^2+0.111766
   g+0.989164\right)^4} ,\NN\\
&&\scriptscriptstyle C_{2,3}^{(-2)}(g) 
=\frac{0.000108752 \left(939.545 g^2+271.353 g+529 \right)^2}{\left(g^2-0.55488
   g+1.02344\right)^2 \left(g^2+2.84369 g+2.69513\right)^2} ,\quad
C_{3,2}^{(-2)}(g) 
=\frac{0.0482704 \left(44.5959 g^2+7.30306 g+27.\right)^2}{\left(g^2-0.70521 g+1.00504\right)^2
   \left(g^2+2.86897 g+2.95114\right)^2} ,\NN\\
&&\scriptscriptstyle C_{3,4}^{(-1)}(g) 
= \frac{96 \left(4638 g^2+446 g+969\right)}{4638 g^6+18998 g^5+30581 g^4+25104 g^3+64800
   g^2+10704 g+23256} ,\NN\\
&&\scriptscriptstyle C_{4,3}^{(-2)}(g) 
=\frac{96 (g+0.0993445)^2 \left(g^2+0.279722 g+0.55003\right)^2}{(g+0.0993806)^2
   \left(g^2-0.553882 g+1.00883\right)^2 \left(g^2+2.83357 g+2.67002\right)^2} ,\NN\\
&&\scriptscriptstyle C_{4,3}^{(-1)}(g) 
=   \frac{96 \left(8393 g^2+638 g+1546\right)}{8393 g^6+34210 g^5+54456 g^4+43584 g^3+127224
   g^2+15312 g+37104} ,\NN\\
&&\scriptscriptstyle C_{5,4}^{(-1)}(g) 
= \frac{96 \left(2297148 g^3+607943 g^2+509356 g+71294\right) }
 { 2297148 g^7+9796535 g^6+16724016 g^5+14944968 g^4+34104576 g^3 
+11168520 g^2+12224544 g+1711056 }, \NN\\
&&\scriptscriptstyle C_{6,7}^{(-1)}(g) 
= 32 \left( 34967691 g^5-12313050 g^4+33661002 g^3+15122952 g^2+8312739
   g+2651546\right)
\times (11655897 g^9+42519238 g^8 \NN\\
&&\scriptscriptstyle  \ \ \ \ \ \ \ \ \ \ \ 
+64738316 g^7+71919808 g^6+217595516 g^5 
+36474424 g^4+178708928 g^3 +78558880 g^2+66501912 g+21212368 )^{-1} ,\NN\\
&&\scriptscriptstyle C_{7,6}^{(-1)}(g) 
=32 (107262402 g^5+163890609 g^4+195374448 g^3 
 +115458702 g^2 +43155430g+11655897)
\times (35754134 g^9+197646739 g^8 \NN\\
&&\scriptscriptstyle  \ \ \ \ \ \ \ \ \ \ \ 
+498170432 g^7+769783252 g^6+1218567440g^5
+1226615044 g^4+1059003056 g^3+737175264 g^2+345243440 g+93247176 )^{-1} \NN\\
\end{\eqa}

\subsubsection{$5\times 5$ lattice}
\label{app:L5}
\begin{\eqa}
&&\scriptscriptstyle C_{1,1}^{(-4)}(g)
=\frac{64 \left(g+2 \sqrt[4]{2}-1\right)^4}{\left(g^2+2 \sqrt[4]{2} g+4 \sqrt{2}-2 \sqrt[4]{2}\right)^4} ,\quad 
 C_{1,1}^{(-4/3)}(g)
=64 \sqrt[3]{2} \left( \sqrt[4]{2} g^3+3 \sqrt[4]{2} g^2+16\right)^{-4/3} ,\quad
 C_{1,2}^{(-1)}(g)
=  \frac{128}{2 g^4+8 g^3+3 g^2+64} ,\NN\\
&&\scriptscriptstyle C_{1,2}^{(-2)}(g)
=\frac{128 \left(8 \left(16 \sqrt{2}-5\right) g+487\right)^2}{\left(-8 \left(5 \sqrt{2}-32\right)
   g^3+\left(512+407 \sqrt{2}\right) g^2+64 \left(16 \sqrt{2}-5\right) g+3896\right)^2} ,\NN\\
&&\scriptscriptstyle C_{2,2}^{(-4)}(g)
=\frac{319.017 \left(0.669255 g^2+0.179256 g+1.\right)^4}{(g+1.90917)^4 \left(g^2-0.641328
   g+1.86145\right)^4} ,\quad
 C_{2,3}^{(-2)}(g)
=  \frac{0.00132573 \left(175629. g^2+68676.4 g+237169.\right)^2}{\left(799.342 g^4+1911.25
   g^3+705.392 g^2+1768.15 g+6106.19\right)^2} ,\NN\\
&&\scriptscriptstyle C_{3,2}^{(-2)}(g)
=\frac{128 \left(12048.5 g^2+2709.7 g+17153\right)^2}{\left(17039.2 g^4+37910.5 g^3+10623.2
   g^2+21677.6 g+137224\right)^2} ,\quad
C_{3,3}^{(-4)}(g)
=  \frac{2 \left(0.431032 g^3+0.608595 g^2+0.580552 g+1.\right)^4}{\left(0.181227 g^4+0.437109
   g^3+0.296095 g^2+0.580552 g+1\right)^4} ,\NN\\
&&\scriptscriptstyle C_{3,3}^{(-4/7)}(g)
=  128 2^{6/7} \left(8 \sqrt[4]{2} g^7+56 \sqrt[4]{2} g^6+105 \sqrt[4]{2} g^5-35 \sqrt[4]{2}
   g^4+5376 g^3-8960 g^2+4096\right)^{-4/7} ,\NN\\
&&\scriptscriptstyle C_{3,4}^{(-2)}(g)
= \frac{64. (g+8.47078)^2 \left(g^2+0.21507 g+1.39409\right)^2}{(g+8.49496)^2 \left(g^2-1.44691
   g+1.94352\right)^2 \left(g^2+3.63781 g+4.04611\right)^2} ,\NN\\
&&\scriptscriptstyle C_{3,4}^{(-1)}(g)
=   \frac{32 \left(12978 g^2+6980 g+11239\right)}{6489 g^6+29446 g^5+29313 g^4-4732 g^3-17132
   g^2+111680 g+179824} ,\NN\\
&&\scriptscriptstyle C_{4,3}^{(-1)}(g)
=    \frac{32 \left(4090 g^2+3620 g+6489\right)}{2045 g^6+9990 g^5+13552 g^4+5468 g^3-64340 g^2+57920
   g+103824} ,\NN\\
&& \scriptscriptstyle C_{4,4}^{(-4/9)}(g)
= 512\left( 64 2^{3/4} g^9+576 2^{3/4} g^8+1656 2^{3/4} g^7+840 2^{3/4} g^6-\frac{6795
   g^5}{\sqrt[4]{2}} 
+870912 g^4+442368 g^3-737280 g^2+262144 \right)^{-4/9} ,\NN\\
&&\scriptscriptstyle C_{4,5}^{(-1)}(g)
= \frac{ 32 \left( 230402 g^3+305620 g^2+360351 g+288280\right) }
{115201 g^7+613614 g^6+964217g^5+518052 g^4-61228 g^3 
-875680 g^2+5765616 g+4612480 } ,\NN\\
&&\scriptscriptstyle  C_{6,5}^{(-1)}(g)
= 32 \left(277152946 g^4-7310192 g^3+260742483 g^2+47677554 g+45898498\right)
\times (138576473g^8+550650796 g^7 \NN\\
&&\scriptscriptstyle \ \ \ \ \ \ \ \ \ \ \ \ 
+323615567 g^6-153041266 g^5+158916108 g^4 
-519732176 g^3+3253909768g^2+762840864 g+734375968 )^{-1} ,\NN\\
&&\scriptscriptstyle C_{6,7}^{(-1)}(g)
= 32 (19064616590 g^5+6619951212 g^4+4651960211 g^3 
+13827726376 g^2+12299524053g+5632465646)\NN\\
&&\scriptscriptstyle \ \  \ \ \ \ \ \ \ \ \ \times 
(9532308295 g^9+41439208786 g^8+29864344972 g^7-26478794456 g^6 
+8828921538g^5+22018705676 g^4 \NN\\
&&\scriptscriptstyle \ \ \ \ \ \ \ \ \ \ \ 
-103969529932 g^3+108594309096 g^2 +196792384848 g+90119450336 )^{-1} ,\NN\\
&&\scriptscriptstyle C_{7,6}^{(-1)}(g)
=32 (77240285964 g^5+4144676522 g^4+19445518862 g^3 
+34689643011 g^2+45930642372g+19064616590) \NN\\
&&\scriptscriptstyle \ \ \ \ \ \ \ \ \ \ \times 
(38620142982 g^9+156552910189 g^8+75942326948 g^7-133756348289 g^6 
+48291876322 g^5+76206605316 g^4 \NN\\
&&\scriptscriptstyle \ \ \ \ \ \ \ \ \ \ \ 
 -378709146568 g^3+173741956376 g^2 +734890277952g +305033865440 )^{-1}
\end{\eqa}

\subsubsection{$8\times 8$ lattice}
\label{app:L8}
\begin{\eqa}
&&\scriptscriptstyle C_{1,1}^{(-4)}(g)
=\frac{64 \left(g+2 \sqrt[4]{2}-1\right)^4}{\left(g^2+2 \sqrt[4]{2} g+4 \sqrt{2}-2 \sqrt[4]{2}\right)^4} ,\quad
C_{1,1}^{(-4/3)}(g)
=64 \sqrt[3]{2} \left(\sqrt[4]{2} g^3+3 \sqrt[4]{2} g^2+16\right)^{-4/3} ,\quad
C_{1,2}^{(-1)}(g)  =\frac{128}{2 g^4+8 g^3+3 g^2+64} , \NN\\
&&\scriptscriptstyle C_{1,2}^{(-2)}(g)
=\frac{128 \left(8 \left(16 \sqrt{2}-5\right) g+487\right)^2}{\left(-8 \left(5 \sqrt{2}-32\right) g^3+\left(512+407
   \sqrt{2}\right) g^2+64 \left(16 \sqrt{2}-5\right) g+3896\right)^2}, \quad
 C_{2,2}^{(-4)}(g)
= \frac{2. \left(0.669255 g^2+0.179256 g+1.\right)^4}{\left(0.281387 g^3+0.356755 g^2+0.179256 g+1.\right)^4} ,\NN\\
&&\scriptscriptstyle C_{2,3}^{(-2)}(g)
=\frac{0.00132573 \left(175629. g^2+68676.4 g+237169.\right)^2}{\left(799.342 g^4+1911.25 g^3+705.392 g^2+1768.15
   g+6106.19\right)^2} ,\quad
C_{2,3}^{(-1)}(g)
=\frac{128 (83 g+10)}{166 g^5+684 g^4+329 g^3-800 g^2+5312 g+640} ,\NN\\
&&\scriptscriptstyle C_{3,2}^{(-1)}(g)
=\frac{64 (72 g+83)}{72 g^5+371 g^4+440 g^3-3320 g^2+2304 g+2656} ,\quad
 C_{3,3}^{(-4)}(g)
=\frac{2. \left(0.0716093 g^3+0.659177 g^2+0.245925 g+1.\right)^4}{\left(0.030108 g^4+0.307258 g^3+0.346677 g^2+0.245925
   g+1.\right)^4} ,\NN\\
&&\scriptscriptstyle C_{3,3}^{(-4/7)}(g)
=   128 2^{6/7} 
\left(8 \sqrt[4]{2} g^7+56 \sqrt[4]{2} g^6+105 \sqrt[4]{2} g^5-35 \sqrt[4]{2} g^4
+8960 g^3-8960g^2+4096\right)^{-4/7} ,\NN\\
&& \scriptscriptstyle C_{3,4}^{(-1)}(g)
=\frac{32 \left(12338 g^2+11236 g+11239\right)}{6169 g^6+30294 g^5+37345 g^4+60 g^3-27372 g^2+179776 g+179824} ,\NN\\
&&\scriptscriptstyle  C_{4,3}^{(-1)}(g)
=\frac{32 \left(3690 g^2+5796 g+6169\right)}{1845 g^6+10278 g^5+17444 g^4+7460 g^3-64340 g^2+92736 g+98704} ,\NN\\
&&\scriptscriptstyle C_{4,4}^{(-4/9)}(g)
=512 \left( 64 2^{3/4} g^9+576 2^{3/4} g^8+1656 2^{3/4} g^7+840 2^{3/4} g^6-\frac{6795 g^5}{\sqrt[4]{2}} 
+428544g^4+737280 g^3-737280 g^2+262144\right)^{-4/9} ,\NN\\
&&\scriptscriptstyle C_{4,5}^{(-1)}(g)
=\frac{32 \left(208682 g^3+447124 g^2+593899 g+429792\right)}
{104341 g^7+640926 g^6+1347709 g^5+1216332 g^4 
 +56772 g^3-1441856g^2+9502384 g+6876672 } ,\NN\\
&&\scriptscriptstyle C_{5,6}^{(-1)}(g)
=\frac{32 \left(8223938 g^4+2925016 g^3+23633941 g^2+24931938 g+22756915\right) } 
 {4111969 g^8+17910384 g^7+23834956 g^6+41367768
   g^5+66515288 g^4 
+3299796 g^3-76995244 g^2+398911008 g+364110640} ,\NN\\
&&\scriptscriptstyle C_{6,7}^{(-1)}(g)
=32 (2868358258 g^5-534538208 g^4+2620749213 g^3 
  +4306817710 g^2+5981505947 g+3547958864 ) 
\times (1434179129g^9+5469447412 g^8 \NN\\
&&\scriptscriptstyle \ \ \ \ \ \ \ \ \ 
+2392566884 g^7-176892020 g^6 +13113571912 g^5+10749317956 g^4 
-6738954252 g^3-2050093920 g^2 +95704095152g+56767341824 )^{-1} ,\NN\\
&&\scriptscriptstyle C_{7,6}^{(-1)}(g)
=32 ( 3729267560 g^5-1187460246 g^4+1626376328 g^3 
+4209185323 g^2 +7157300308 g+5736716516 )
\times (1864633780   g^9+6864804997 g^8 \NN\\
&&\scriptscriptstyle \ \ \ \ \ \ \ \ \ \ 
+1235218342 g^7-4856418767 g^6 
+13854661436 g^5+17016069700 g^4 
-2389654592 g^3-47387365152g^2 +114516804928 g+91787464256 )^{-1} ,\NN\\
&&\scriptscriptstyle C_{7,8}^{(-1)}(g)
=32 (195993105538 g^6+811071518032 g^5-90814453839 g^4+663928537486 g^3 
+1365385463933 g^2+1869156822440g+1303853612472)  \NN\\
&&\scriptscriptstyle \ \ \ \ \ \ \ \ \ \ \times 
(97996552769 g^{10}+797521970092 g^9+1723730638298 g^8+268656235744 g^7 
-207735519482 g^6 +3883412178028 g^5 \NN\\
&&\scriptscriptstyle \ \ \ \ \ \ \ \ \ \  
+3406981458828 g^4-683207599584 g^3 
-4230904826512 g^2+29906509159040 g+20861657799552 )^{-1} ,\NN\\
&&\scriptscriptstyle C_{8,7}^{(-1)}(g)
=32 (1881323333564 g^6-605415982824 g^5+1800057804060 g^4 
+2713663149372 g^3+3635590280823 g^2+1838011785324g-391986211076) \NN\\
&&\scriptscriptstyle \ \ \ \ \ \ \ \ \ \times 
( 940661666782 g^{10}+3459938675716 g^9+1100189436555 g^8-200423138222 g^7 
+8932877665783g^6 +6103674295452g^5 \NN\\
&&\scriptscriptstyle \ \ \ \ \ \ \ \ \ \  
-5582700200716 g^4-1181349538048 g^3 
+66009168714688 g^2+29408188565184 g-6271779377216 )^{-1} ,\NN\\
&&\scriptscriptstyle C_{8,9}^{(-1)}(g)
=32 ( 440983411412506 g^7+298415770089536 g^6+286897659450525 g^5
+33283071804262 g^4+870265241542217g^3 \NN\\
&&\scriptscriptstyle \ \ \ \ \ \ \ \ \ \ \ \ \  
+1255703738099444 g^2 +1442315420615496 g+318054473744432 ) 
\times (220491705706253 g^{11}+1031174707869780g^{10} \NN\\
&&\scriptscriptstyle \ \ \ \ \ \ \ \ \ \ \ 
 +1071017928463714 g^9 -288209846160932 g^8  
-304782048389130 g^7 +1489590651055080 g^6 
+2713217114957772 g^5 \NN\\
&&\scriptscriptstyle \ \ \ \ \ \ \ \ \ \ \  
+2992544904211504 g^4 -8560975072745808 g^3 
+13730170334702464 g^2+23077046729847936 g+5088871579910912 )^{-1} ,\NN\\
&&\scriptscriptstyle C_{9,10}^{(-1)}(g)
=32 (238383963504470290 g^8+872536488680167328 g^7-142166027096683727 g^6
+731239697144276110 g^5 \NN\\
&&\scriptscriptstyle \ \ \ \ \ \ \ \ \ \  
 -220790772453261619g^4+959382749783352236 g^3 +1300379502502145742 g^2
+1737488555921690408 g+675174681476983256 ) \NN\\
&&\scriptscriptstyle \ \ \ \ \ \ \ \ \ \ \times 
(119191981752235145g^{12}+913036171349024244 g^{11}+1852777936440345510 g^{10}
+139730252127720372 g^9 \NN\\
&&\scriptscriptstyle \ \ \ \ \ \ \ \ \ \ \   
 -1084871711151303636 g^8+396619365159964700 g^7 
+172449770824598210 g^6+3467719807273727176 g^5 \NN\\
&&\scriptscriptstyle \ \ \ \ \ \ \ \ \ \ \  
 +2508829983230774392 g^4 -5896153492360507264g^3 
+7302578410494666752 g^2 \NN\\
&&\scriptscriptstyle \ \ \ \ \ \ \ \ \ \ \  
+27799816894747046528 g+10802794903631732096 )^{-1}
\end{\eqa}

\subsubsection{Infinite lattice}
\label{app:Linf}

\begin{\eqa}
&& \scriptscriptstyle C_{1,1}^{(-2/3)}(g)
=96\left(6 \sqrt{6} g^3+81 \sqrt{\frac{3}{2}} g^2+44 g+64\right)^{-2/3} ,\quad
C_{2,1}^{(-1)}(g) 
=\frac{32 (265 g+708)}{530 g^3+3801 g^2+3144 g+3776} ,\NN\\
&& \scriptscriptstyle C_{2,2}^{(-2/5)}(g)
=192 \sqrt[5]{2} 3^{4/5}
\left( 2592 \sqrt{6} g^5+29160 \sqrt{6} g^4+163215 \sqrt{6} g^3
+182960 g^2 +84480   g+73728\right)^{-2/5} ,\NN\\
&& \scriptscriptstyle C_{2,3}^{(-1)}(g)
=\frac{32 \left(68224 g^2-74327 g+209220\right)}{136448 g^4+465362 g^3+1113977 g^2+115016 g+1115840} ,\NN\\
&& \scriptscriptstyle C_{3,4}^{(-1)}(g)
= \frac{32 \left(216459904 g^3-556333856 g^2+1591853427 g+563647212\right)}{432919808 g^5+835471424 g^4+2505900230
   g^3+3434401043 g^2+9867689240 g+3006118464} ,\NN\\
&& \scriptscriptstyle C_{4,4}^{(-2/9)}(g)
= \frac{16}{\left((g+0.464717) \left(g^2-0.729826 g+0.274268\right) \left(g^2+0.284047 g+0.122015\right) \left(g^2+3.24318
   g+58.9783\right) \left(g^2+16.9879 g+90.0307\right)\right)^{2/9}} ,\NN\\
&& \scriptscriptstyle C_{4,5}^{(-1)}(g)
=\frac{16. \left(g^2-3.13804 g+8.35966\right) \left(g^2+0.378281 g+0.0674372\right)}{\left(g^2-1.70712 g+4.26849\right)
   \left(g^2+0.370646 g+0.0673414\right) \left(g^2+3.07672 g+5.22999\right)} ,\NN\\
&& \scriptscriptstyle C_{5,5}^{(-2/11)}(g)
=192 2^{5/11} 3^{10/11}
(483729408 \sqrt{6} g^{11}+11972302848 \sqrt{6} g^{10}
+147824683776 \sqrt{6}g^9+1169669045952 \sqrt{6} g^8 \NN\\
&&\scriptscriptstyle \ \ \ \ \ \ \ \ \ \ \  
+6339773446380 \sqrt{6} g^7+22461400927053 \sqrt{6} g^6
-7156938909824 g^5+852953401344g^4+744245821440 g^3 \NN\\
&&\scriptscriptstyle \ \ \ \ \ \ \ \ \ \ \  
+1876659535872 g^2+657733976064 g+260919263232 )^{-2/11}  ,\NN\\
&& \scriptscriptstyle C_{5,6}^{(-1)}(g)
= \frac{16. (g+0.593244) \left(g^2-3.20416 g+7.38955\right) \left(g^2-0.324953 g+0.547816\right)}{(g+0.775198)
   \left(g^2-1.93029 g+2.75727\right) \left(g^2-0.0123326 g+0.600863\right) \left(g^2+2.73156 g+4.98641\right)},\NN\\
&& \scriptscriptstyle C_{5,6}^{(-2/13)}(g)
= 192 2^{7/13} 3^{12/13}
(34828517376 \sqrt{6} g^{13}+1018734133248 \sqrt{6} g^{12}+14870688528384 \sqrt{6}
   g^{11}+140659593085440 \sqrt{6} g^{10} \NN\\
&&\scriptscriptstyle \ \ \ \ \ \ \ \ \ \ \  
+937732168125792 \sqrt{6} g^9+4426291424759256 \sqrt{6} g^8+13142668430464443
   \sqrt{6} g^7-371479076823296 g^6 -1699299709526016 g^5\NN\\
&&\scriptscriptstyle \ \ \ \ \ \ \ \ \ \ \  
+375050965745664 g^4+340534744842240 g^3 
 +460033275985920g^2+149245818568704 g+50096498540544 )^{-2/13} ,\NN\\
&& \scriptscriptstyle C_{6,7}^{(-1)}(g)
=\frac{16. \left(g^2-2.95821 g+6.25399\right) \left(g^2-1.0231 g+1.32814\right) \left(g^2+1.95021
   g+0.959357\right)}{\left(g^2-2.39834 g+2.10061\right) \left(g^2+0.659846 g+2.72912\right) \left(g^2+1.32657
   g+0.931903\right) \left(g^2+2.88082 g+3.97752\right)} ,\NN\\
&& \scriptscriptstyle C_{7,6}^{(-1)}(g)
=\frac{16. (g+0.810667) (g+1.59844) \left(g^2-2.90106 g+5.78544\right) \left(g^2-1.07657 g+1.60638\right)}
{\left(g^2-2.58038
   g+2.22288\right) \left(g^2+1.02861 g+3.69991\right) \left(g^2+1.17219 g+0.975858\right) \left(g^2+3.31106
   g+4.00128\right)} ,\NN\\
&& \scriptscriptstyle C_{7,7}^{(-2/15)}(g)
=192 2^{7/15} 3^{4/5}
( 417942208512 \sqrt{6} g^{15}+14105549537280 \sqrt{6} g^{14}+237639327621120 \sqrt{6}
   g^{13}+2612318387742720 \sqrt{6} g^{12} \NN\\
&&\scriptscriptstyle \ \ \ \ \ \ \ \ \ \ \  
+20579350882757760 \sqrt{6} g^{11}+119353976853772896 \sqrt{6}
   g^{10}+491242721481413460 \sqrt{6} g^9+1139708471239040805 \sqrt{6} g^8 \NN\\
&&\scriptscriptstyle \ \ \ \ \ \ \ \ \ \ \  
-194072561423700480 g^7-37105930238402560
   g^6-60627299960881152 g^5+23255899837562880 g^4 \NN\\
&&\scriptscriptstyle \ \ \ \ \ \ \ \ \ \ \  
+19297624947425280 g^3+18248693270446080 g^2+5510614839459840
   g+1603087953297408 )^{-2/15} ,\NN\\
&& \scriptscriptstyle C_{7,8}^{(-1)}(g)
=\frac{16. (g-0.096973) \left(g^2-2.96364 g+6.26443\right) \left(g^2-1.00752 g+1.31953\right) \left(g^2+1.97223
   g+0.973597\right)}{(g-0.096973) \left(g^2-2.39445 g+2.10634\right) \left(g^2+0.684941 g+2.71423\right) \left(g^2+1.33568
   g+0.952398\right) \left(g^2+2.8749 g+3.94144\right)} ,\NN\\
&& \scriptscriptstyle C_{9,8}^{(-1)}(g)
=\frac{16. (g+0.873142) (g+1.15878) \left(g^2-2.95221 g+6.28731\right) \left(g^2-1.30535 g+1.00527\right) \left(g^2-0.805193
   g+1.07716\right)}
{\left(g^2-2.34581 g+1.94148\right) \left(g^2-1.12973 g+0.941604\right) \left(g^2+0.736498
   g+2.54333\right) \left(g^2+1.37219 g+1.03371\right) \left(g^2+2.83602 g+3.82186\right)} ,\NN\\
&& \scriptscriptstyle C_{9,10}^{(-1)}(g)
=16. (g+2.02265) \left(g^2-2.92912 g+6.43755\right) \left(g^2-2.28951 g+2.16034\right) \left(g^2-0.689709
   g+1.28544\right) \left(g^2+1.90831 g+0.911862\right) \NN\\
&&\scriptscriptstyle \ \ \ \ \ \ \ \ \ \ \  
\bigl[
(g+1.92569) \left(g^2-2.62295 g+2.06628\right) \left(g^2-1.76646
   g+2.49362\right) \left(g^2+0.893195 g+2.26641\right) \NN\\
&&\scriptscriptstyle \ \ \ \ \ \ \ \ \ \ \  
\left( g^2+1.45963 g+1.11812\right) \left( g^2 +2.63351g+3.49684\right) \bigr]^{-1} ,\NN\\
&& \scriptscriptstyle C_{10,9}^{(-1)}(g)
=16 (g+1.41271) \left(g^2-2.97662 g+6.55427\right) \left(g^2-2.78547 g+2.62556\right) \left(g^2-0.755849
   g+1.3408\right) \left(g^2+1.66102 g+0.701554\right) \NN\\
&&\scriptscriptstyle \ \ \ \ \ \ \ \ \ \ \  
\bigl[ (g+0.939731) \left(g^2-2.83704 g+2.26043\right) \left(g^2-2.19723
   g+2.86593\right) \left(g^2+0.992018 g+2.48976\right)  \NN\\
&&\scriptscriptstyle \ \ \ \ \ \ \ \ \ \ \  
\left(g^2+1.41627 g+1.15284\right) \left( g^2+2.74204 g+3.48984\right) \bigr]^{-1}
\end{\eqa}

\subsection{Four-dimensional $SU(3)$ pure Yang-Mills theory on lattice}
\label{app:YM}
\begin{\eqa}
&&\scriptscriptstyle P_{1,1}^{(-1)} (\beta )
=\frac{1600000 \pi ^2 (4 \beta +9)+43377282}{800000 \pi ^2 \left(4 \beta ^2+9 \beta +18\right)+2409849 (\beta +18)},\quad
P_{1,1}^{(-1/3)} (\beta )
=\left( \frac{\beta ^3}{8}-\frac{7229547 \beta ^2}{3200000 \pi ^2}+\frac{\beta }{6}+1\right)^{-1/3} ,\NN\\
&&\scriptscriptstyle P_{2,2}^{(-1)} (\beta )
=\frac{2 \beta ^2+6.23458 \beta +20.8014}{\beta ^3+2.50687 \beta ^2+7.39021 \beta +20.8014},\quad
P_{3,3}^{(-1)} (\beta )
=\frac{2 (\beta +3.3955) \left(\beta ^2-0.198893 \beta +12.0174\right)}{\left(\beta ^2-2.09682 \beta +9.61319\right) \left(\beta
   ^2+4.683 \beta +8.48936\right)} ,\NN\\
&&\scriptscriptstyle P_{4,4}^{(-1)} (\beta )
=\frac{2 \left(\beta ^2-2.50678 \beta +13.4054\right) \left(\beta ^2+5.7244 \beta +12.6824\right)}{(\beta +3.09342) \left(\beta
   ^2-3.54558 \beta +11.2267\right) \left(\beta ^2+3.05935 \beta +9.79082\right)},\NN\\
&&\scriptscriptstyle P_{5,5}^{(-1)} (\beta )
=\frac{2 (\beta +3.68567) \left(\beta ^2-4.08079 \beta +14.5847\right) \left(\beta ^2+3.61475 \beta +13.7196\right)}{\left(\beta
   ^2-4.63748 \beta +12.6358\right) \left(\beta ^2+1.38581 \beta +10.9814\right) \left(\beta ^2+5.86088 \beta +10.6298\right)} ,\NN\\
&&\scriptscriptstyle P_{6,6}^{(-1)} (\beta )
=\frac{2 \left(\beta ^2-5.18648 \beta +15.5937\right) \left(\beta ^2+1.581 \beta +14.6612\right) \left(\beta ^2+6.83275 \beta
   +14.4495\right)}{(\beta +3.39479) \left(\beta ^2-5.47262 \beta +13.8545\right) \left(\beta ^2-0.0916253 \beta +12.0825\right)
   \left(\beta ^2+4.78629 \beta +11.6262\right)} ,\NN\\
&&\scriptscriptstyle P_{7,7}^{(-1)} (\beta )
=\frac{2 (\beta +3.89835) \left(\beta ^2-5.9913 \beta +16.4715\right) \left(\beta ^2-0.140455 \beta +15.5324\right) \left(\beta
   ^2+5.47891 \beta +15.258\right)}{\left(\beta ^2-6.12185 \beta +14.911\right) \left(\beta ^2-1.34364 \beta +13.1167\right)
   \left(\beta ^2+3.49201 \beta +12.5578\right) \left(\beta ^2+6.60856 \beta +12.3918\right)} ,\NN\\
&&\scriptscriptstyle P_{8,8}^{(-1)} (\beta )
=\frac{2 \left(\beta ^2-6.59846 \beta +17.2502\right) \left(\beta ^2-1.55339 \beta +16.3433\right) \left(\beta ^2+3.93023 \beta
   +16.0062\right) \left(\beta ^2+7.48658 \beta +15.9063\right)}{(\beta +3.62668) \left(\beta ^2-6.63649 \beta +15.8369\right)
   \left(\beta ^2-2.39994 \beta +14.0937\right) \left(\beta ^2+2.20409 \beta +13.4255\right) \left(\beta ^2+5.86018 \beta
   +13.2095\right)} ,\NN\\
&&\scriptscriptstyle P_{9,9}^{(-1)} (\beta )
=\frac{2 (\beta +4.0654) \left(\beta ^2-7.07259 \beta +17.9526\right) \left(\beta ^2-2.71136 \beta +17.0902\right) \left(\beta
   ^2+2.43467 \beta +16.6901\right) \left(\beta ^2+6.55819 \beta +16.5611\right)}{\left(\beta ^2-7.05373 \beta +16.6601\right)
   \left(\beta ^2-3.29763 \beta +15.0097\right) \left(\beta ^2+1.00968 \beta +14.2272\right) \left(\beta ^2+4.8619 \beta
   +13.9689\right) \left(\beta ^2+7.14368 \beta +13.8747\right)} ,\NN\\
&&\scriptscriptstyle P_{10,10}^{(-1)} (\beta )
= 2 \left(\beta ^2-7.45432 \beta +18.593\right) \left(\beta ^2-3.66556 \beta +17.7643\right) \left(\beta ^2+1.08036 \beta
   +17.3105\right) \left(\beta ^2+5.38063 \beta +17.1591\right)  \NN\\
 &&\scriptscriptstyle \ \ \ \ \ \ \ \ \ \  \left(\beta ^2+7.9308 \beta +17.1034\right) \times
\bigl[ (\beta +3.80954)
   \left(\beta ^2-7.39967 \beta +17.4015\right) \left(\beta ^2-4.06586 \beta +15.8524\right) \left(\beta ^2-0.0631074 \beta
   +14.9646\right) \NN\\
 &&\scriptscriptstyle \ \ \ \ \ \ \ \ \ \ \  
\left(\beta ^2+3.78748 \beta +14.6692\right) \left(\beta ^2+6.5931 \beta +14.5475\right) \bigr]^{-1} ,\NN\\
&&\scriptscriptstyle P_{11,11}^{(-1)} (\beta )
=2 (\beta +4.19731) \left(\beta ^2-7.76923 \beta +19.1795\right) \left(\beta ^2-4.45532 \beta +18.3621\right) \left(\beta
   ^2-0.111216 \beta +17.8786\right) \left(\beta ^2+4.14811 \beta +17.709\right) 
  \NN\\
 &&\scriptscriptstyle \ \ \ \ \ \ \ \ \ \  
\left(\beta ^2+7.25733 \beta   +17.6376\right)
 \times
\bigl[ \left(\beta ^2-7.69195 \beta +18.0747\right) \left(\beta ^2-4.72437 \beta +16.6111\right) \left(\beta
   ^2-1.01157 \beta +15.6487\right) \NN\\
 &&\scriptscriptstyle \ \ \ \ \ \ \ \ \ \  
 \left(\beta ^2+2.72805 \beta +15.3187\right)
 \left(\beta ^2+5.80926 \beta +15.1747\right)
   \left(\beta ^2+7.54715 \beta +15.1161\right) \bigr]^{-1} ,\NN\\
&&\scriptscriptstyle P_{12,12}^{(-1)} (\beta )
=2 \left(\beta ^2-8.03357 \beta +19.7176\right) \left(\beta ^2-5.11101 \beta +18.8907\right) \left(\beta ^2-1.14735 \beta
   +18.4125\right) \left(\beta ^2+2.95613 \beta +18.224\right) \NN\\
 &&\scriptscriptstyle \ \ \ \ \ \ \ \ \ \    
 \left(\beta ^2+6.34655 \beta +18.1394\right) \left(\beta
   ^2+8.25925 \beta +18.105\right) 
\times 
\bigl[ (\beta +3.95793) \left(\beta ^2-7.94228 \beta +18.6885\right) \left(\beta ^2-5.28747 \beta
   +17.2852\right) \NN\\
 &&\scriptscriptstyle \ \ \ \ \ \ \ \ \ \  
\left(\beta ^2-1.84571 \beta +16.2982\right) \left(\beta ^2+1.72728 \beta +15.9291\right) \left(\beta
   ^2+4.92134 \beta +15.7648\right) \left(\beta ^2+7.12847 \beta +15.6882\right) \bigr]^{-1},\NN\\
 &&\scriptscriptstyle P_{13,13}^{(-1)} (\beta )
= 2 (\beta +4.30767) \left(\beta ^2-8.2584 \beta +20.2116\right) \left(\beta ^2-5.65839 \beta +19.3661\right) \left(\beta
   ^2-2.04777 \beta +18.9269\right) \left(\beta ^2+1.84563 \beta +18.712\right) \NN\\
 &&\scriptscriptstyle \ \ \ \ \ \ \ \ \ \  
\left(\beta ^2+5.33917 \beta +18.6147\right)
   \left(\beta ^2+7.75383 \beta +18.5694\right)
\times
\bigl[ \left(\beta ^2-8.15878 \beta +19.2496\right) \left(\beta ^2-5.76856 \beta
   +17.8857\right) \NN\\
 &&\scriptscriptstyle \ \ \ \ \ \ \ \ \ \  
 \left(\beta ^2-2.58215 \beta +16.9294\right) \left(\beta ^2+0.801766 \beta +16.5083\right) \left(\beta
   ^2+4.00525 \beta +16.3235\right) 
\left(\beta ^2+6.50354 \beta +16.232\right) \left(\beta ^2+7.87025 \beta +16.1925\right) \bigr]^{-1} ,\NN\\
 &&\scriptscriptstyle P_{14,14}^{(-1)} (\beta )
 = 9.53730\times 10^{-10} ((\beta -8.45194) \beta +20.6669) ((\beta -6.12027) \beta +19.8053)
   ((\beta -2.83501) \beta +19.4247) (\beta  (\beta +0.829692)+19.1718) \NN\\
 &&\scriptscriptstyle \ \ \ \ \ \ \ \ \ \  
(\beta  (\beta +4.31495)+19.0624) (\beta  (\beta
   +7.03398)+19.0077) (\beta  (\beta +8.52139)+18.9842) 
 \NN\\
 &&\scriptscriptstyle \ \ \ \ \ \ \ \ \  
  \times \bigl[
4.76865 \times 10^{-10} \beta  ((\beta
   -8.26905) \beta +19.634)
((\beta -5.95963) \beta +18.4991) ((\beta -2.75599) \beta +17.9289) (\beta  (\beta
   +0.765819)+17.5293)  \NN\\
 &&\scriptscriptstyle \ \ \ \ \ \ \ \ \ \  
(\beta  (\beta +4.09727)+17.3444)  (\beta  (\beta +6.69248)+17.248)(\beta  (\beta +8.11148)+17.2058)+1 \bigr]^{-1} ,\NN\\
 &&\scriptscriptstyle P_{15,15}^{(-1)} (\beta )
=(1.86410\times 10^{-10} \beta  ((\beta -8.50134) \beta +20.8591) ((\beta -6.18268) \beta
   +20.0391) ((\beta -2.91328) \beta +19.8405) (\beta  (\beta +0.812334)+19.6054) \NN\\
 &&\scriptscriptstyle \ \ \ \ \ \ \ \ \ \  
(\beta  (\beta +4.3502)+19.5008) (\beta  (\beta
   +7.10958)+19.4469) (\beta  (\beta +8.61881)+19.4234)+1 )  \NN\\
 &&\scriptscriptstyle \ \ \ \ \ \ \ \ \  
\times \bigl[
\beta  (9.32052\times 10^{-11} \beta    ((\beta -8.33651) \beta +19.8869)
((\beta -6.05378) \beta +18.8207) ((\beta -2.87389) \beta +18.5005)   \NN\\
 &&\scriptscriptstyle \ \ \ \ \ \ \ \ \ \  
(\beta  (\beta   +0.734611)+18.149)  (\beta  (\beta +4.14726)+17.9777)
(\beta  (\beta +6.80597)+17.8853) (\beta  (\beta +8.25954)+17.8441)+0.27881)+1 \bigr]^{-1}
\end{\eqa}

\subsection{$c=1$ string theory at self-dual radius}
\label{app:string}

\begin{\eqa}
&&\scriptscriptstyle F_{1,1}^{(-1)} (\mu )
= \frac{0.0701003 \mu +0.0845789}{0.841203 \mu ^2+1.81409 \mu +1} ,\quad
 F_{1,1}^{(-1/3)} (\mu )
=0.0833333 \left( (\mu +0.839474) \left(\mu ^2+2.01053 \mu +1.13937\right) \right)^{-1/3} ,\NN\\
&&\scriptscriptstyle F_{2,2}^{(-1)} (\mu )
 =\frac{0.0833333 \mu ^2+0.217718 \mu +0.160546}{\mu ^3+3.56261 \mu ^2+4.44436 \mu +1.89818} ,\quad
 F_{2,2}^{(-1/3)} (\mu )
= 0.144274
 \left( \frac{(\mu +0.827577) (\mu +5.19357) \left(\mu ^2+2.01816 \mu
   +1.1548\right)}{0.192704 \mu +1} \right)^{-1/3} ,\NN\\
&&\scriptscriptstyle F_{2,2}^{(-1/5)} (\mu )
=0.0833333\left( (\mu +1.40603) \left(\mu ^2+1.42212 \mu +0.535002\right) \left(\mu ^2+1.92185 \mu   +1.23434\right) \right)^{-1/5}  ,\NN\\
&&\scriptscriptstyle F_{3,3}^{(-1)} (\mu )
=\frac{0.0833333 (\mu +1.7262) \left(\mu ^2+2.93648 \mu +2.57422\right)}{(\mu +1.0795) (\mu +1.65115) \left(\mu
   ^2+2.88203 \mu +2.45633\right)} ,\NN\\
&&\scriptscriptstyle   F_{3,3}^{(-1/3)} (\mu )
   =0.105268
\bigl[ \frac{(\mu +0.880493) \left(\mu ^2+2.0225 \mu +1.1097\right) \left(\mu ^2+2.22788 \mu
   +1.97319\right)}{0.496101 \mu ^2+1.13154 \mu +1} \bigr]^{-1/3}  ,\NN\\
&&\scriptscriptstyle   F_{3,3}^{(-1/7)} (\mu )
=0.0833333\left( (\mu +0.646175) \left(\mu ^2+1.39726 \mu +0.57416\right) \left(\mu ^2+1.92491 \mu   +1.3455\right) \left(\mu ^2+2.68166 \mu +1.80564\right) \right)^{-1/7} ,\NN\\
&&\scriptscriptstyle   F_{4,4}^{(-1)} (\mu )
=\frac{0.0833333 \left(\mu ^2+2.74502 \mu +2.0421\right) \left(\mu ^2+2.94932 \mu +3.46359\right)}{(\mu
   +1.06811) \left(\mu ^2+2.6273 \mu +1.8885\right) \left(\mu ^2+2.94893 \mu +3.45485\right)} ,\NN\\
&&\scriptscriptstyle   F_{4,4}^{(-1/3)} (\mu )  
=0.0833333 \bigl[ \frac{\left(\mu ^2+1.65869 \mu +0.693832\right) \left(\mu ^2+2.07696 \mu
   +1.18885\right) \left(\mu ^2+2.44411 \mu +2.36393\right)}{(\mu +0.840978) \left(\mu ^2+2.48878 \mu
   +2.42416\right)} \bigr]^{-1/3},\NN\\
&&\scriptscriptstyle   F_{4,4}^{(-1/9)} (\mu )  
=\frac{0.0833333}{\sqrt[9]{(\mu +0.596352) \left(\mu ^2+1.23613 \mu +0.430642\right) \left(\mu ^2+1.42961 \mu
   +0.746351\right) \left(\mu ^2+2.01326 \mu +1.62233\right) \left(\mu ^2+3.27465 \mu +2.81391\right)}} ,\NN\\
&&\scriptscriptstyle   F_{5,5}^{(-1)} (\mu )  
=\frac{0.0833333 (\mu +0.0152634) \left(\mu ^2+2.7458 \mu +2.04335\right) \left(\mu ^2+2.95129 \mu
   +3.46744\right)}{(\mu +0.0152634) (\mu +1.06814) \left(\mu ^2+2.62801 \mu +1.88957\right) \left(\mu
   ^2+2.95094 \mu +3.45875\right)} ,\NN\\
&&\scriptscriptstyle   F_{5,5}^{(-1/11)} (\mu )  
=0.0833333
 \bigl[ (\mu +1.77794) \left(\mu ^2+1.07909 \mu +0.300369\right) \left(\mu ^2+1.13316 \mu
   +0.416661\right) \left(\mu ^2+1.33834 \mu +0.785355\right) \NN\\
 &&\scriptscriptstyle \ \ \ \ \ \ \ \ \ \  
\left(\mu ^2+1.96496 \mu +1.7284\right)
   \left(\mu ^2+3.15651 \mu +2.8123\right) \bigr]^{-\frac{1}{11}}
\end{\eqa}

\providecommand{\href}[2]{#2}\begingroup\raggedright\endgroup

\end{document}